\documentclass[11pt,letterpaper]{article}

\usepackage{latexsym,multirow}
\usepackage{amssymb,amsmath, bm}
\usepackage{graphicx}
\usepackage{booktabs}
\usepackage[e]{esvect}

\usepackage{enumerate}

\usepackage{natbib}
\usepackage[pdftex, bookmarksopen=true, bookmarksnumbered=true,
pdfstartview=FitH, breaklinks=true, urlbordercolor={0 1 0}, citebordercolor={0 0 1}]{hyperref}
\usepackage{colortbl}
\usepackage{subfigure}

\usepackage{dcolumn}
\newcolumntype{.}{D{.}{.}{-1}}
\newcolumntype{d}[1]{D{.}{.}{#1}}
\usepackage{theorem}
\theoremstyle{plain}
\theoremheaderfont{\scshape}
\newtheorem{assumption}{Assumption}
\newtheorem{example}{Example}

\newtheorem{corollary}{Corollary}

\newtheorem{proposition}{Proposition}
\newtheorem{theorem}{Theorem}
\newtheorem{remark}{Remark}

\newtheorem{lemma}{Lemma}
\newtheorem{constraint}{Constraint}

\usepackage{rotating}

\usepackage[compact]{titlesec}

\allowdisplaybreaks

\newcommand\spacingset[1]{\renewcommand{\baselinestretch}%
{#1}\small\normalsize}

\newcommand{\blind}{0}

\newcommand*{\QEDB}{\hfill\ensuremath{\square}}

\newcommand{\bone}{\mathbf{1}}

\newcommand{\pr}{\mathbb{P}}
\newcommand{\var}{\textnormal{var}}
\newcommand{\argmin}{\textnormal{argmin}}
\newcommand{\cov}{\textnormal{cov}}

\newcommand{\bW}{W}
\newcommand{\bw}{w}

\newcommand{\oY}{\bar{Y}}
\newcommand{\wY}{\widehat{Y}}

\newcommand{\oS}{\vv{Z}}
\newcommand{\os}{\vv{z}}
\newcommand{\bS}{S}
\newcommand{\bY}{Y}
\newcommand{\obY}{\bar{\bY}}
\newcommand{\wbY}{\widehat{\bY}}
\newcommand{\btheta}{\theta}

\newcommand{\E}{\mathbb{E}}
\newcommand{\bX}{X}

\newcommand{\tbW}{\widetilde{W}}
\newcommand{\tw}{\tilde{w}}

\newcommand{\bgamma}{\bm{\gamma}}

\newcommand{\wbtheta}{\widehat{\btheta}}

\begin{document} 

\newcommand{\tit}{Principled analysis of crossover designs: causal effects, efficient estimation, and robust inference
}
%
%
\spacingset{1.25}

\if0\blind

{\title{\bf\tit}

  \author{
     Zhichao Jiang\thanks{School of Mathematics, Sun Yat-sen University, Guangzhou,  Guangdong 510275, China. Email: \href{mailto:jiangzhch7@mail.sysu.edu.cn}{jiangzhch7@mail.sysu.edu.cn} 
      } \hspace{.5in}
      Peng Ding \thanks{Department of Statistics, University of California, Berkeley, CA 94720, USA. Email: \href{mailto:pengdingpku@berkeley.edu}{pengdingpku@berkeley.edu}}
       }

\date{
\today
}

\maketitle

}\fi

\if1\blind
\title{\bf \tit}

\maketitle
\fi

\pdfbookmark[1]{Title Page}{Title Page}

\thispagestyle{empty}
\setcounter{page}{0}
         
\begin{abstract}
Crossover designs randomly assign each unit to receive a sequence of treatments.  By comparing outcomes within the same unit, these designs can effectively eliminate between-unit variation and facilitate the identification of both instantaneous effects of current treatments and carryover effects from past treatments. They are widely used in traditional biomedical studies and are increasingly adopted in modern digital platforms. However, standard analyses of crossover designs often rely on strong parametric models, making inference vulnerable to model misspecification. This paper adopts a design-based framework to analyze general crossover designs.  We make two main contributions. First, we use potential outcomes to formally define the causal estimands and assumptions on the data-generating process. For any given type of crossover design and assumptions on potential outcomes, we outline a procedure for identification and estimation, emphasizing the central role of the treatment assignment mechanism in design-based inference. Second, we unify the analysis of crossover designs using least squares, with restrictions on the coefficients and weights on the units.
Based on the theory, we recommend the specification of the regression function, weighting scheme, and coefficient restrictions to assess identifiability, construct efficient estimators, and estimate variances in a unified fashion. Crucially, the least squares procedure is simple to implement, and yields not only consistent and efficient point estimates but also valid variance estimates even when the working regression model is misspecified.

\end{abstract}


\clearpage
\spacingset{1.5}

\section{Introduction}
The crossover design, also known as a repeated measurements design, change-over design, or switchback design, is a type of experimental design in which each unit receives the treatments at multiple time periods in a predetermined sequence \citep{cochran1939long,cochran1950experimental, senn2002cross,jones2003design, milliken2009analysis,basse2023minimax,bojinov2023design,arkhangelsky2024causal}.
Because each unit is exposed to different treatments over time, these designs enable within-unit comparisons that eliminate between-unit variation, thereby enhancing statistical efficiency and allowing for the estimation of various causal effects.
Crossover designs are widely used in medical research, as well as in digital experimentation \citep[e.g.,][]{brown1980crossover,cook2007introduction}. Mills et al. \cite{mills2009design} review 519 randomized experiments published in December 2000 and find 166 of them employed crossover designs.
Despite their advantages, the sequential administration of treatments introduces analytical challenges, including the issues of carryover effects and within-unit correlation. 
Failing to account for these issues may lead to invalid statistical analyses.

The statistical analysis of crossover designs has traditionally relied on parametric models to address these challenges. Depending on the type of outcome, researchers often adopt linear or generalized linear models that incorporate three main components: (1) random effects to capture within-unit correlations, (2) period effects to adjust for time trends, and (3) treatment effects. The treatment effects typically include parameters for the overall treatment mean, instantaneous effects at each time period, and carryover effects from previously administered treatments, depending on the order of carryover effects specified by the analyst.  
This modeling approach has led to a substantial body of literature, including book-length treatments \citep{senn2002cross,jones2003design}, dedicated book chapters \citep{hinkelmann2007design,milliken2009analysis,patterson2017bioequivalence}, and a special issue at the journal {\it Statistical Methods in Medical Research} published in 1994 \citep{murray1994editorial}. This line of work examines crossover designs with varying numbers of periods and treatment conditions, different outcome types, and complex data structures, and also develops optimal design strategies under various inferential goals \citep[e.g.,][]{kershner1981two,hedayat2006optimal,kunert2008optimal,xu2018incorporating,jankar2021optimal,wang2021analysis,basse2023minimax,madsen2023unbiased}.
While widely used due to their simplicity for implementation, these model-based approaches present three main limitations. 
First, they are sensitive to model misspecification, which may lead to biased estimates and invalid inference when the model is misspecified. 
Second, the causal estimand is defined through the model rather than independently of it. This conflates causal assumptions with modeling assumptions used for estimation, making the causal interpretation of the estimand dependent on the correctness of the model specification.
Third, the models often lack the flexibility to accommodate a broader range of scientific assumptions or to adapt to diverse experimental configurations. For instance, many models assume that treatment effects are constant across periods, which is often too strong in practice.

In contrast, modern causal inference has increasingly adopted the design-based framework to analyze complex randomized experiments. This framework conditions on all potential outcomes and covariates, and treats randomization as the basis for valid inference \citep{neyman1990application,imbens2015causal,li2017general}. 
A key strength of the design-based framework is its robustness to model misspecification, as it avoids strong parametric assumptions on the data-generating process. 
It also enables researchers to incorporate scientific knowledge through explicitly stated assumptions on the potential outcomes. 
 The design-based framework has been applied to a variety of experimental settings \citep[e.g.,][]{lin2013agnostic,dasgupta2015causal,fogarty2018regression,liu2020regression,bojinov2021panel,imai2021causal,pashley2021insights,pashley2022block,zhao2022reconciling,zhao2022regression,bojinov2023design,zhao2023covariate}, and has emerged as a principled foundation for analyzing randomized designs.

We adopt the design-based framework to analyze crossover designs. Within this framework, we use potential outcomes to formally define both instantaneous and carryover effects at each period and explicitly state the assumptions commonly adopted in crossover studies (Section~\ref{sec::setup}).
These assumptions include the absence of certain orders of carryover effects and the invariance of treatment effects across time periods. Our definitions and assumptions parallel those used in the design-based analyses of time-series and panel experiments \citep[e.g.,][]{bojinov2021panel,bojinov2023design,lin2025unifying,ni2025decision,ni2025enhancing}.

Crossover designs present unique inferential challenges. While causal estimands are defined with respect to potential outcomes under all possible treatment sequences, experimental designs may  implement only a subset of these sequences. Consequently, both identification and estimation depend on the specific design and the set of assumptions imposed. This paper develops a unified strategy for analyzing general crossover designs under various sets of assumptions. This strategy addresses three fundamental questions: (1) Does a linear unbiased estimator exist for the causal quantity of interest? (2) If the answer to question (1) is ``yes,'' what is the most efficient estimator? (3) How should valid inference be conducted?

To build intuition for identification and estimation, we begin with two-period crossover designs: one with four treatment sequences (Section~\ref{sec::2p4t}) and the other with two sequences (Section~\ref{sec::2p2t}). The setup in Section~\ref{sec::2p2t} is similar to Shi and Ye \cite{shi2024behavioral}, which focuses on the two-period, two-sequence design under a super-population framework. 
Through the analysis of these two designs, we outline a general procedure for studying the identification and estimation of causal effects under the design-based framework. For each causal estimand, we assess whether a linear unbiased estimator exists under the stated assumptions. When such estimators exist, we derive the most efficient one by optimally combining all the unbiased estimators.

The analyses in Sections~\ref{sec::2p4t} and~\ref{sec::2p2t} reveal the substantial differences in identifiability and efficiency across design types and assumptions. These differences illustrate the challenge of relying on case-by-case derivations and motivate the need for a unified analytical approach.
To address this difficulty, Section~\ref{sec::reg-general} introduces a regression-based approach grounded in the design-based framework. This approach uses restricted weighted least squares to unify identification, estimation, and variance estimation. The regression formulation encodes the design structure through the regressor matrix (we do not use the standard terminology ``design matrix'' to avoid confusion under the ``design-based'' framework), while the assumptions are incorporated as linear restrictions on the regression coefficients. The weight matrix is chosen to optimize efficiency and can be consistently estimated from the observed data.

We establish a sufficient condition for the identification of all causal effects of interest based on the regressor matrix and the restriction matrix. Under this condition,  the restricted weighted least squares estimator is consistent for the causal effects. For variance estimation, we propose the Eicker--Huber--White (EHW) variance estimator under restricted weighted least squares, and show that it provides a conservative estimate of the design-based variance, and yields consistent inference when individual causal effects are constant across units.
Importantly, all results rely entirely on the design-based framework. The regression model serves solely as a tool for constructing estimators, rather than as a set of assumptions on the outcome generation process. This distinction aligns with earlier work on regression adjustment in randomized experiments  \citep[see, e.g., ][]{freedman2008regression,lin2013agnostic,miratrix2013adjusting,imbens2015causal,bloniarz2016lasso,fogarty2018regression,liu2020regression,guo2023generalized}.

Section~\ref{sec::extension} extends the regression-based analysis in three directions: arbitrary weighting schemes, population-level versions of the assumptions, and testing procedures for the assumptions.
Section~\ref{sec::app} applies the proposed approach to data from a three-period crossover experiment. Section S3 of the online supplementary material develops the theory of restricted weighted least squares in detail. These results extend classical theory on weighted and restricted least squares, including closed-form expressions for the estimators and their variances, as well as a generalized version of the Gauss–Markov theorem under linear restrictions. 
Because the outcome for each unit is vector-valued, the restricted weighted least squares regression can be interpreted as a generalized estimating equations approach with parameter restrictions \citep{liang1986longitudinal}. It can also be viewed as a form of seemingly unrelated regression that incorporates both coefficient restrictions and unit-specific weights \citep{zellner1962efficient}.
The results on restricted weighted least squares are not only crucial for analyzing crossover designs in this paper but also of independent interest for other statistical problems.

\section{Notation and Assumptions}
\label{sec::setup}
 We consider the crossover designs involving two treatment conditions $A$ and $B$, although the concepts naturally extend to settings with more than two treatment conditions.  
For unit $i$,  let $\oS_i$ be the sequence of treatments assigned to unit $i$ over a total of $T$ periods.  
For example, $\oS_i=AB$  indicates that treatment $A$ is assigned to unit $i$ in period 1 and treatment $B$ is assigned to the same unit in period 2, whereas $\oS_i=AAA$ denotes that the same treatment $A$ is administered in all three periods.
In crossover designs, units are randomly assigned to one of the treatment sequences.  We use  $N$ to denote the total number of units and $N_{\os}$ to denote the number of units assigned to treatment sequence $\os \in \mathcal{S}^\textup{obs}$, where $\mathcal{S}^\textup{obs}$ is the set of all treatment sequences implemented in the experiment.  With $T$ treatment periods, there are $2^T$ possible treatment sequences in total, and $\mathcal{S}^\textup{obs}$ is a subset of the complete set $\mathcal{S}=\{(z_1\cdots z_T): z_1,\ldots,z_T \in \{A,B\}\}$. While it is possible for all $2^T$ sequences to be implemented in some settings, this becomes practically infeasible as $T$ increases. Consequently, in applications with large $T$, the implemented set $\mathcal{S}^\textup{obs}$ typically includes only a small subset of the full set.
We focus on {\it complete randomization}, in which $N_{\os}$ is fixed in design for all $\os$. Under complete randomization, the joint treatment probability of the treatment sequences for all units is given by
 \begin{equation*}
      \pr(\oS_1 = \os_1,\ldots,\oS_N = \os_N) \ = \ \frac{ \prod_{\os \in \mathcal{S}^\textup{obs}} (N_{\os} !) }{N!}
    \end{equation*}
    for all $(\os_1,\ldots,\os_N)$ such that $\sum_{i=1}^N \bone(\oS_i=\os) = N_{\os}$ for each $\os \in \mathcal{S}^\textup{obs}$.

Let $Y_{it}$ be the observed outcome of unit $i$ at period $t$.
 We adopt the potential outcomes framework to define causal effects \citep{neyman1990application}. We assume no interference across units, meaning that the treatment sequence of one unit does not affect the outcomes of other units. However, we allow for within-unit carryover effects,  so that a unit’s outcome at one time point may depend on treatments received at other time points.  Under this setup,
 we define the potential outcome $Y_{it}(\os)$  as the outcome that unit $i$ would experience at time $t$ if it were assigned treatment sequence $\os$. The observed outcome is a deterministic function of the potential outcomes and the treatment sequence, i.e., $Y_{it}= Y_{it}(\os)$ if $\oS_i=\os$. 
Let $\oY_{t}(\os)$ be the average of the potential outcomes at period $t$ under treatment sequence $\os$, i.e.,
\begin{eqnarray*}
\oY_{t}(\os) = \frac{1}{N} \sum_{i=1}^N Y_{it}(\os).
\end{eqnarray*}
Let $\os_{[t_1,t_2]}$ be the subsequence of $\os$ from period $t_1$ to period $t_2$ with $\os_{[t_1,t_2]}=\emptyset$ if $t_1 > t_2$. For example, for  $\os=ABBA$, $\os_{[1,3]}=ABB$,  $\os_{[2,2]}=B$, and  $\os_{[3,4]}=BA$. We assume the potential outcomes do not depend on treatments in later periods.
\begin{assumption}[No anticipation]
\label{asm::noanticipation}
$Y_{it}(\os)=Y_{it}(\os')$ if $\os_{[1,t]} = \os'_{[1,t]} $.
\end{assumption}
Assumption~\ref{asm::noanticipation} is plausible when the units do not know what treatment they will receive at later periods. It would be violated if the units can predict the future treatments.
Assumption~\ref{asm::noanticipation} is commonly adopted for studies with time-varying treatment \citep[e.g.,][]{bojinov2019time, bojinov2023design,jiang2023instrumental}.
Throughout the paper, we focus on the scenarios in which Assumption \ref{asm::noanticipation} is plausible. 
Under Assumption~\ref{asm::noanticipation},
we can  simplify the individual potential outcome $Y_{it}(\os)$ as $Y_{it}(\os_{[1,t]})$ and the average potential outcome $\oY_{t}(\os)$ as $\oY_{t}(\os_{[1,t]})$.

We define the average treatment effect at period 1 as
\begin{eqnarray*}
\tau_1=\oY_{1}(A)- \oY_{1}(B),
\end{eqnarray*}
which compares the potential outcomes at period 1 under treatment conditions $A$ and $B$.
For $t \geq 2$, we then define the conditional
average treatment effect at period $t$  as
\begin{eqnarray*}
\tau_t(\os_{[1,t-1]})&=&\oY_{t}(\os_{[1,t-1]}A)- \oY_{t}(\os_{[1,t-1]}B),
\end{eqnarray*}
which compares the potential outcomes at period $t$ with different treatment conditions at period $t$ while fixing the treatment sequence from period 1 to period $t-1$. Consequently, we have $2^{t-1}$ conditional treatment effects at period $t$ and $\sum_{t=1}^T 2^{t-1} = 2^T-1$ conditional treatment effects across all $T$ periods.  
This definition is analogous to the conditional causal effect in the context of factorial experiments \citep{wu2011experiments, zhao2022regression}.

In crossover designs, another quantity of interest is the carryover effect, which quantifies the impact of a previous treatment on the outcome in subsequent periods.
We 
define the $k$-th order   conditional carryover effect at period $t$
as
\begin{eqnarray*}
\tau_t^k( \os_{[1,t-k-1]}, \os_{[t-k+1,t]})&=& \oY_t(\os_{[1,t-k-1]}A\os_{[t-k+1,t]})- \oY_t(\os_{[1,t-k-1]}B\os_{[t-k+1,t]}).
\end{eqnarray*}
It compares the potential outcomes at period $t$ under two treatment sequences that differ only in the treatment condition at period 
$t-k$. Consequently, there are $2^{t-1}$ different $k$-th order carryover effects at period $t$.  
When $k=0$, the $0$-th order carryover effect reduces to the treatment effect at $t$, i.e., $\tau_t^0(\os_{[1,t-1]})=\tau_t(\os_{[1,t-1]})$. 

Given the large number of causal quantities in crossover designs, practitioners often invoke assumptions to reduce the number of parameters to facilitate estimation and interpretation. We consider two assumptions that 
are commonly used or implicitly imposed in many analyses of crossover designs.
\begin{assumption}[No more than $k$-th order carryover effect]
\label{asm::nocarryover}
$Y_{it}(\os)=Y_{it}(\os')$ if  $\os_{[t-k+1,t]}=\os'_{[t-k+1,t]}$,  that is, $z_{t-j+1}=z'_{t-j+1}$ for $j=1,\ldots,k$.
\end{assumption}
Assumption~\ref{asm::nocarryover} posits that the carryover effects beyond order $k$ are zero.
This assumption holds if the effect of treatments wears off after $k$ time periods. 
In crossover designs,  a washout period is often included, during which no treatment is given \citep[e.g.,][]{sun2025dynamic}. 
This period allows the effect of previous treatments to dissipate before subsequent treatments begin, reducing the extent of carryover effects.
 Assumption~\ref{asm::nocarryover} is more credible when treatment effects decay rapidly relative to the spacing between treatment periods, such as in settings with short-acting interventions or sufficiently long washout periods. However, it may be violated when treatments have persistent biological or behavioral effects, such as cumulative exposure, delayed responses, or learning and adaptation over time. In such cases, higher-order carryover effects may remain even when washout periods are implemented.
 
 Under Assumption~\ref{asm::nocarryover} with order $k$, we can simplify $Y_{it}(\os)$ as $Y_{it}(\os_{[t-k+1,t]})$, which depends only on the treatment conditions from period $t-k+1$ to $t$.  When $k=1$, Assumption~\ref{asm::nocarryover}  rules out all carryover effects, simplifying the potential outcomes to $Y_{it}(z_t)$.
 
The restriction in Assumption~\ref{asm::nocarryover} leverages the temporal ordering of treatments, based on the idea that more recent treatments are more likely to influence the outcome than those administered further in the past. This assumption effectively reduces the number of potential outcomes at time $t$ from $2^t$ to $2^k$ for $t > k$.

\begin{assumption}[Time invariant effect of treatment sequences with length $k$]
\label{asm::invarianteffect}
Under Assumption~\ref{asm::nocarryover} with no more than $k$-th order, 
$$Y_{it}(\os_{[t-k+1,t]})-Y_{it}(\os'_{[t-k+1,t]})= Y_{it'}(\os_{[t'-k+1,t']})-Y_{it'}(\os'_{[t'-k+1,t']})$$  
 for any periods $t'>t \geq k$,
if  $\os_{[t-k+1,t]}=\os_{[t'-k+1,t']}$ and $\os'_{[t-k+1,t]}=\os'_{[t'-k+1,t']}$.
\end{assumption}
Assumption~\ref{asm::invarianteffect}  requires that $Y_{it}(\os_{[t-k+1,t]})-Y_{it}(\os'_{[t-k+1,t]})$ does not depend on $t$.
 In other words, for any two time points $t$ and $t’$ such that  $t'>t \geq k$, if the treatment sequences leading up to $t$ and $t’$ are identical, then the corresponding effects are also identical. 
Assumption~\ref{asm::invarianteffect} requires Assumption~\ref{asm::nocarryover} to hold with the same order $k$, so that treatment effects at different periods are defined based on treatment histories of the same length. The two orders can differ technically, but we adopt a common $k$ for coherence and interpretability. 
Assumption~\ref{asm::invarianteffect} is often implicitly imposed in traditional model-based analyses of crossover designs \citep[e.g.,][]{senn2002cross,jones2003design}. Assumption~\ref{asm::invarianteffect} with $k=1$ has also been used in the context of staggered treatment adoption to facilitate the identification and estimation of treatment effects \citep{athey2022design}. Assumption~\ref{asm::invarianteffect} is more plausible when the outcome process is stationary and the effect of treatment does not interact with time. It may be violated in the presence of time trends, period effects, or dynamic responses to treatment, such as fatigue, habituation, or changing baseline conditions. 
We give two examples to illustrate Assumption~\ref{asm::invarianteffect}.
\begin{example}
Consider a crossover design with two periods. Suppose that Assumption~\ref{asm::nocarryover} holds with order $1$, so that $Y_{i1}(\os) = Y_{i1}(z_1)$ and $Y_{i2}(\os) = Y_{i2}(z_2)$ for any treatment sequence $\os=z_1z_2$.  Then, Assumption~\ref{asm::invarianteffect} means that $Y_{i1}(A)-Y_{i1}(B)=Y_{i2}(A)-Y_{i2}(B)$, which implies that the effect of treatment condition $A$ versus $B$ is the same for each unit $i$ at time periods $1$ and $2$.
\end{example}

\begin{example}
Consider a crossover design with $T$ periods. Suppose that Assumption~\ref{asm::nocarryover} holds with order $2$, so that $Y_{it}(\os) = Y_{it}(\os_{[t-1,t]})$ for any treatment sequence $\os=z_1\cdots z_T$ and $t\leq T$. Then, Assumption~\ref{asm::invarianteffect} means that $Y_{it}(z_1z_2)-Y_{it}(z_1'z_2')=Y_{it'}(z_1z_2)-Y_{it'}(z_1'z_2')$ for any  $z_1z_2, z_1'z_2' \in \{AA,AB,BA,BB\}$ and any $t,t' \leq T$, which implies that the effect of treatment sequence  $z_1z_2$ versus $z_1'z_2'$ for each unit $i$ is the same across different periods.
\end{example}

Assumptions~\ref{asm::nocarryover}~and~\ref{asm::invarianteffect} are often jointly imposed implicitly in model-based analyses of crossover designs. However, they concern different aspects of the potential outcomes.
Assumption~\ref{asm::nocarryover} restricts the impact of earlier treatments on current outcomes, thereby ruling out carryover effects. In contrast, Assumption~\ref{asm::invarianteffect} imposes temporal invariance in the effect of treatment sequences, requiring that these effects remain stable over time.  Assumption~\ref{asm::nocarryover} is plausible if previous treatments do not have any lingering effects, while Assumption~\ref{asm::invarianteffect} is plausible if the effect of treatment sequences does not vary systematically over time.



In this paper, we study the identification, estimation, and inference of treatment effects in crossover designs within the design-based framework. 
We use Assumptions~\ref{asm::nocarryover}~and~\ref{asm::invarianteffect} to  illustrate how restrictions on potential outcomes enable identification and estimation, while allowing for more general sets of assumptions within the proposed framework. We  provide a comprehensive and general set of results under various combinations of the assumptions.
We start with two crossover designs with two periods and then establish the theories for general designs.

 \section{Crossover designs with two periods and four sequences}
 \label{sec::2p4t}
The crossover design with two periods and four treatment sequences is referred to as the $2^2$ crossover design \citep{balaam1968two}. 
 In this case, units are randomly assigned to one of the four treatment sequences: $AA$, $AB$, $BA$, and $BB$. 
Each unit has eight potential outcomes $Y_{i1}(AA)$, $Y_{i1}(AB)$, $Y_{i1}(BA)$, $Y_{i1}(BB)$, $Y_{i2}(AA)$,  $Y_{i2}(AB)$,  $Y_{i2}(BA)$,  and $Y_{i2}(BB)$.  
Under Assumption~\ref{asm::noanticipation}, the four potential outcomes at period 1 are not affected by the treatment at the second period and depend only on the treatment at period 1. Therefore, we can simplify $Y_{i1}(AA)$ and $Y_{i1}(AB)$ as $Y_{i1}(A)$, and $Y_{i1}(BA)$ and $Y_{i1}(BB)$ as $Y_{i1}(B)$. 
 In contrast, the latter four potential outcomes depend on the treatments at both periods. In experiments with two time periods, we observe only two of them for each unit.  As shown in Table~\ref{tab::po}, for units assigned to $AA$, we observe $Y_{i1}=Y_{i1}(A)$ and $Y_{i2}=Y_{i2}(AA)$; for units  assigned to $AB$, we observe $Y_{i1}=Y_{i1}(A)$ and $Y_{i2}=Y_{i2}(AB)$; for units assigned to $BA$, we observe $Y_{i1}=Y_{i1}(B)$ and $Y_{i2}=Y_{i2}(BA)$; and for units assigned to $BB$, we observe $Y_{i1}=Y_{i1}(B)$ and $Y_{i2}=Y_{i2}(BB)$.
\begin{table}[htp]
\caption{Potential outcomes under Assumption~\ref{asm::noanticipation} and the observed ones under each treatment sequence, where \checkmark indicates observed potential outcomes and ? indicates unobserved potential outcomes.}
\begin{center}
\begin{tabular}{cccccccc}
\hline
  & Treatment     &  \multicolumn{6}{c}{Potential outcomes} \\
Units & sequence &  $Y_{i1}(A)$&  $Y_{i1}(B)$ &  $Y_{i2}(AA)$ &  $Y_{i2}(AB)$&  $Y_{i2}(BA)$ &  $Y_{i2}(BB)$\\ \hline
1 &  $AA$                &  \checkmark   & ?  & \checkmark & ? & ? & ?  \\
2 &  $AB$                &  \checkmark   & ?   & ?  & \checkmark& ? & ?  \\
3 &  $BA$                & ?    &  \checkmark      & ?  & ? & \checkmark & ?  \\
\vdots & \vdots  & \vdots& \vdots& \vdots& \vdots& \vdots& \vdots\\
$N$ &  $BB$                & ?    &  \checkmark      & ?  & ?  & ? & \checkmark  \\  \hline
\end{tabular}
\end{center}
\label{tab::po}
\end{table}%

Applying the definitions in Section~\ref{sec::setup} to the $2^2$ design, the average treatment effect at period 1 is $\tau_1= \oY_1(A)- \oY_1(B)$,  the conditional average treatment effect at period 2 is  $\tau_2(z_1)= \oY_2(z_1A)- \oY_2(z_1B)$ for $z_1=A,B$, and the first-order  conditional average carryover effect is $\tau_2^1(z_2)=\oY_2(Az_2)- \oY_2(Bz_2)$ for $z_2=A,B$. These quantities take the form
\begin{eqnarray*}
\tau(\bw)&=& \sum_{t=1}^2\sum_{\os\in \mathcal{S}} w_{t}(\os) \oY_{t}(\os),
\end{eqnarray*}
where $\mathcal{S}=\{AA,AB,BA,BB\}$ and $w_{t}(\os)$ represents the weight associated with the average potential outcome $\oY_{t}(\os)$ for $t=1,2$ and $\os \in \mathcal{S}$ with $\bw=\{w_{t}(\os): t=1,2, \os=AA,AB,BA,BB\}$.
Although the expression of $\tau(\bw)$ can be simplified by writing $Y_{i1}(AA)=Y_{i1}(AB)=Y_{i1}(A)$ and $Y_{i1}(BA)=Y_{i1}(BB)=Y_{i1}(B)$ under Assumption~\ref{asm::noanticipation}, we retain the current notation to preserve generality and facilitate the development of the methodology. As a result, the same quantity may be represented by different sets of weights. This non-uniqueness does not affect the identification or estimation procedures introduced below.
We focus on the linear estimators in the following form:
\begin{eqnarray}
\label{eqn::general-est}
\widehat{\tau}(\tw)&=& \sum_{t=1}^2\sum_{\os\in \mathcal{S}^\textup{obs}} \tw_t(\os) \wY_{t}(\os),
\end{eqnarray}
where
 \begin{eqnarray*}
\wY_t(\os)= \frac{1}{N_{\os}}  \sum_{i=1}^N  Y_{it} \bone(\oS_i=\os)
\end{eqnarray*}
 is the average of the observed outcome for units under treatment sequence $\os$ and $ \tw_{t}(\os)$  is the corresponding weight. 
 We distinguish between the set of observed treatment sequences, $\mathcal{S}^\textup{obs}$, and the potentially broader set $\mathcal{S}$, which may include treatment sequences not observed in the experiment but of inferential interest. In the specific $2^2$ crossover designs, $\mathcal{S}=\mathcal{S}^\textup{obs}=\{AA,AB,BA,BB\}$. 
 We assume  $\mathcal{S}^\textup{obs}$ is fixed by the design.
 The weight  $ \tw_{t}(\os)$  may differ from $w_{t}(\os) $, as the summation in~\eqref{eqn::general-est} is taken over the observed sequences only.
 
We focus on point estimation under fixed weights $\tw_t(\os)$ in two-period crossover designs, and defer results involving estimated weights to the general framework in Section~\ref{sec::reg-general}. Variance and covariance analysis, including variance expressions and the construction of conservative variance estimators, is provided in the online supplementary material.

\subsection{Without Assumption~\ref{asm::nocarryover}~or~\ref{asm::invarianteffect}}
 We first consider the scenario with only Assumption~\ref{asm::noanticipation}.
Under complete randomization,  $\widehat\tau(\tw)$ is unbiased for 
\begin{eqnarray*}
\E\left\{ \widehat\tau(\tw)\right\}&=&\sum_{t=1}^2\sum_{\os\in \mathcal{S}^\textup{obs}} \tw_{t}(\os) \oY_{t}(\os).
\end{eqnarray*}
To ensure  that $\widehat\tau(\tw)$ is unbiased for the target quantity $\tau(\bw)$, the weights must satisfy 
\begin{eqnarray*}
\sum_{t=1}^2\sum_{\os\in \mathcal{S}^\textup{obs}} \tw_{t}(\os) \oY_{t}(\os)&=& \sum_{t=1}^2\sum_{\os\in \mathcal{S}} w_{t}(\os) \oY_{t}(\os)
\end{eqnarray*}
for all possible values of $\oY_{t}(\os)$ consistent with the assumptions on the data generating mechanism. This requirement imposes a system of equations on the weights
 $\tw_{t}(\os)$.
 If these equations have a unique solution, then the corresponding  $\widehat\tau(\tw)$  is the unique unbiased estimator in the form of~\eqref{eqn::general-est}. If multiple solutions exist, then there are multiple unbiased estimators. We can derive the 
 the best linear unbiased estimator (BLUE) by optimally combining these unbiased estimators to minimize variance.

 

The following proposition gives the  BLUEs  for $\tau_1$, and unbiased estimators for $\tau_2(z_1)$ and $\tau_2^1(z_2)$.

\begin{proposition}
\label{prop::2p4t}
Suppose that Assumption~\ref{asm::noanticipation} holds.  The BLUE for $\tau_1$ is  $\widehat{\tau}_1 =  \widehat{Y}_1(A)- \widehat{Y}_1(B)$, where
\begin{eqnarray*}
 \widehat{Y}_1(z) =
\frac{N_{zA}}{N_{zA}+N_{zB}}\cdot \widehat{Y}_1(zA)+ \frac{N_{zB}}{N_{zA}+N_{zB}}\cdot \widehat{Y}_1(zB);
\end{eqnarray*}
an unbiased estimator for $\tau_2(z_1)$ is $\widehat{\tau}_2(z_1)=\widehat{Y}_2(z_1A)-  \widehat{Y}_2(z_1B)$; and an unbiased estimator for  $\tau_2^1( z_2)$ is   $\widehat{\tau}_2^1(z_2)=\widehat{Y}_2(Az_2)-  \widehat{Y}_2(Bz_2)$. 
\end{proposition}
Proposition~\ref{prop::2p4t} shows that all the treatment effects and carryover effects are identifiable in two-period crossover designs with four treatment sequences.   
For $\tau_1$, multiple unbiased estimators arise from different sequence comparisons. The BLUE admits a simple closed-form expression in this case and reduces to the difference between the average outcomes under treatment A and B at period 1. For $\tau_2(z_1)$ and $\tau_2^1(z_2)$, multiple unbiased estimators can be constructed by adding zero-mean contrasts implied by Assumption~\ref{asm::noanticipation}. For example, under Assumption~\ref{asm::noanticipation},
$
\E\{\widehat{Y}_1(AA)-\widehat{Y}_1(BA)\}
-
\E\{\widehat{Y}_1(AB)-\widehat{Y}_1(BB)\}=0,
$
which can be used to construct alternative unbiased estimators.  The derivation of their BLUEs follows from the general framework developed in Section~\ref{sec::reg-general}.

\subsection{With Assumption~\ref{asm::nocarryover}}
We then consider the analysis under Assumption~\ref{asm::noanticipation}, and Assumption~\ref{asm::nocarryover} with $k=1$. Assumption~\ref{asm::nocarryover} with $k=1$ implies that the treatment received at period 1 does not affect the outcome at period 2, i.e., $Y_{i2}(AA)=Y_{i2}(BA)=Y_{i2}(A)$ and  $Y_{i2}(AB)=Y_{i2}(BB)=Y_{i2}(B)$. Therefore, 
the carryover effect $\tau_2^1(z_1)$ vanishes and the two average causal effects at period 2, $
\tau_2(A)$ and $\tau_2(B)$, become the same.

\begin{proposition}
\label{prop::2p4t-nocarryover}
Suppose that Assumption~\ref{asm::noanticipation} holds, and Assumption~\ref{asm::nocarryover} holds with $k=1$. An unbiased estimator for $\tau_1$ is $\widehat \tau_1$ in Proposition~\ref{prop::2p4t}; an unbiased estimator for $\tau_2$ is  $\widehat{\tau}_2=\widehat{Y}_2(A) -\widehat{Y}_2(B) $, where
\begin{eqnarray*}
  \widehat{Y}_2(z)  
= \frac{N_{Az}}{N_{Az}+N_{Bz}}\cdot \widehat{Y}_2(Az)+ \frac{N_{Bz}}{N_{Az}+N_{Bz}}\cdot \widehat{Y}_2(Bz);
\end{eqnarray*}
and the carryover effect vanishes.
\end{proposition}

Under the additional restriction imposed by  Assumption~\ref{asm::nocarryover} with $k=1$, the BLUE for $\tau_1$ may differ from that in Proposition~\ref{prop::2p4t}, because the restriction allows information from period 2 outcomes to be incorporated into the estimation of $\tau_1$, thereby improving efficiency. 

\subsection{With Assumptions~\ref{asm::nocarryover}~and~\ref{asm::invarianteffect}}

Finally, we consider the analysis under  Assumption~\ref{asm::noanticipation}, and Assumptions~\ref{asm::nocarryover}~and~\ref{asm::invarianteffect}  with $k=1$.
 In this case, the average causal effects at all periods become the same, i.e., $\tau_1=\tau_2(A)=\tau_2(B)$. We can simplify 
 them as $\tau$. 
 Because $\widehat \tau_1$, $\widehat \tau_2(A)$, and $\widehat \tau_2(B)$ are unbiased for $\tau$, we can obtain a more efficient estimator by linearly combining them.
 
To derive the optimal combination, we formulate a variance minimization problem and introduce the following notation.
Define 
\begin{eqnarray*}
S^2_{t}(\os) &=& \frac{1}{N-1} \sum_{i=1}^N \{Y_{it}(\os)-\oY_t(\os)\}^2,\\
S_{12}(\os)&=& \frac{1}{N-1} \sum_{i=1}^N \{Y_{i1}(\os)-\oY_1(\os)\}\{Y_{i2}(\os)-\oY_2(\os)\},
\end{eqnarray*}
where $S^2_{t}(\os) $ is the variance of potential outcome $Y_{it}(\os)$ and $S_{12}(\os)$ is the covariance between $Y_{i1}(\os)$ and $Y_{i2}(\os)$.
Under Assumption~\ref{asm::noanticipation}, $S^2_1(zA)=S^2_1(zB)$ for $z=A,B$, which we denote by $S^2_1(z)$. Under  Assumption~\ref{asm::nocarryover} with $k=1$,  $S^2_2(Az)=S^2_2(Bz)$ for $z=A,B$, which we denote by $S^2_2(z)$.

 \begin{proposition}
\label{prop::2p4t-invariant}
Suppose that Assumption~\ref{asm::noanticipation} holds, and Assumptions~\ref{asm::nocarryover}~and~\ref{asm::invarianteffect}  hold with $k=1$.
The BLUE for $\tau$ is 
\begin{eqnarray*}
\sum_{t=1}^2\sum_{\os\in \mathcal{S}^\textup{obs}} \tw^\ast_{t}(\os) \wY_{t}(\os),
\end{eqnarray*}
 where  the weights $\tw^\ast_{t}(\os)$'s minimize
\begin{eqnarray*}
f(\tw )&=&\sum_{z_1=A,B} \left\{\frac{\tw^2_{1}(z_1A)}{N_{z_1A}}+\frac{\tw^2_{1}(z_1B)}{N_{z_1B}}\right\} S^2_{1}(z_1)+\sum_{z_2=A,B} \left\{\frac{\tw^2_{2}(Az_2)}{N_{Az_2}}+\frac{\tw^2_{2}(Bz_2)}{N_{Bz_2}}\right\} S^2_{2}(z_2)\\
&& +\sum_{z_1,z_2=A,B}\frac{2\tw_{1}(z_1z_2)\tw_{2}(z_1z_2)  S_{12}(z_1z_2) }{N_{z_1z_2}}
\end{eqnarray*}
subject to the following linear constraints:
\begin{eqnarray*}
\label{eqn::restriction2p4t-res1}
\tw_{1}(AA)+\tw_{1}(AB) + \tw_{1}(BA)+ \tw_{1}(BB)&=&0,\\
\label{eqn::restriction2p4t-res2} \tw_{2}(AA)+\tw_{2}(AB) + \tw_{2}(BA)+ \tw_{2}(BB)&=&0,\\
\label{eqn::restriction2p4t-res3} \tw_{1}(AA)+\tw_{1}(AB)+\tw_{2}(AA)+\tw_{2}(BA)&=&1.
\end{eqnarray*}
\end{proposition}
Because $f(\tw)$ is quadratic in $\tw$, we can explicitly solve for $\tw$ using the Lagrange multiplier method.

 \section{Crossover designs with two periods and two sequences}
 \label{sec::2p2t}
We now consider the crossover design with two periods and two treatment sequences, $AB$ and $BA$. 
In this scenario, we do not have data from treatment sequences $AA$ and $BB$, which are present in the design with four treatment sequences. 
Consequently, we cannot calculate $\widehat{Y}_1(AA)$, $\widehat{Y}_2(AA)$, $\widehat{Y}_1(BB)$, and $\widehat{Y}_2(BB)$.  The estimator in~\eqref{eqn::general-est} becomes
\begin{eqnarray*}
\widehat\tau(\tw)&=&\tw_{1}(AB)\wY_1(AB)+\tw_{2}(AB)\wY_2(AB)+\tw_{1}(BA)\wY_1(BA)+\tw_{2}(BA)\wY_2(BA).
\end{eqnarray*}
However, the causal quantities of interest remain the same as those in Section~\ref{sec::2p4t}  and depend on the potential outcomes under $AA$ and $BB$.
As a result,  $\mathcal{S}^\textup{obs} \neq \mathcal{S}$ because $\mathcal{S}=\{AA,AB,BA,BB\}$ but $\mathcal{S}^\textup{obs}=\{AB,BA\}$. Similar to Section~\ref{sec::2p4t}, we restrict attention to point estimation.

 \subsection{Without Assumption~\ref{asm::nocarryover}~or~\ref{asm::invarianteffect}}
\label{sec::2p2t-noasm}
We first consider the analysis under only Assumption~\ref{asm::noanticipation}.

\begin{proposition}
\label{prop::2p2t}
Suppose that Assumption~\ref{asm::noanticipation} holds. The only unbiased estimator for $\tau_1$  is $\widehat{\tau}_1= \widehat{Y}_1(AB)- \widehat{Y}_1(BA)$,
and no unbiased estimator exists for $\tau_2(z_1)$ or $\tau_2^1(z_2)$.
\end{proposition}
Proposition~\ref{prop::2p2t} shows  that  no unbiased estimator exists for the treatment effect at period 2 and the carryover effect. 
This echoes the perspective in Jones and Kenward \cite{jones2003design} on effect aliasing, indicating that the crossover design with two treatment sequences cannot disentangle the carryover effect from the temporal changes in the instantaneous effects.

\subsection{With Assumptions~\ref{asm::nocarryover}}
\label{sec::2p2t-asm2}
We then consider the analysis under Assumption~\ref{asm::noanticipation} and Assumption~\ref{asm::nocarryover} with $k=1$. 
In this case, the carryover effect $\tau_2^1(z_2)=0$ for $z_2=A,B$. Thus, we no longer have the problem of effect aliasing discussed in Section~\ref{sec::2p2t-noasm} and can unbiasedly estimate the treatment effect at period 2.
\begin{proposition}
\label{prop::2p2t-nocarryover}
Suppose that Assumption~\ref{asm::noanticipation} holds, and Assumption~\ref{asm::nocarryover} holds with $k=1$.
The only unbiased estimator for $\tau_1$ is $\widehat{\tau}_1$ in Proposition~\ref{prop::2p2t}; the only unbiased estimator for $\tau_2=\tau_2(A)=\tau_2(B)$ is 
  $ \widehat \tau_2= \widehat{Y}_2(BA)- \widehat{Y}_2(AB)$.
\end{proposition}
%


\subsection{With Assumptions~\ref{asm::nocarryover}~and~\ref{asm::invarianteffect}}
Finally, we consider the analysis under  Assumption~\ref{asm::noanticipation}, and Assumptions~\ref{asm::nocarryover}~and~\ref{asm::invarianteffect}  with $k=1$.
 In this case,
$\widehat \tau_1$ and $\widehat \tau_2$  are unbiased for $\tau=\tau_1=\tau_2(A)=\tau_2(B)$, and thus we can obtain a more efficient estimator by linearly combining them.
In particular, for any constant $p$, $p \widehat{\tau}_1+ (1-p) \widehat{\tau}_2$
is an unbiased estimator. 
The following proposition gives the optimal value of $p$ that minimizes the variance.
 \begin{proposition}
 \label{prop::2p2t-invariant}
Suppose that Assumption~\ref{asm::noanticipation} holds, and Assumptions~\ref{asm::nocarryover}~and~\ref{asm::invarianteffect} hold with $k=1$.
 The BLUE for $\tau$ is $\widehat \tau=p\widehat\tau_1+(1-p) \widehat \tau_2$, where
\begin{eqnarray*}
p&=& \frac{  \frac{S^2_{2}(AB)+S_{12}(AB)}{N_{AB}  }+ \frac{S^2_{2}(BA)+S_{12}(BA)}{N_{BA}  } }{  \frac{S^2_{1}(AB)+S^2_{2}(AB)+2S_{12}(AB)}{N_{AB}  }+ \frac{S^2_{1}(BA)+S^2_{2}(BA)+2S_{12}(BA)}{N_{BA}  } }.
\end{eqnarray*}
 \end{proposition}

\section{A unified regression-based analysis of crossover designs}
\label{sec::reg-general}

Sections~\ref{sec::2p4t} and~\ref{sec::2p2t} detail the analyses of two specific crossover designs. While these examples help illustrate the core concepts and intuition behind our approach, they are not general to accommodate all possible crossover designs. To address this limitation, we develop a unified framework in this section. A key feature of our recommended method is its simplicity in implementation via least squares, with properly specified regression function, weights, restrictions on the coefficients, and covariance estimation. We present the details below. 

\subsection{Setup}
We now consider a general crossover design with $N$ units and $T$ periods. We begin by generalizing the notation.
 Let $\mathcal{S}^\textup{obs} \subset \mathcal{S}$ be the set of all the treatment sequences in the experiment.  Designs with the same number of periods may contain different sets of treatment sequences.
 For example, in Section~\ref{sec::2p4t}, a crossover experiment with two periods and four treatment sequences has $\mathcal{S}^\textup{obs}=\{AA,AB,BA,BB\}$, whereas in Section~\ref{sec::2p2t}, a crossover experiment with two periods and two treatment sequences has $\mathcal{S}^\textup{obs}=\{AB,BA\}$.

We use $\bY_i = (Y_{i1},\ldots,Y_{iT})$ to denote the $T$-dimensional observed outcome, and
 $$\bY_i(\os) =(Y_{i1}(\os),Y_{i2}(\os),\ldots, Y_{iT}(\os))^\top $$
  to denote the $T$-dimensional potential outcome of unit $i$ under treatment sequence $\os$ with the corresponding average potential outcome vector 
  $$\obY(\os) =(\oY_{1}(\os),\oY_{2}(\os),\ldots,\oY_{T}(\os) )^\top  =  \sum_{i=1}^N \bY_i(\os)/N.$$
The set of treatment sequences considered in the definition of potential outcomes can be larger than $\mathcal{S}^\textup{obs}$. For example, we have  $\mathcal{S}^\textup{obs}=\{AB,BA\}$ in a crossover experiment with two periods and two treatment sequences, but we are interested in the treatment effect $\oY_2(AA)-\oY_2(AB)$, with $\os=AA$  not in $\mathcal{S}^\textup{obs}$. To make such a distinction, we use $\mathcal{S}$ to denote the set of sequences for defining the potential outcomes.

We focus on causal effects on the difference scale.
For any $K\geq 1$ and coefficient matrix $\bW(\os) \in \mathbb{R}^{K\times T}$,
define the individual causal effect $\btheta_i(\bW) = \sum_{\os \in \mathcal{S}} \bW(\os) \bY_i(\os)$, where $\bW = \{\bW(\os): \os \in \mathcal{S}\}$. 
The corresponding  average causal effect is 
\begin{eqnarray*}
\btheta(\bW)&=&  \sum_{\os \in \mathcal{S}} \bW(\os) \obY(\os).
\end{eqnarray*}
This definition encompasses all causal quantities in crossover designs that are linear in the average potential outcomes $\obY(\os)$.
For example, in two-period crossover designs with $\mathcal{S}=\{AA,AB,BA,BB\}$,  the average potential outcome vector under $\os=z_1z_2 \in \mathcal{S} $ is 
$$\obY(z_1z_2)=(\oY_1(z_1z_2),\oY_2(z_1z_2))^\top.$$
We can express the conditional carryover effect $\tau_2^1(B)$ as $\btheta(\bW)$ with $\bW(AA)=\bW(BA)=(0,0)$, $\bW(AB)=(0,1)$, and $\bW(BB)=(0,-1)$. 
We can also consider vector causal quantities. For example, the joint carryover effect $(\tau_2^1(A),\tau_2^1(B))^\top$ can be expressed in terms of  $\btheta(\bW)$ with
\begin{eqnarray*}
\bW(AA)&=&\begin{pmatrix}
0&1\\
0&0
\end{pmatrix}, \quad \bW(AB)\ =\ \begin{pmatrix}
0&0\\
0&1
\end{pmatrix}, \quad \bW(BA)\ =\ \begin{pmatrix}
0&-1\\
0&0
\end{pmatrix}, \quad \bW(BB)\ =\ \begin{pmatrix}
0&0\\
0&-1
\end{pmatrix}.
\end{eqnarray*}
Depending on the assumptions on the potential outcomes, there may be multiple ways to express the causal quantities in terms of $\btheta(\bW)$. For example, under Assumption~\ref{asm::noanticipation}, the treatment effect $\tau_1$ is  equal to $\oY_1(Az)-\oY_1(Bz)$ for both $z=A,B$, which correspond to different $\bW$ in $\btheta(\bW)$. This non-uniqueness does not affect the identification or estimation procedures developed below.

An intuitive estimator for $\bY(\os)$ is the average outcome under $\os$:
 $\widehat \bY(\os)=\sum_{i=1}^N \bY_i\bm{1}(\oS_i=\os)/N_{\os} $, where $N_{\os}$ is the number of units satisfying $\oS_i=\os$ with $\sum_{\os \in \mathcal{S}^\textup{obs}}N_{\os}=N$.
 Under complete randomization, $\widehat \bY(\os)$ is unbiased for $\obY(\os)$. 
If $\btheta(\bW)$  does not involve the potential outcomes for $\os \notin \mathcal{S}^\textup{obs}$, then an unbiased estimator can be obtained by directly substituting $\widehat \bY(\os)$ into the expression of $\btheta(\bW)$. If multiple unbiased estimators exist, then the main research question is to find the BLUE.
 However, difficulties arise when $\btheta(\bW)$ involves treatment sequences not in $\mathcal{S}^\textup{obs}$. In this case, the estimation of $\btheta(\bW)$ relies on additional assumptions. Such a situation is common in practice because the sample size is often limited compared with the number of  possible treatment sequences when $T$ is large.
 We  focus on the linear estimator in the form of 
\begin{eqnarray}
\label{eqn::est-form}
\wbtheta(\tbW)&=& \sum_{\os \in \mathcal{S}^\textup{obs}} \tbW(\os) \wbY(\os),
\end{eqnarray}
where $\tbW(\os) \in \mathbb{R}^{K\times T}$ is the weight matrix associated with $\wbY(\os)$ and $\tbW = \{\tbW(\os): \os \in \mathcal{S}^\textup{obs}\}$. 
Because $\tbW(\os)$ is defined only for the observed treatment sequences, it may differ from the target weight $\bW(\os)$, which is defined for all sequences in $\mathcal{S}$.
While the weight  $\tbW(\os)$  may in general depend on the observed data,   we begin by treating it as fixed to facilitate both estimator construction and theoretical analysis.  We then obtain the final estimator by replacing $\tbW(z)$ with its consistent estimate.

  We investigate three fundamental questions for analyzing general crossover designs under various sets of assumptions:
\begin{itemize}
\item[Q1.] Can we obtain a linear unbiased estimator in the form~\eqref{eqn::est-form}  for the causal quantity $\btheta(\bW)$?
\item[Q2.]  If the answer to Q1 is ``yes'', what is the BLUE?
\item[Q3.] How should we conduct inference when  $\tbW(z)$  is replaced with its consistent estimate in $\wbtheta(\tbW)$?
\end{itemize}

These questions are nontrivial, even for a given estimand $\btheta(\bW)$ under a specific crossover design. For Q1, 
if $\btheta(\bW)$  involves a treatment sequence $\os'$ not implemented in the experiment, then we must examine whether $\obY(\os')$ can be expressed in terms of $\{\obY(\os): \os \in \mathcal{S}^\textup{obs}\}$ under the imposed assumptions.  For Q2, 
when multiple linear unbiased estimators of $\btheta(\bW)$  are available, we can consider linear combinations of all unbiased estimators and seek the combination that minimizes the variance. However, the BLUE may not have a closed form, necessitating optimization techniques for its calculation. For Q3, the asymptotic properties of $\wbtheta(\tbW)$ might change when  $\tbW(z)$ is replaced by its consistent estimate. Analyzing these properties becomes particularly challenging when the efficient estimator lacks a closed-form representation.

Moreover, the answers to these three questions can vary substantially depending on the choice of the causal estimand and the structure of the crossover design. This complication is illustrated in our analysis of two specific crossover designs in Sections~\ref{sec::2p4t} and~\ref{sec::2p2t}. These challenges underscore the need for a unified and principled approach to the analysis of general crossover designs. Therefore, we propose a simple-to-implement regression-based approach in the following subsection, which, perhaps surprisingly, yields unified, simultaneous solutions to all three questions.

%

\subsection{Restricted weighted least squares for analyzing crossover designs}
\label{sec::reg-general-theory}
We propose a regression-based approach that simultaneously yields both the estimator and the associated inference procedure. This approach offers several key advantages.
First, it provides a principled framework for implementation based on restricted weighted least squares. This ensures the validity of both estimation and inference under the design-based framework, without requiring correct specification of the regression models. Second, it accommodates a broad class of causal estimands and is applicable to crossover designs with arbitrary structures.
Third, it is easy to implement, requiring only minor modifications to standard least squares routines.

We begin with the {\it derived linear model} \citep{kempthorne1952design,hinkelmann2007design,zhao2023covariate} to provide some intuition for the regression-based approach.
For each $i$ and $\os$, we decompose $Y_{it}(\os)=\oY_t(\os)+\epsilon_{it}(\os)$, where $\epsilon_{it}(\os) = Y_{it}-\oY_t(\os) $. This leads to the following derived linear model of the observed outcome:
\begin{eqnarray}
\label{eqn::derivedlinear}
Y_{it}&=&   \sum_{\os \in \mathcal{S}}   Y_{it}(\os)  \bm{1}(\oS_i=\os) \ =\ \sum_{\os \in \mathcal{S}} \oY_t(\os) \bm{1}(\oS_i=\os)+\epsilon_{it},
\end{eqnarray}
with the error term given by
\begin{eqnarray*}
\epsilon_{it}&=& \sum_{\os \in \mathcal{S}}   \left\{Y_{it}(\os)-\oY_t(\os) \right\}\bm{1}(\oS_i=\os).
\end{eqnarray*}
Although some treatment sequences in $\mathcal{S}$ may not be observed in the data, we still include them in the decomposition in~\eqref{eqn::derivedlinear}. This is necessary because $\btheta(\bW)$ may involve potential outcomes corresponding to unobserved sequences. Including the full set of sequences in~\eqref{eqn::derivedlinear} ensures that the analysis can accommodate such cases.

The derived linear model in~\eqref{eqn::derivedlinear} motivates the  following regression specification for the observed outcome at each period $t$:
\begin{eqnarray}
\label{eqn::reg-general-t}
Y_{it} &=& \sum_{\os \in \mathcal{S}}   \gamma_{t,\os}\bm{1}(\oS_i=\os)+e_{it},
\end{eqnarray}
where the regressors are the indicator functions for the assigned treatment sequences.  Stacking the regression in~\eqref{eqn::reg-general-t} for all time periods result in the joint regression of the vector outcome $\bY_i=(Y_{i1},\ldots,Y_{iT})^\top$:
\begin{eqnarray}
\label{eqn::reg-general}
\begin{split}
Y_{i1} &=& \sum_{\os \in \mathcal{S}}   \gamma_{1,\os}\bm{1}(\oS_i=\os)+e_{i1},\\
\vdots && \\
Y_{iT} &=& \sum_{\os \in \mathcal{S}}   \gamma_{T,\os}\bm{1}(\oS_i=\os)+e_{iT}.
\end{split}
\end{eqnarray}
Define the vector error as $e_i = (e_{i1}, \ldots, e_{iT})^\top$, and  vertically stack the vectors $\bgamma_{\os} = (\gamma_{1,\os}, \ldots, \gamma_{T,\os})^\top$ for all $\os \in \mathcal{S}$ to define the coefficient vector $\bgamma = \{ \bgamma_{\os}^\top: \os \in \mathcal{S} \}^\top$.
Because the outcomes at different time periods are correlated, we introduce weights in the regression to capture the covariance structure of the outcome vectors.
Define
\begin{eqnarray*}
 \bS(\os,\os') &=&\frac{1}{N-1} \sum_{i=1}^N \{\bY_i(\os)-\obY(\os)\}\{\bY_i(\os')-\obY(\os')\}^\top
\end{eqnarray*}
as the finite-population covariance matrix between the potential outcome vectors $\bY_i(\os)$ and $\bY_i(\os')$. When $\os=\os'$,  $ \bS(\os,\os') $ reduces to $ \bS^2(\os)$.
 For units with $\oS_i=\os$, the residual in the derived linear model~\eqref{eqn::derivedlinear} is equal to $\bY_i(\os)-\obY(\os)$. This motivates using the covariance matrix of $\bY_i(\os)$, $\bS^2(\os)$, as the weight  matrix for units with $\oS_i=\os$.  
 
 The regression in~\eqref{eqn::reg-general} with weights $\bS^2(\os)$ serves as the foundation for analyzing general crossover designs under various sets of assumptions. 
To facilitate the exposition, we introduce additional notation to express the regression in matrix form.  Let $X_i$ denote the regressor matrix for unit $i$. 
From~\eqref{eqn::reg-general}, we can write
\begin{eqnarray*}
X_i&=& \{\bm{1}(\oS_i=\os):\os \in \mathcal{S}\} \otimes I_T,
\end{eqnarray*}
where $ \{\bm{1}(\oS_i=\os):\os \in \mathcal{S}\}$ is a row vector whose entries are the indicators $\bm{1}(\oS_i=\os)$ for each $\os \in \mathcal{S}$, $I_T$ is the $T$-dimensional identity matrix, and $\otimes$ denotes the Kronecker product.
The regressor matrix $X_i$ consists of $T$ rows and $T\times |\mathcal{S}|$ columns, where $|\mathcal{S}|$ represents the number of elements in $\mathcal{S}$. For the $t$-th row of $X_i$, the entries corresponding to $\{\gamma_{t,\os}: \os \in \mathcal{S}  \}$ are given by $\{\bm{1}(\oS_i=\os): \os \in \mathcal{S}  \}$, and the entries corresponding to $\{\gamma_{t',\os}: \os \in \mathcal{S}, t' \neq t  \}$ are all zero. 
We define the stacked outcome vector as $\bY = (\bY_1^\top, \ldots, \bY_n^\top)^\top$, the stacked error vector as $e = (e_1^\top, \ldots, e_n^\top)^\top$, and the full regressor matrix as $\bX = (X_1^\top, \ldots, X_n^\top)^\top$.
The weighted regression in ~\eqref{eqn::reg-general}   can then be written in the following matrix form:
\begin{eqnarray}
\label{eqn::reg-matrix}
\bY &=& \bX \bgamma+ \bm{e},   \quad \cov(\bm{e})\ = \ \text{diag}\left(\bS^2(\oS_i): i=1,\ldots,n \right),
\end{eqnarray}
where $\bS^2(\oS_i)=\bS^2(\os)$ for units with $\oS_i=\os$.

Fitting the regression in~\eqref{eqn::reg-matrix} requires knowing the weight matrices. In practice, however, the weight matrices are typically unknown and must be estimated first. In what follows, we will first consider the case with known weight matrices and then extend the results to settings with consistently estimated weight matrices.


Assumptions~\ref{asm::noanticipation}~to~\ref{asm::invarianteffect} imply constraints on the means of potential outcomes $\oY(\os)$. Leveraging the link between the derived linear model and the regression formulation, we translate these constraints into restrictions on the coefficients in~\eqref{eqn::reg-matrix}.
 We state the three restrictions corresponding to the three assumptions below.
\begin{constraint}
\label{cons::noanticipation}
For all $t$, 
$\gamma_{t,\os} = \gamma_{t,\os'}$ if   $\os_{[1,t]}= \os'_{[1,t]}$.
\end{constraint}

\begin{constraint}
\label{cons::carryover}
For $t\geq k$, 
$\gamma_{t,\os} = \gamma_{t,\os'}$ if   $\os_{[t-k+1,t]}= \os'_{[t-k+1,t]}$.
\end{constraint}

\begin{constraint}
\label{cons::invarianteffect}
For $t'>t\geq k$, 
$\gamma_{t,\os}-\gamma_{t,\os'} =\gamma_{t',\os}-\gamma_{t',\os'} $ if  $\os_{[t-k+1,t]}=\os_{[t'-k+1,t']}$ and $\os'_{[t-k+1,t]}=\os'_{[t'-k+1,t']}$.
\end{constraint} 
These constraints are obtained by replacing $Y_{it}(\os)$ with $\gamma_{t,\os}$ in Assumptions~\ref{asm::noanticipation}~to~\ref{asm::invarianteffect}. They can all be written in terms of 
$$C \bgamma=0,$$ 
where $C$ is a constant restriction matrix. We consider the weighted least squares estimation of~\eqref{eqn::reg-matrix} with weight matrix $\bS^2(\oS_i)$ and restriction $C \bgamma=0$, where different specifications of $C$ correspond to the analysis of crossover designs with distinct sets of assumptions imposed:
\begin{itemize}
\item[Scenario (a):] $C=C_1$ represents Constraint~\ref{cons::noanticipation}, which corresponds to Assumption~\ref{asm::noanticipation};
\item[Scenario (b):]   $C=C_2$ represents Constraints~\ref{cons::noanticipation}~and~\ref{cons::carryover}, which corresponds to Assumption~\ref{asm::noanticipation} and Assumption~\ref{asm::nocarryover} with order $k$;
\item[Scenario (c):]   $C=C_3$ represents Constraints~\ref{cons::noanticipation}~to~\ref{cons::invarianteffect}, which corresponds to Assumption~\ref{asm::noanticipation}, and Assumptions~\ref{asm::nocarryover}~and~\ref{asm::invarianteffect} with order $k$.
\end{itemize}
These matrices $C_1$, $C_2$, $C_3$ can be constructed for any given crossover design. 
In what follows, we use $C$ to denote the restriction matrix corresponding to the scenario under consideration.
With the regressor matrix and corresponding restriction matrix specified, the following theorem provides a unified framework for analyzing crossover designs via restricted weighted least squares. 
\begin{theorem}
\label{thm::reg-general}
Consider the restricted weighted least squares estimation in~\eqref{eqn::reg-matrix} with known weight matrix $\bS^2(\oS_i)$ for unit $i$ and restriction $C \bgamma=0$.
Suppose that $X^\top X + C^\top C $ is of full rank. The following results hold for Scenarios (a), (b), and (c).
\begin{itemize}
\item[(a)] $\btheta(\bW)$ can be unbiasedly estimated for any $\bW$.
\item[(b)] The restricted weighted least squares estimation has a unique solution, denoted by $\widehat \bgamma^\textup{wls}$.
The estimator $ \sum_{\os \in \mathcal{S}} \bW(\os) \widehat \bgamma_{\os}^\textup{wls}$  is 
the BLUE for $\btheta(\bW)$ in the form of~\eqref{eqn::est-form}.
\item[(c)]  The BLUE from (b) can be expressed as the following linear combination of $\widehat \bY(\os)$ for $\os \in \mathcal{S}^\textup{obs}$:
\begin{eqnarray}
\label{eqn::linearcombintation}
 \widehat{\theta}(\tbW^\textup{wls})&=&\sum_{\os \in \mathcal{S}^\textup{obs}} \tbW^\textup{wls}(\os) \widehat \bY(\os),
\end{eqnarray}
where $\tbW^\textup{wls}(\os)$ is determined by $\bW$ and $\widehat \bgamma_{\os}^\textup{wls}$ from (b).
Its variance is given by
\begin{eqnarray*}
\cov \{ \widehat{\theta}(\tbW^\textup{wls})\} &=&\sum_{\os \in \mathcal{S}^\textup{obs}}  \frac{\tbW^\textup{wls}(\os) \bS^2(\os)\tbW^\textup{wls}(\os)^\top}{N_{\os}} -\frac{1}{N}S^2(\btheta(\bW)),
\end{eqnarray*}
where
\begin{eqnarray*}
\bS^2(\btheta(\bW)) =\frac{1}{N-1} \sum_{i=1}^N \{\btheta_i(\bW)-\btheta(\bW)\}\{\btheta_i(\bW)-\btheta(\bW)\}^\top
\end{eqnarray*}
is the variance of the individual causal effect $\btheta_i(\bW)$.
\end{itemize}
\end{theorem}

\subsection{Interpretations and proof strategy for the main Theorem~\ref{thm::reg-general}}
\label{sec::interpretation}
We provide interpretations of Theorem~\ref{thm::reg-general} and outline the main ideas behind its proof.

The rank condition of the matrix $X^\top X + C^\top C $ is a sufficient condition for the existence of an unbiased estimator for $\bY(\os)$ with any $\os \in \mathcal{S}$. Consequently, it also guarantees the existence of an unbiased estimator for all $\btheta(\bW)$. 
 However,  this rank condition may not be necessary for the existence of an unbiased estimator of
$\btheta(\bW)$ for a specific $\bW$. The necessary condition can vary depending on the particular choice of 
$\bW$, the number of periods and sequences in the design, and the assumptions imposed. For this reason, we present Theorem~\ref{thm::reg-general}(a) as a simple and broadly applicable sufficient condition, and relegate more nuanced sufficient and necessary conditions to Section~\ref{app::necessary} in the online supplementary material.

The full rank condition can be interpreted in terms of the amount of information provided by the design and the imposed constraints. In a $T$-period design with all $2^T$ treatment sequences, the number of regression coefficients is $T\cdot 2^T$. When all treatment sequences are observed, the design matrix $X$ provides enough information to identify all coefficients. In this case, $X^\top X$ is of full rank, and the rank condition holds under any set of constraints. When only a subset $\mathcal{S}^{\textup{obs}}$ is observed, the information contained in $X$ is limited: each observed sequence contributes information for $T$ coefficients, so the design alone effectively identifies at most $T\cdot |\mathcal{S}^{\textup{obs}}|$ coefficients. In this case, the linear constraints encoded in $C$ play a crucial role in recovering the missing information. The full rank condition requires that the information from the observed sequences, together with the imposed constraints, is sufficient to identify all regression coefficients.

Theorem~\ref{thm::reg-general}(b) establishes that the restricted weighted least squares estimator $\widehat \bgamma^\textup{wls}$ is the BLUE for the average  potential outcome vector $\{\obY(\os): \os \in \mathcal{S}\}$.
By replacing each $\obY(\os)$ with $\widehat \bgamma_{\os}^\textup{wls}$,  we obtain the BLUE for $\btheta(\bW)$.
This offers a simple procedure to compute the optimal estimator for $\btheta(\bW)$.

Theorem~\ref{thm::reg-general}(c) provides an alternative expression of the BLUE for $\btheta(\bW)$ to facilitate the derivation of its variance formula. 
Because the units with the same treatment sequence share the same regressor matrix and weight matrix in the restricted weighted least squares estimation in~\eqref{eqn::reg-matrix}, the restricted least squares estimator $\widehat \bgamma_{\os}^\textup{wls}$ can be expressed as a linear combination of $\wbY(\os)$ for $\os \in \mathcal{S}^\textup{obs}$. This 
implies that the estimator in  Theorem~\ref{thm::reg-general}(b) can also be written as a linear combination of $\wbY(\os)$, as denoted by $ \widehat{\theta}(\tbW^\textup{wls})$  in~\eqref{eqn::linearcombintation}. The weights for the linear combination are functions of $\bW$, the regressor matrix, and weight matrix in the regression, and are therefore fixed. From~\eqref{eqn::linearcombintation}, we can apply the result in Li and Ding \cite{li2017general} to obtain the variance formula of $ \widehat{\theta}(\tbW^\textup{wls})$ under the design-based framework.

The proof of Theorem~\ref{thm::reg-general} is non-trivial. The main challenge lies in the fact that the estimator is constructed via restricted weighted least squares, yet its properties are analyzed under the design-based framework, without assuming correct specification of the underlying linear model. We summarize the key steps of the proof below.

We begin by developing the theoretical foundation for restricted weighted least squares estimation in Section~\ref{app::rwls} in the online supplementary material, which extends the classical results on restricted least squares   \citep{theil1971principles,rao1973linear,greene1991restricted,zhao2023covariate} to settings with vector outcomes and heteroskedasticity. Importantly, we generalize the classical Gauss--Markov Theorem and show that the restricted weighted least squares estimator $\widehat \bgamma_{\os}^\textup{wls}$ is the BLUE for the coefficient vector $\bgamma$. However, this optimality property requires the corresponding restricted weighted linear model to be correctly specified. 

To eliminate this requirement of a correct linear model, we restate the generalized Gauss--Markov Theorem in a purely algebraic form. This reformulation allows us to interpret $\widehat \bgamma_{\os}^\textup{wls}$ as a solution to a constrained optimization problem. Under our setup of  restricted least squares estimation, the constraint coincides with the unbiasedness property for $\obY$ under the design-based framework, and the objective function matches the design-based variance of the estimator of the form~\eqref{eqn::est-form}, up to a constant. Consequently, this constrained minimization perspective shows that $\widehat \bgamma_{\os}^\textup{wls}$ is the BLUE for $\obY$ under the design-based framework. This directly implies the property of $ \widehat{\theta}(\tbW^\textup{wls})$ stated in Theorem~\ref{thm::reg-general}.

%
%

In some scenarios, the BLUE may not be unique. It is possible that multiple unbiased estimators can achieve the minimal variance. Nonetheless, we focus on the restricted weighted least squares estimator for its ease of computation and universal applicability across different settings.

\subsection{Regression-based inference and design-based properties}
\label{sec::regression-property}
We then consider the inference procedure for $\btheta(\bW)$. From Theorem~\ref{thm::reg-general}, $\widehat \btheta(\bW)$ can be written as a linear combination of $\widehat \bY(\os)$. 
This structure allows us to apply the finite-population central limit theorem in Li and Ding \cite{li2017general} to establish the asymptotic normality of $\widehat{\btheta}(\bW)$, as stated in the following corollary.
\begin{corollary}
\label{thm::inference}
Suppose that  for any $\os,\os' \in  \mathcal{S}^\textup{obs}$, $\bS(\os,\os')$ has a limiting value, $N_{\os}/N$ has a positive limiting value, and $\max_{\os \in \mathcal{S}^\textup{obs}}\max_{1\leq i \leq N}||\bY_i(\os)-\obY(\os)||^2_2/N \rightarrow 0$. If the conditions in Theorem~\ref{thm::reg-general} hold,  then $N\cov\{ \widehat{\theta}(\tbW^\textup{wls})\}$ has a limiting value, denoted by $\bm{V}$, and 
\begin{eqnarray*}
N^{1/2}\{ \widehat{\theta}(\tbW^\textup{wls})-\btheta(\bW)\} \stackrel{d}{\rightarrow} N(0,\bm{V}).
\end{eqnarray*}
\end{corollary}
The condition $\max_{\os \in \mathcal{S}^\textup{obs}}\max_{1\leq i \leq N}||\bY_i(\os)-\obY(\os)||^2_2/N \rightarrow 0$ holds if the outcome is bounded or $\bY_i(\os)$  is drawn from a superpopulation with more than two moments \citep{li2017general}.
Corollary~\ref{thm::inference} serves as the basis for developing a statistical inference procedure for $\btheta(\bW)$.
However, two challenges remain. First, the estimation process requires the weights $\bS^2(\os)$ to be known, which is unrealistic in practice. Second, 
the variance formula in Theorem~\ref{thm::reg-general}(c) involves the variance of the individual causal effect, which cannot be estimated from the observed data. 

In the following, we demonstrate that it is valid to use consistently estimated weights in the restricted weighted least squares estimation when the true weights are unknown. Additionally, we provide an asymptotically conservative variance estimator by extending the EHW variance to the restricted weighted least squares setting.

\begin{theorem}
\label{thm::reg-omegahat}
Suppose that the conditions in Corollary~\ref{thm::inference} hold.
  Define
\begin{eqnarray*}
 \widehat{\bS}^2(\os) &=&\frac{1}{N_{\os}-1} \sum_{i=1}^N \bm{1}(\oS_i=\os)\{\bY_i- \widehat \bY(\os)\}\{\bY_i  - \widehat \bY(\os)\}^\top
\end{eqnarray*}
as the sample covariance of $\bY_i(\os)$ for $\os \in \mathcal{S}^\textup{obs}$. 
 Let $\widehat{\bgamma}^\textup{fwls}$ be the feasible restricted weighted least squares estimator with weights $ \widehat{\bS}^2(\os)$. We have  $N^{1/2}(\widehat{\bgamma}^\textup{fwls}-\widehat \bgamma^\textup{wls})$ converges to zero in probability.
\end{theorem}
Theorem~\ref{thm::reg-omegahat}  demonstrates that the weighted least squares estimator with these estimated weights is asymptotically equivalent to the one using true weights. 
It motivates a two-step procedure for obtaining the BLUE. First, we compute the sample covariances to estimate the true weights. Then, we perform the restricted weighted least squares using these estimated weights to obtain $\widehat{\bgamma}^\textup{fwls}$. The estimator solves the following optimization problem:
\begin{eqnarray*}
&&\underset{\beta}{\argmin} (Y-X\beta)^\top \widehat{\Omega}^{-1} (Y-X\beta), \quad \text{subject to }\ C \beta \ =\  0,
\end{eqnarray*}
where $ \widehat{\Omega} = \text{diag}(\widehat{\bS}^2(\oS_i):  i=1,\ldots,n)$ is the  estimated weight matrix. 
In applying the method of Lagrange multipliers, we need the following matrix (see the first order condition in~\eqref{eqn::firstorder}):
\begin{eqnarray}
\label{eqn::U-hat}
\widehat{U} &=&\begin{pmatrix}
X^\top \widehat{\Omega}^{-1}X   &C^\top \\
C & 0  
\end{pmatrix}^{-1} \ = \ \begin{pmatrix}
\widehat{U}_{11}   &\widehat{U}_{12} \\
\widehat{U}_{21} & \widehat{U}_{22}
\end{pmatrix},
\end{eqnarray}
where the inverse exists  under the  
 full rank condition of $X^\top X + C^\top C $ in Theorem~\ref{thm::reg-general}.
Lemma~\ref{lem::rwls-unbias} in the online supplementary material shows that
\begin{eqnarray}
\label{eqn::fwls}
\widehat{\bgamma}^\textup{fwls}=\widehat{U}_{11}X^\top \widehat{\Omega}^{-1}Y.
\end{eqnarray}
The final estimator for $\btheta(\bW)$ is then given by $ \sum_{\os \in \mathcal{S}} \bW(\os) \widehat{\bgamma}_{\os}^\textup{fwls}$, which replaces $\widehat \bgamma_{\os}^\textup{wls}$ with $\widehat{\bgamma}_{\os}^\textup{fwls}$ in the expression of $\widehat \btheta(W)$. 

For inference, we derive an asymptotically conservative estimator for the design-based variance of  $\widehat{\bgamma}^\textup{fwls} $.  To motivate a covariance estimator, we return to the model-based framework. In particular, we temporarily
treat $\widehat{U}_{11}X^\top \widehat{\Omega}^{-1}$ as fixed. The covariance matrix of $\widehat{\bgamma}^\textup{fwls} $ under the linear model is then $\widehat{U}_{11}X^\top \widehat \Omega^{-1}  \cov(Y) \widehat \Omega^{-1}X \widehat{U}_{11}$. 
 We estimate  $\cov(Y) $ using the diagonal matrix $\widehat{\Sigma} = \text{diag}\{\widehat{\Sigma}_i: i=1,\ldots,N\}$, where each diagonal block is
 $\widehat{\Sigma}_i = (Y_i -X_i\widehat{\bgamma}^\textup{fwls} )(Y_i -X_i \widehat{\bgamma}^\textup{fwls} )^\top$. Substituting $\widehat{\Sigma}$ yields the EHW-type variance estimator for $\widehat{\bgamma}^\textup{fwls}$:
\begin{eqnarray}
\label{eqn::fwls-ehw}
\widehat{V}_{\textsc{EHW}}&=&\widehat{U}_{11}X^\top \widehat \Omega^{-1}  \widehat{\Sigma} \widehat \Omega^{-1}X \widehat{U}_{11}.
\end{eqnarray}
With a vector outcome and a block-diagonal weight matrix, this estimator can also be viewed as the cluster-robust variance estimator in generalized estimation equations \citep{liang1986longitudinal}.
The following theorem formally shows that the EHW-type variance estimator is asymptotically conservative for the design-based variance of $\widehat \bgamma^\textup{fwls}$.
%
\begin{theorem}
\label{thm::reg-variance}
Under the conditions in Corollary~\ref{thm::inference}, 
we have  $N\{\widehat{V}_{\textsc{EHW}}-\var (\widehat{\bgamma}^\textup{fwls}  )\} = \bS^2(\obY(\cdot)) + o_{\mathbb{P}}(1) $ for Scenarios (a), (b), and (c), where $\obY(\cdot) =\{\obY(\os): \os \in \mathcal{S}\} $ is the vector of average potential outcomees and $\bS^2(\obY(\cdot))$ is the variance of the individual potential outcomes $\{\bY_i(\os): \os \in \mathcal{S}\}$.
\end{theorem}
Because  $\widehat{\theta}(\tbW^\textup{fwls})= \sum_{\os \in \mathcal{S}} \bW(\os) \widehat{\bgamma}_{\os}^\textup{fwls}$ is a linear combination of  $\widehat{\bgamma}^\textup{fwls} $, Theorem~\ref{thm::reg-variance} implies that its variance can be conservatively estimated by $W \widehat{V}_{\textsc{EHW}} W^\top$, where $W = \{W(\os): \os \in \mathcal{S}\}$.

\subsection{Practical implementation of the regression-based approach}
\label{sec::reg-general-practice}
The results in Section~\ref{sec::reg-general-theory} suggest a general workflow for analyzing crossover designs using a regression-based approach:

\begin{enumerate}
\item[Step 1.]  Express the target causal effect in terms of $\btheta(\bW)$ and specify the regression equation as in~\eqref{eqn::reg-general}. This requires specifying the set $\mathcal{S}$, which is a superset of $\mathcal{S}^\textup{obs}$ that includes all treatment sequences relevant to $\btheta(\bW)$. 
The default choice is to let $\mathcal{S}$ contain all possible treatment sequences. 
However, $\mathcal{S}$ can vary depending on the choice of $\btheta(\bW)$.

\item[Step 2.]  Formulate the constraint $C\bgamma=0$ based on the assumptions imposed in the analysis. These constraints follow directly from the assumed structural properties of the potential outcomes, which should be informed by background knowledge or expert opinion.

\item[Step 3.] Verify that $X^\top X + C^\top C $ is full column rank, which guarantees the identifiability of $\btheta(\bW)$. Checking this condition can be challenging due to the high dimensionality of $X$ and the complexity of the constraints. 
However, simplifications are often possible by exploiting the structure of the data and the constraints. 
Because $X_i = X_{i'}$ for any $i,i'$ with $\oS_i=\oS_{i'}$, we can write $X_i$ as $X_{\os}$ for units with $\oS_i=\os$. 
In addition, the constraints can often be partitioned into subsets of the form $A_l\bgamma=0$ for $l=1,\ldots,L$, with $C = (A_1^\top,\ldots,A_L^\top)^\top$. 
Using these simplifications, we obtain
\begin{eqnarray*}
X^\top X+C^\top C &=& \sum_{\os \in \oS} N_{\os} X_{\os}^\top X_{\os}+ \sum_{l=1}^L A_l^\top A_l,
\end{eqnarray*}
which facilitates computation. 
This formulation also implies that identifiability improves as more treatment sequences are implemented in the experiment or as additional assumptions are imposed in the analysis.

\item[Step 4.] If the full column rank condition in Step 3 holds, then compute the sample variance $\widehat{\bS}^2(\os)$ for each treatment sequence using the observed data.
The variance estimator need not be the sample variance;
for example, a pooled variance estimator may be used if multiple treatment sequences are assumed to share the same variance. 
The asymptotic theory remains valid as long as the variance estimator is consistent.

\item[Step 5.] Perform the restricted weighted least squares with the estimated weights $\widehat{\bS}^2(\os)$ and constraint $C \gamma = 0$ to obtain the BLUE for $\btheta(\bW)$ in~\eqref{eqn::fwls}. Additionally, compute the EHW variance in~\eqref{eqn::fwls-ehw} to obtain the conservative variance estimator $W \widehat{V}_{\textsc{EHW}} W^\top$ for inference.
\end{enumerate}

In the supplementary material, we provide computational details, including explicit constructions of the constraint matrix $C$, strategies to facilitate matrix computations, and an illustration of the approach for two-period crossover designs.

\section{Extensions of the regression-based analysis}
\label{sec::extension}
We extend the regression-based analysis in three directions: allowing for arbitrary weight matrices, relaxing the assumptions to hold only at the population level, and developing a procedure for testing the assumptions.

\subsection{Regression-based analysis with arbitrary weight matrices}
\label{sec::extension-weight}
First, we consider the restricted weighted least squares estimation in~\eqref{eqn::reg-matrix} with weight matrix $\Gamma=\mathrm{diag}(\Gamma_i: i=1,\ldots,n)$ and restriction $C\bgamma=0$. We assume that $\Gamma_i=\Gamma_j$ whenever $\oS_i=\oS_j$. This setup includes the unweighted case as a special case,  where $\Gamma$ is the identity matrix.   The following theorem summarizes this extension.

\begin{theorem}
\label{thm::reg-general-constantweight}
Consider the restricted weighted least squares estimation in~\eqref{eqn::reg-matrix} with weight matrix $\Gamma$ and restriction $C\bgamma=0$. Suppose that $X^\top X + C^\top C$ is of full rank. Then, for Scenarios (a), (b), and (c), the estimator has a unique solution, is unbiased for $\btheta(\bW)$, and admits a linear representation based on $\widehat{\bY}(\os)$ for $\os \in \mathcal{S}^\textup{obs}$, with variance given analogously to Theorem~\ref{thm::reg-general}(c). Moreover, Corollary~\ref{thm::inference} continues to hold, that is, the EHW-type variance estimator remains asymptotically conservative for the design-based variance. In general, however, the restricted weighted least squares estimator is not the BLUE for $\btheta(\bW)$.
\end{theorem}

\subsection{Relaxing assumptions to the population level}
\label{sec::extension-pop}
We relax Assumptions~\ref{asm::noanticipation}~to~\ref{asm::invarianteffect} to hold at the population level. 
These relaxed assumptions allow for heterogeneity in individual potential outcomes, as long as the corresponding restrictions hold on average over units.
We consider the same estimator as in Theorem~\ref{thm::reg-general} and summarize the result in the following theorem.

\begin{theorem}
\label{thm::reg-general-average}
Consider the restricted weighted least squares estimation in~\eqref{eqn::reg-matrix} with known weight matrix $\bS^2(\oS_i)$ and restriction $C\bgamma=0$. Suppose that $X^\top X + C^\top C$ is of full rank. Then, for Scenarios (a), (b), and (c), with the corresponding assumptions replaced by their population-level counterparts, the estimator has a unique solution, is unbiased for $\btheta(\bW)$, and admits a linear representation based on $\widehat{\bY}(\os)$ for $\os \in \mathcal{S}^\textup{obs}$. 
Its variance is given by
\begin{eqnarray*}
\var \{ \widehat{\theta}(\tbW^\textup{wls})\} &=&\sum_{\os \in \mathcal{S}^\textup{obs}}  \frac{\tbW^\textup{wls}(\os) \bS^2(\os)\tbW^\textup{wls}(\os)^\top}{N_{\os}} -\frac{1}{N}S^2(\btheta(\tbW^\textup{wls}) ),
\end{eqnarray*}
where
\begin{eqnarray*}
S^2(\btheta(\tbW^\textup{wls}) ) =\frac{1}{N-1} \sum_{i=1}^N \{\btheta_i(\tbW^\textup{wls})-\btheta(\bW)\}\{\btheta_i(\tbW^\textup{wls})-\btheta(\bW)\}^\top
\end{eqnarray*}
is the covariance of the individual causal effect $\btheta_i(\tbW^\textup{wls}) =\sum_{\os \in \mathcal{S}^\textup{obs}} \tbW^\textup{wls}(\os)  \bY_i(\os) $.
Moreover, Corollary~\ref{thm::inference} continues to hold, that is, the EHW-type variance estimator remains asymptotically conservative for the design-based variance. In general, however, the restricted weighted least squares estimator is not  the BLUE for $\btheta(\bW)$.
\end{theorem}

Compared with Theorem~\ref{thm::reg-general}, the variance formula in Theorem~\ref{thm::reg-general-average} changes under the relaxed assumptions. In particular, the individual causal effect $\btheta_i(\tbW^\textup{wls})$ may differ from $\btheta_i(\bW)$ in Theorem~\ref{thm::reg-general} and $\tbW^\textup{wls}$ affects both the first and second terms of the variance formula. As a result, the restricted weighted least squares estimator no longer has the BLUE property. Nevertheless, the EHW-type variance estimator remains asymptotically conservative, since $S^2(\btheta(\tbW^\textup{wls}) ) $ is positive semidefinite. Theorem~\ref{thm::reg-general-average} also holds for the restricted weighted least squares estimation with an arbitrary weight matrix $\Gamma$, as Theorem~\ref{thm::reg-general-constantweight}. To avoid repetition, we omit the details. 

Although our proposed estimator no longer achieves the BLUE property under the population-level assumptions, it retains an optimality property in terms of the estimated variance, as shown in the following theorem.

\begin{theorem}
\label{thm::var-compare}
Let $\widehat{V}_{\textsc{EHW}}^\textup{wls2}$ be the EHW-type variance estimator for the restricted weighted least squares estimation with an arbitrary weight matrix. Then
\[
\widehat{V}_{\textsc{EHW}}^\textup{wls2}- \widehat{V}_{\textsc{EHW}} = C + o_{\mathbb{P}}(N^{-1}),
\]
where $C$ is positive semi-definite.
\end{theorem}

Theorem~\ref{thm::var-compare} shows that, among all restricted weighted least squares estimators with arbitrary weight matrices, the corresponding EHW-type variance estimator in Theorem~\ref{thm::reg-general-average} is asymptotically optimal.
In other words, under the population-level versions of Assumptions~\ref{asm::noanticipation}–\ref{asm::invarianteffect}, the estimator in Theorem~\ref{thm::reg-general-average}  may not achieve optimal sampling precision, but its EHW-type variance estimator attains the smallest estimated precesion within this class.  These results are reminiscent of the discussion of ``sampling precision'' and ``estimated precision'' in Li and Ding \cite{li2020rerandomization} on regression adjustment in rerandomization.

\subsection{Testing assumptions}
\label{sec::extension-testing}

We consider testing the assumptions imposed in our framework by testing the corresponding linear restrictions in the regression-based analysis. Classical approaches rely on the unrestricted estimator, which requires the regression coefficients to be identifiable without imposing restrictions. In our setting, this condition may fail because $X$ may not have full column rank. Therefore, We need a different approach.  The following theorem provides the basis for testing these assumptions in this setting.

\begin{theorem}
\label{thm::rwls-testing}
Suppose that  $X^\top X + C^\top C$ is of full rank and the conditions in Corollary~\ref{thm::inference} hold. 
For Scenarios (a), (b), and (c), the statistic
\begin{eqnarray}
\label{eqn::teststat}
(Y-X\widehat{\bgamma}^\textup{fwls})^\top  \widehat \Omega^{-1}  X  (C^\top \widehat U_{21} X^\top \widehat \Omega^{-1} \widehat{\Sigma}  \widehat \Omega^{-1} X \widehat U_{12}C)^+  X^\top \widehat \Omega^{-1}   (Y-X \widehat{\bgamma}^\textup{fwls}) 
\end{eqnarray}
is asymptotically stochastically dominated by a chi-squared distribution with degrees of freedom $\text{rank} (X \widehat U_{12})$, where the superscript $+$ denotes the Moore--Penrose inverse.
\end{theorem}
The statistic in~\eqref{eqn::teststat} takes the form of a score statistic. The term $X^\top \widehat \Omega^{-1}   (Y-X \widehat{\bgamma}^\textup{fwls})$ is asymptotically normal. However, similar to the variance estimation of $ \widehat{\bgamma}^\textup{fwls}$, we can only construct a conservative estimator of its covariance matrix, given by the middle term of~\eqref{eqn::teststat}. This leads to the stochastic dominance by a chi-squared distribution. 
As a result, 
Theorem~\ref{thm::rwls-testing} yields  a conservative test based on the above statistic and the corresponding chi-squared distribution.

Interestingly, even the model-based analogue of Theorem~\ref{thm::rwls-testing} is not well documented in the literature. We provide a corresponding discussion in the online supplementary material for comparison with Theorem~\ref{thm::rwls-testing} .


\section{Application}
\label{sec::app}
Before presenting the empirical application, we briefly summarize the main findings from the simulation studies in the online supplementary material. The regression-based estimators are approximately unbiased across all scenarios, and their variance estimators are conservative in general. When the weights are estimated from the data, finite-sample bias in the estimated variances may lead to mild undercoverage, particularly in small samples, whereas this issue diminishes with larger sample sizes or when unweighted estimators are used. In contrast, conventional model-based approaches can exhibit substantial bias and undercoverage when their implicit assumptions are violated. 

We then reanalyze the data from Jones and Kenward \cite{jones1987modelling} using the proposed regression-based approach. The data come from a three-period crossover trial evaluating the effect of an analgesic on dysmenorrhea. The treatment has three conditions: a placebo and two active doses of the analgesic.
For simplicity and to better illustrate our methodology, we combine the two active treatments into a single condition, denoted by treatment $A$, and label the placebo as condition $B$. This results in a simplified crossover design with three observed treatment sequences: $AAB$, $ABA$, and $BAA$. The outcome is a binary indicator of pain relief. The original analysis in Jones and Kenward \cite{jones1987modelling} employs a logistic regression model, and thus the validity of the results depends on correct model specification.

In contrast, our regression-based approach enables valid inference without requiring the outcome model to be correctly specified. Following this principle, we apply linear regression even though the outcome is binary \citep{freedman2008randomization,basse2023minimax}.
 With three time periods, we are interested in estimating the following effects:
\begin{itemize}
 \item instantaneous effects: $\tau_1$, $\tau_2(z_1)$, and $\tau_3(z_1z_2)$ for $z_1,z_2=A,B$;
 \item  first order carryover effects: $\tau_2^1(z_1)$, $\tau_3^1(z_1,z_3)$ for $z_1,z_3=A,B$; 
 \item  second order carryover effects: $\tau_3^2(z_1z_2)$ for $z_1,z_2=A,B$.
 \end{itemize}

Based on the identifiability conditions in Theorem~\ref{thm::reg-general}, the causal effects are not identifiable under Scenario (a), Scenario (b) with $k=2$, or Scenario (c) with $k=2$.
Therefore, we focus on  the following two scenarios:
\begin{itemize}
\item Scenario (b1), which corresponds to Assumption~\ref{asm::noanticipation}, and Assumption~\ref{asm::nocarryover} with order $k=1$.
\item Scenario (c1), which corresponds to Assumption~\ref{asm::noanticipation}, and Assumptions~\ref{asm::nocarryover}~and~\ref{asm::invarianteffect} with order $k=1$.
\end{itemize}
We assess the plausibility of the imposed assumptions using the proposed test in Theorem~\ref{thm::rwls-testing}. The resulting p-values are $0.056$ for Scenario (b1) and $0.177$ for Scenario (c1). 

We compute point estimates and corresponding 95\% confidence intervals for the identifiable causal effects using restricted weighted least squares. Under both scenarios, all carryover effects are assumed to be zero, and the instantaneous causal effects do not depend on treatment history. Therefore, we report only the three instantaneous causal effects in Table~\ref{tab::realdata}.
For comparison, we also fit the linear mixed-effects model in~\eqref{eqn::lme} without carryover effects. The estimated treatment effect is $0.280$ with a 95\% confidence interval of $(0.160, 0.400)$. This result is broadly consistent with the design-based results, leading to similar substantive conclusions.

Under Scenario (b1), the estimated effects are all positive and statistically significant, suggesting that the analgesic is effective at relieving pain across all three periods. Moreover, the effects are similar across time, indicating no substantial variation in treatment efficacy over the course of the study. Consequently, when the additional constraint of time-invariant effects is imposed in Scenario (c1), the point estimates remain similar, but the confidence intervals become narrower, reflecting improved estimation efficiency.
Overall, the findings under both scenarios are consistent with those reported in the original logistic regression analysis in \cite{jones1987modelling}.

\begin{table}[htpb]
\centering
\caption{Point estimates and 95\% confidence intervals for the three instantaneous causal effects under Scenarios (b1) and (c1).}
\begin{tabular}{lccc}
\toprule
\textbf{Scenario} & $\tau_1$ & $\tau_2$ & $\tau_3$ \\
\midrule
(b1) & 0.552 (0.368, 0.737) & 0.509 (0.320, 0.698) & 0.582 (0.392, 0.772) \\
(c1) & 0.547 (0.421, 0.673) & 0.547 (0.421, 0.673) & 0.547 (0.421, 0.673) \\
\bottomrule
\end{tabular}
\label{tab::realdata}
\end{table}

\section{Discussion}
\label{sec::discussion}

We discuss several features of the framework and directions for further development.

First, the framework accommodates a broad class of outcome types. The proposed estimators  apply directly to continuous, discrete, and binary outcomes, as they target well-defined causal estimands and yield unbiased estimates under randomization without requiring additional modeling assumptions. In contrast, nonlinear models rely on additional assumptions that are not justified by the randomization and may not align with the causal estimands of interest \citep{freedman2008randomization,guo2023generalized}.

Extending the framework to time-to-event outcomes is more challenging due to the presence of censoring. In crossover designs, censoring may arise from dropout or death, preventing subsequent treatments from being implemented and thereby altering the effective treatment assignment mechanism. This complicates both the design and the definition of causal estimands. Existing approaches based on inverse probability weighting can address censoring in simpler settings \citep{cheng2022addressing}, and recent work has begun to study related issues in sequential randomized experiments with dropout \citep{jiang2024longitudinal}. Extending our framework to accommodate time-to-event outcomes remains an important direction for future research.


Second, the framework applies to a broad class of experimental designs. In the absence of interference between units, the design is represented through the regressor matrix, while assumptions on potential outcomes are encoded as constraints on the regression coefficients. Randomization then guarantees the design-based properties of the resulting estimators. This representation allows the framework to accommodate a wide range of treatment assignment mechanisms, including those studied in \cite{bajari2021multiple,bajari2023experimental,missault2025robust}. Among these, Missault et al. \cite{missault2025robust} propose a class of designs that includes crossover designs as a special case, although their analysis focuses on a more limited set of estimands and does not adopt a unified regression-based formulation.
 The framework can also accommodate clustered settings, including clustered crossover designs, by appropriately redefining the unit of analysis, and is related to recent work on staggered rollout designs that similarly index potential outcomes by treatment sequences and address anticipation and time-varying effects under clustered assignment \citep{chen2025model1,chen2025model2}.

The framework may also be extended to observational longitudinal studies. A natural approach is to incorporate inverse probability weights based on propensity scores, paralleling marginal structural models \citep{robins2000marginal}. In addition, the restrictions in our framework may help improve efficiency or relax overlap requirements under suitable conditions. Developing these extensions requires additional assumptions and careful theoretical analysis, which we leave for future research.

Third, our framework allows for covariate adjustment to improve efficiency in randomized experiments. A standard approach is to include interactions between covariates and treatment indicators, which preserves identifiability and improves efficiency asymptotically. However, extending this approach to crossover designs introduces additional challenges. The number of parameters can grow quickly when allowing for interactions with treatment sequences, which may be problematic in small samples, and the presence of carryover effects and imposed restrictions further complicates the role of covariates. An alternative is to use inverse probability weighting based on estimated treatment assignment probabilities conditional on covariates, followed by restricted weighted least squares. In simple settings, this approach is equivalent to regression adjustment \citep{robins2007comment}, but its extension to crossover designs remains to be developed.


\section*{Acknowledgement}

We thank Iavor Bojinov,  Yuanhong Cai, Zhihao Jin, Pengfei Tian, and Panos Toulis for their helpful comments and suggestions.

\pdfbookmark[1]{References}{References}
\spacingset{1.45}
\bibliographystyle{Chicago}
\bibliography{crossover-ref}

\newpage

\pagenumbering{arabic} 
\renewcommand*{\thepage}{S\arabic{page}}

\setcounter{equation}{0}
\renewcommand {\theequation} {S\arabic{equation}}
\renewcommand {\thelemma} {S\arabic{lemma}}
\renewcommand {\theproposition} {S\arabic{proposition}}
\setcounter{section}{0}
\setcounter{proposition}{0}
\renewcommand {\thesection} {S\arabic{section}}

\begin{center}
  \LARGE {\bf Supplementary Material}
\end{center}

Section~\ref{app::additional} presents additional results. Section~\ref{app::necessary} outlines identification conditions for the instantaneous and carryover effects that do not rely on the rank condition in Theorem~\ref{thm::reg-general}, as discussed in Section~\ref{sec::interpretation}.
  Section~\ref{app::computation} details the computational procedures for implementing the point and variance estimator, as described in Section~\ref{sec::reg-general-practice}. 
  
 Section~\ref{app::simulation} provides  simulation results.

Section~\ref{app::rwls} develops the theoretical foundation for restricted weighted least squares estimation. The results are not only the building blocks for our proofs but also of independent interest. 

Section~\ref{app::proof} presents the proofs of all results.

\section{Additional results}
\label{app::additional}

\subsection{Weaker identification conditions in general crossover designs}
\label{app::necessary}

Section~\ref{sec::reg-general} presents a unified regression-based framework for analyzing general crossover designs, which requires that the matrix $X^\top X + C^\top C$ be of full rank. When this condition is violated, more specialized procedures are needed to investigate the estimation of the causal effect.

 We first provide the result for the scenario with only  Assumption~\ref{asm::noanticipation}. 
 \begin{proposition}
\label{prop::rand-general}
Suppose that Assumption~\ref{asm::noanticipation} holds. 
Consider any $\os' \in \mathcal{S}$ and $t=1,\ldots,T$. If there exists a treatment sequence $\os \in \mathcal{S}^\textup{obs}$ such that $\os_{[1,t]} = \os'_{[1,t]}$, then $\wY_t(\os)$ is unbiased for $\oY_{t}(\os'_{[1,t]})$.
\end{proposition}
Assumption~\ref{asm::noanticipation} implies $\oY_{t}(\os') = \oY_{t}(\os)$ when $\os_{[1,t]} = \os'_{[1,t]}$. Thus the unbiased estimator for $\oY_{t}(\os)$ is also unbiased for $\oY_{t}(\os')$. We can show that the condition in Proposition~\ref{prop::rand-general} is also necessary in the sense that if an unbiased estimator exists for any value of $\oY_{t}(\os')$, then there must exist a treatment sequence $\os \in \mathcal{S}^\textup{obs}$ such that $\os_{[1,t]} = \os'_{[1,t]}$.

The condition in Proposition~\ref{prop::rand-general} can be relaxed if Assumption~\ref{asm::nocarryover} holds in addition, as stated in Proposition~\ref{prop::rand-nocarryover}.
\begin{proposition}
\label{prop::rand-nocarryover}
Suppose that  Assumption~\ref{asm::noanticipation} holds, and Assumption~\ref{asm::nocarryover} holds with order $k < T$.
Consider any $\os' \in \mathcal{S}$.   
When $t\leq k$, if there exists a treatment sequence $\os \in \mathcal{S}^\textup{obs}$ such that $\os_{[1,t]} = \os'_{[1,t]}$, then $\wY_t(\os)$ is unbiased for $\oY_{t}(\os'_{[1,t]})$; when $t>k$, if there exists a treatment sequence $\os \in \mathcal{S}^\textup{obs}$ such that $\os_{[t-k+1,t]} = \os'_{[t-k+1,t]}$, then $\wY_t(\os)$ is unbiased for $\oY_{t}(\os'_{[1,t]})$.
\end{proposition}
When $t\leq k$, the condition in Proposition~\ref{prop::rand-nocarryover} is identical to that in Proposition~\ref{prop::rand-general}. When $t>k$, Proposition~\ref{prop::rand-nocarryover}  only requires that  $\os$ and $\os'$ share the same  treatment conditions from period $t-k+1$ to $t$.  We provide an example to illustrate Proposition~\ref{prop::rand-nocarryover}.
Consider a crossover design with $T=3$ under Assumptions~\ref{asm::noanticipation}, and Assumption~\ref{asm::nocarryover} with $k=2$.
We have $\oY_2(AA)=\oY_3(z_1AA)$ for any $z_1=A,B$. If the treatment sequence $AAA$ or $BAA$ is implemented in the experiment,  then
we can obtain an unbiased estimator for $\oY_2(AA)$.
Similar to Proposition~\ref{prop::rand-general}, we can also show that these conditions are necessary.

Finally, we consider the scenario with both Assumptions~\ref{asm::nocarryover}~and~\ref{asm::invarianteffect}.
\begin{proposition}
\label{prop::rand-invarianteffect}
Suppose that  Assumption~\ref{asm::noanticipation} holds, and Assumptions~\ref{asm::nocarryover}~and~\ref{asm::invarianteffect} hold with order $k < T$
Consider any $\os' \in \mathcal{S}$.   
Then $ \oY_{t}(\os')$ can be unbiasedly estimated under either of the following conditions:
\begin{itemize}
\item[(a)]  Conditions in  Proposition~\ref{prop::rand-nocarryover} hold;
\item[(b)]  If there exist three treatment sequences $\os'_1, \os'_2,\os'_3 \in \mathcal{S}$ and $t' \geq k$ such that $\os_{1,[t'-k+1,t']} = \os'_{[t-k+1,t]}$ and $\os_{2,[t-k+1,t]}=\os_{3,[t'-k+1,t']}$, and $\oY_{t'}(\os'_1)$, $\oY_t(\os'_2)$, and $\oY_{t'}(\os'_3)$ can be unbiasedly estimated.
\end{itemize}
\end{proposition}
Proposition~\ref{prop::rand-invarianteffect} relaxes the conditions in Proposition~\ref{prop::rand-nocarryover} with an additional condition in Proposition~\ref{prop::rand-invarianteffect}(b), which implies  $ \oY_t(\os')-\oY_t(\os'_2) =\oY_{t'}(\os'_1)-\oY_{t'}(\os'_3) $  under Assumption~\ref{asm::invarianteffect}. 
The condition is similar to the parallel trends assumption in the difference-in-differences method in the sense that it assumes the effect of treatment sequence $\os_{1,[t'-k+1,t']} = \os'_{[t-k+1,t]}$ versus $\os_{2,[t-k+1,t]}=\os_{3,[t'-k+1,t']}$ is the same at time $t$ and $t'$.
Under this condition, we can use the unbiased estimators of  $\oY_{t'}(\os_1)$, $\oY_{t'}(\os'_3)$, and $\wY_{t}(\os'_2)  $ to obtain an unbiased estimator of  $ \oY_t(\os')$.
We provide an example to illustrate Proposition~\ref{prop::rand-invarianteffect}(b).
Consider a crossover design under Assumption~\ref{asm::nocarryover}~and~\ref{asm::invarianteffect}  with $k=2$.
We have $\oY_2(AB)-\oY_2(AA)=\oY_3(AB)-\oY_3(AA)$. If $\oY_2(AB)$, $\oY_3(AB)$, and $\oY_3(AA)$  can be unbiasedly estimated, then
we can obtain an unbiased estimator of $\oY_2(AA)$.

\subsection{Computational details}
\label{app::computation}

\subsubsection{Forms of the constraint matrix}

We provide the forms of $C_1$, $C_2$, and $C_3$ for the general $T$-period crossover design. 
The coefficient vector $\gamma$ contains $T\cdot 2^T$ parameters. The constraint matrices can be written as
\[
C_1 = A_1, \quad 
C_2 = (A_1^\top, A_{2k}^\top)^\top, \quad 
C_3 = (A_1^\top, A_{2k}^\top, A_{3k}^\top)^\top,
\]
where $A_1 \bgamma = 0$ corresponds to Constraint~\ref{cons::noanticipation}, $A_{2k} \bgamma = 0$ corresponds to Constraint~\ref{cons::carryover} with order $k$, and $A_{3k} \bgamma = 0$ corresponds to Constraint~\ref{cons::invarianteffect} with order $k$.

We first construct $A_1$. For each period $t$, partition $\mathcal{S}$ according to the prefix $\os_{[1,t]}$. At period $t$, each equivalence class consists of full treatment sequences that share the same treatment history up to period $t$ but may differ in future treatment assignments. Under Constraint~\ref{cons::noanticipation}, the coefficients $\gamma_{t,\os}$ must be invariant to future treatments and are therefore equal within each class.

For each equivalence class, choose one reference sequence $\os^\ast$ within the class, and impose
\begin{eqnarray*}
\gamma_{t,\os}-\gamma_{t,\os^\ast}=0
\end{eqnarray*}
for every other sequence $\os$ in that class. Collecting these rows over all $t$ and all equivalence classes gives the matrix $A_1$.

At period $t$, there are $2^t$ equivalence classes, each containing $2^{T-t}$ sequences. Therefore, the number of independent constraints contributed by period $t$ is $2^t \cdot (2^{T-t}-1)=2^T-2^t$. Summing over $t=1,\ldots,T$, the number of rows in matrix $A_1$ is
\begin{eqnarray*}
\sum_{t=1}^T (2^T-2^t)= (T-2)\cdot 2^T+2.
\end{eqnarray*}

We next construct $A_{2k}$ as additional constraints on top of $A_1$. Under $A_1$, the coefficients $\gamma_{t,\os}$ depend only on the prefix $\os_{[1,t]}$, so that at period $t$ there are $2^t$ distinct groups.

Constraint~\ref{cons::carryover} further requires that, for $t\geq k$, the coefficients depend only on the most recent $k$ treatments. Therefore, the additional independent constraints can be expressed at the prefix level: for each period $t\geq k$, partition the set of prefixes $\os_{[1,t]}$ according to their last $k$ treatments $\os_{[t-k+1,t]}$. Each equivalence class consists of prefixes that share the same most recent $k$ treatments but may differ in earlier treatments.

For each equivalence class, choose one reference prefix $\os^\ast_{[1,t]}$ within the class, and impose
\begin{eqnarray*}
\gamma_{t,\os_{[1,t]}}-\gamma_{t,\os^\ast_{[1,t]}}=0
\end{eqnarray*}
for every other prefix $\os_{[1,t]}$ in that class.

To construct the matrix $A_{2k}$, the rows must be written in terms of the original coefficient vector $\bgamma$, whose entries are indexed by full treatment sequences. For each such prefix-level constraint, choose arbitrary full-sequence extensions $\os$ and $\os^\ast$ consistent with the prefixes $\os_{[1,t]}$ and $\os^\ast_{[1,t]}$, and impose
\begin{eqnarray*}
\gamma_{t,\os}-\gamma_{t,\os^\ast}=0.
\end{eqnarray*}
This is well defined because, under $A_1$, the choice of treatments beyond period $t$ does not affect the coefficient.

We now count the number of independent constraints. Under $A_1$, there are $2^t$ distinct groups at period $t$, whereas Constraint~\ref{cons::carryover} reduces this to $2^k$ groups. Therefore, the number of additional independent constraints contributed by period $t$ is $2^t-2^k$. Summing over $t=k,\ldots,T$, the number of additional rows in matrix $A_{2k}$ is
\begin{eqnarray*}
\sum_{t=k}^T (2^t-2^k) = 2^{T+1}-(T-k+2)\cdot 2^k.
\end{eqnarray*}

We finally construct $A_{3k}$. Under $A_{2k}$, the coefficients $\gamma_{t,\os}$ for $t\geq k$ depend only on the most recent $k$ treatments. Therefore, for each period $t\geq k$, we can partition $\mathcal{S}$ according to $\os_{[t-k+1,t]}$, with each equivalence class consisting of full sequences that share the same treatment history over the most recent $k$ periods.

Constraint~\ref{cons::invarianteffect} requires that for any two such histories, the difference in coefficients is invariant across periods. Therefore, the additional independent constraints can be expressed at the reduced level as follows. Fix a reference history $h^\ast \in \{A,B\}^k$, where $\{A,B\}^k$ denotes the set of all length-$k$ treatment sequences.. For each $t>k$, and for each history $h\neq h^\ast$, we impose
\begin{eqnarray*}
(\gamma_{t,\os_{[t-k+1,t]}=h}-\gamma_{t,\os_{[t-k+1,t]}=h^\ast})
-
(\gamma_{k,\os_{[1,k]}=h}-\gamma_{k,\os_{[1,k]}=h^\ast})
=0,
\end{eqnarray*}
where the notation indicates coefficients corresponding to sequences with the specified recent treatment histories.

To construct the matrix $A_{3k}$, the rows must be written in terms of the original coefficient vector $\bgamma$, whose entries are indexed by full treatment sequences. For each such constraint, choose arbitrary full-sequence extensions $\os^1, \os^2,  \os^3,  \os^4$ such that
\begin{eqnarray*}
\os^1_{[t-k+1,t]} = h,\quad 
\os^2_{[t-k+1,t]} = h^\ast,\quad
\os^3_{[1,k]} = h,\quad 
\os^4_{[1,k]} = h^\ast,
\end{eqnarray*}
and impose
\begin{eqnarray*}
(\gamma_{t,\os^1}-\gamma_{t,\os^2})
-
(\gamma_{k,\os^3}-\gamma_{k,\os^4})
=0.
\end{eqnarray*}
This is well defined because, under $A_1$ and $A_{2k}$, the coefficients depend only on the most recent $k$ treatments.

We now count the number of independent constraints. For each period $t>k$, there are $2^k-1$ independent contrasts relative to the reference history $h^\ast$. Therefore, the number of additional rows in matrix $A_{3k}$ is
\begin{eqnarray*}
\sum_{t=k+1}^T (2^k-1) = (T-k)(2^k-1).
\end{eqnarray*}

\subsubsection{The restricted weighted least squares procedure}
We provide more details for the regression-based estimation for a crossover design with restriction $C \bgamma = 0$.
For each unit, we can write the regressor matrix as 
\begin{eqnarray*}
X_i &=&  (\bm{1}(\oS_i=\os): \os \in \mathcal{S})\otimes I_T
\end{eqnarray*}
and the weight matrix as $ \widehat{\Omega}_i= \widehat{\bS}^2(\oS_i)$.
For units assigned to the same treatment sequence $\oS_i=\os$, the weight matrix is identical and can be denoted by
$
\widehat{\Omega}_{\os}=\widehat{\bS}^2(\os).
$
We have 
\begin{eqnarray}
\nonumber  X^\top \widehat{\Omega}^{-1} X &=& \sum_{i=1}^N ( (\bm{1}(\oS_i=\os): \os \in \mathcal{S})\otimes I_T)^\top \widehat{\Omega}_i^{-1} ((\bm{1}(\oS_i=\os): \os \in \mathcal{S})\otimes I_T)\\
\nonumber  &=&\sum_{\os \in \mathcal{S}^\textup{obs}} \sum_{i: \oS_i=\os} ( (\bm{1}(\oS_i=\os): \os \in \mathcal{S})\otimes I_T)^\top \widehat{\Omega}_i^{-1} ((\bm{1}(\oS_i=\os): \os \in \mathcal{S}^\textup{obs})\otimes I_T)\\
\label{eqn::comp1}&=&\text{diag}( N_{\os} \widehat{\Omega}_{\os}^{-1}: \os  \in \mathcal{S}),\\
\nonumber  X^\top \widehat{\Omega}^{-1} Y &=& \sum_{i=1}^N ( (\bm{1}(\oS_i=\os): \os \in \mathcal{S})\otimes I_T)^\top \widehat{\Omega}_i^{-1} \bY_i\\
\nonumber  &=&\sum_{\os \in \mathcal{S}^\textup{obs}} \sum_{i: \oS_i=\os} ( (\bm{1}(\oS_i=\os): \os \in \mathcal{S})\otimes I_T)^\top \widehat{\Omega}_i^{-1}  \bY_i\\
\label{eqn::comp2}&=& (    N_{\os} \widehat{\Omega}_{\os}^{-1} \wbY(\os): \os  \in \mathcal{S})^\top,
\end{eqnarray}
where, for $\os \notin  \mathcal{S}^\textup{obs}$, we set $N_{\os} \widehat{\Omega}_{\os}^{-1}=0$ and $N_{\os} \widehat{\Omega}_{\os}^{-1} \wbY(\os)=0$.
We can then compute $\widehat{U}_{11}$ from matrix inversion in~\eqref{eqn::U-hat} and use it to obtain the restricted weighted least squares estimator:
$\widehat{\bgamma}^\textup{fwls}  = \widehat{U}_{11}X^\top \widehat{\Omega}^{-1} Y $.

We then compute the EHW variance estimator for $\widehat{\bgamma}^\textup{fwls}$:
\begin{eqnarray*}
\widehat{V}_{\textsc{EHW}}&=&\widehat{U}_{11}X^\top \widehat \Omega^{-1}  \widehat{\Sigma} \widehat \Omega^{-1}X \widehat{U}_{11},
\end{eqnarray*}
where $\widehat{\Sigma} = \text{diag}\{\widehat{\Sigma}_i: i=1,\ldots,N\}$ is a diagonal matrix with diagonal entries
 $\widehat{\Sigma}_i = (Y_i -X_i\widehat{\bgamma}^\textup{fwls} )(Y_i -X_i \widehat{\bgamma}^\textup{fwls} )^\top$. 
 We have 
 \begin{eqnarray}
X^\top \widehat \Omega^{-1}  \widehat{\Sigma} \widehat \Omega^{-1}X 
\nonumber &=&\sum_{\os \in \mathcal{S}^\textup{obs}} \sum_{i: \oS_i=\os}  X_i^\top \widehat{\Omega}_i^{-1} \widehat{\Sigma}_i \widehat{\Omega}_i^{-1} X_i\\
\nonumber &=&\sum_{\os \in \mathcal{S}^\textup{obs}} \left\{  X_i^\top \widehat{\Omega}_{\os}^{-1}\left(\sum_{i: \oS_i=\os}\widehat{\Sigma}_i  \right)\widehat{\Omega}_{\os}^{-1} X_i\right\}\\
\label{eqn::comp3} &=& \text{diag}\left( \widehat{\Omega}_{\os}^{-1}\left(\sum_{i: \oS_i=\os}\widehat{\Sigma}_i  \right)\widehat{\Omega}_{\os}^{-1} : \os  \in \mathcal{S}\right).
\end{eqnarray}
The formulas in~\eqref{eqn::comp1}~to~\eqref{eqn::comp3} facilitate the computation of the restricted least squares procedure.
 

\subsubsection{Illustration of the restricted weighted least squares procedure in two-period crossover designs}
\label{sec::reg-2p}

We apply the regression-based approach to analyze the two typical two-period crossover designs studied in Sections~\ref{sec::2p4t}~and~\ref{sec::2p2t}.
Suppose that we are interested in the two conditional treatment effects at period $2$, $\tau_2(A)$ and $\tau_2(B)$. We first write $(\tau_2(A),\tau_2(B))^\top$ in terms of
\begin{eqnarray*}
\btheta(\bW)&=& \bW(AA) \obY(AA)+\bW(AB) \obY(AB)+\bW(BA) \obY(BA)+\bW(BB) \obY(BB),
\end{eqnarray*}
where $ \obY(z_1z_2)= (\obY_1(z_1z_2),\obY_2(z_1z_2))^\top$ for $z_1,z_2=A,B$ and
\begin{eqnarray*}
\bW(AA) &=&\begin{pmatrix}
0& 1\\
0 & 0 
\end{pmatrix}, \quad \bW(AB) \ =\ \begin{pmatrix}
0& -1\\
0 & 0 
\end{pmatrix}, \quad \bW(BA) \ =\ \begin{pmatrix}
0& 0\\
0 & 1 
\end{pmatrix}, \quad \bW(BB) \ =\ \begin{pmatrix}
0& 0\\
0 & -1 
\end{pmatrix}.
\end{eqnarray*}
Therefore, we have $\mathcal{S}=\{AA,AB,BA,BB\}$.

We then specify the regression equation:
\begin{eqnarray*}
Y_{i1}&=& \sum_{z_1,z_2=A,B} \bm{1}(\oS_i=z_1z_2)\gamma_{1,z_1z_2}+e_{i1},\\
Y_{i2}&=& \sum_{z_1,z_2=A,B} \bm{1}(\oS_i=z_1z_2)\gamma_{2,z_1z_2}+e_{i2}.
\end{eqnarray*}
with parameter $\bgamma = (\gamma_{1,AA},\gamma_{2,AA},\gamma_{1,AB},\gamma_{2,AB},\gamma_{1,BA},\gamma_{2,BA},\gamma_{1,BB},\gamma_{2,BB})^\top$.
These equations can be written  in matrix form as  $\bY_i = X_i\bgamma+\bm{e}_i$, where the regressor matrix $X_i$ is 
 \begin{eqnarray*}
X_i&=& (\bm{1}(\oS_i=AA), \bm{1}(\oS_i=AB),\bm{1}(\oS_i=BA),\bm{1}(\oS_i=BB))\otimes I_2.
\end{eqnarray*}

Next, we write out the constraints $C\bgamma=0$ under different sets of assumptions. 
We illustrate the construction of $C$ in the crossover design with two periods and four treatment sequences: $AA$, $AB$, $BA$, and $BB$. 
The constraint matrices can be written as $C_1=A_1$, $C_2=(A_1^\top,A_2^\top)^\top$, and $C_3=(A_1^\top,A_2^\top,A_3^\top)^\top$, where
$A_1 \bgamma = 0 $ corresponds to Constraint~\ref{cons::noanticipation},
$ A_2 \bgamma = 0$ corresponds to Constraint~\ref{cons::carryover} with order $k = 1$, and 
$ A_3 \bgamma = 0$  corresponds to Constraint~\ref{cons::invarianteffect} with order $k = 1$.
 Constraint~\ref{cons::noanticipation} requires $\gamma_{1,z_1A}=\gamma_{1,z_1B}$ for $z_1=A,B$, which leads to
 \begin{eqnarray*}
A_1&=&\begin{pmatrix}
1 &0&-1 &0&0&0&0&0\\
0 &0 &0&0&1&0&-1&0
\end{pmatrix}.
\end{eqnarray*}
 Constraint~\ref{cons::carryover} requires $\gamma_{2,Az_2}=\gamma_{2,Bz_2}$ for $z_2=A,B$, which leads to 
 \begin{eqnarray*}
A_2&=&\begin{pmatrix}
0 &1 &0 &0&0&-1&0&0\\
0 &0 &0 &1&0&0&0&-1
\end{pmatrix}.
\end{eqnarray*}
Constraint~\ref{cons::invarianteffect} requires $\gamma_{1,Az_2}- \gamma_{1,Bz_2'} =\gamma_{2,z_1A}- \gamma_{2,z_1'B}  $ for $z_1,z_1',z_2,z_2'=A,B$.
After eliminating linearly dependent constraints that already captured by $A_1$ and $A_2$, we retain a single condition: $\gamma_{1,AA}- \gamma_{1,BA} =\gamma_{2,AA}- \gamma_{2,AB}  $. This gives
$$A_3=(1,-1, 0, 1,-1,0,0,0).$$

We then investigate the estimation of $\btheta(\bW)$ under the two designs separately.
For the crossover design with two periods and four sequences,  we have  
$$X^\top X= \text{diag}(N_{AA},N_{AA},N_{AB},N_{AB},N_{BA},N_{BA},N_{BB},N_{BB} ),$$
 which is of full rank.
Therefore, from Theorem~\ref{thm::reg-general},  $\btheta(\bW)$ can be unbiasedly estimated under all of the three scenarios considered in Section~\ref{sec::reg-general}. The corresponding point and variance estimators can be obtained using the regression-based procedure, which yields results equivalent to the analytic expressions derived in Section~\ref{sec::2p4t}.

For the crossover design with two periods and two sequences, $N_{AA}=N_{BB}=0$. Therefore, $X^\top X$ is not of full rank. With Assumption~\ref{asm::noanticipation}, we have  
\begin{eqnarray*}
X^\top X + C_1^\top C_1&=& \begin{pmatrix}
1&0 &-1 &0&0&0&0&0\\
0 &0 &0 &0&0&0&0&0\\
-1&0 & N_{AB}+1 &0&0&0&0&0\\
0 &0 &0 &N_{AB}&0&0&0&0\\
0 &0 &0 &0&N_{BA}+1&0&-1&0\\
0 &0 &0 &0&0&N_{AB}&0&0\\
0 &0 &0 &0&-1&0&1&0\\
0 &0 &0 &0&0&0&0&0
\end{pmatrix},
\end{eqnarray*}
which is not of full rank because it contains two columns of zeros. This rank deficiency necessitates Proposition~\ref{prop::2p2t}  in Section~\ref{sec::2p2t} to study the identification of $\btheta(\bW)$ under  Assumption~\ref{asm::noanticipation}. When Assumption~\ref{asm::nocarryover} with $k=1$ holds in addition, we have 
\begin{eqnarray*}
X^\top X + C_2^\top C_2&=& \begin{pmatrix}
1&0 &-1 &0&0&0&0&0\\
0 &1 &0 &0&0&-1&0&0\\
-1&0 & N_{AB}+1 &0&0&0&0&0\\
0 &0 &0 &N_{AB}+1&0&0&0&-1\\
0 &0 &0 &0&N_{BA}+1&0&-1&0\\
0 &-1 &0 &0&0&N_{BA}+1&0&0\\
0 &0 &0 &0&-1&0&1&0\\
0 &0 &0 &-1&0&0&0&1
\end{pmatrix},
\end{eqnarray*}
which can  be shown to have a non-zero determinant of $N_{AB}^2 N_{BA}^2$. Therefore,  $\btheta(\bW)$ is identifiable in this scenario. 
Because $C_3$ includes additional constraints than $C_2$, this also implies that the matrix $X^\top X + C_3^\top C_3$ is of full rank. Consequently, $\btheta(\bW)$ is identifiable under Assumption~\ref{asm::noanticipation}, and  Assumptions~\ref{asm::nocarryover}~and~\ref{asm::invarianteffect} with $k=1$. 
We can then compute the BLUE of $\btheta(\bW)$ and the corresponding variance estimators using the regression-based procedure under these two scenarios. This yields results equivalent to the analytic expressions derived in Section~\ref{sec::2p2t}.

\section{Simulation studies}
\label{app::simulation}

We conduct simulations to evaluate the finite-sample performance of the regression-based approach under two crossover designs with two time periods. Section~\ref{sec::reg-2p} establishes identification of the causal effects using this framework. Our simulations assess the accuracy of the point estimates and the reliability of the associated variance estimators.

\subsection{Data-generating process with non-constant individual causal effect}

We first generate the potential outcomes  using the following model: 
\begin{eqnarray}
\label{eqn::sim-gen1}
\begin{pmatrix}
Y_{i1}(z_1z_2)\\
Y_{i2}(z_1z_2)
\end{pmatrix}
\sim \textnormal{N} \left( \begin{pmatrix}
\beta_{1z_1z_2}\\
\beta_{2z_1z_2}
\end{pmatrix}, \quad \begin{pmatrix}
1 & \rho \\
\rho & 1
\end{pmatrix}
  \right).
\end{eqnarray}
We then modify the generated potential outcomes to satisfy the conditions of Scenarios (a), (b), and (c):
\begin{itemize}
\item Scenario (a): For each unit and each $z = A, B$, we set $Y_{i1}(zB) = Y_{i1}(zA)$, so that Assumption~\ref{asm::noanticipation} holds.

\item Scenario (b): In addition to the constraints in Scenario (a), we set $Y_{i2}(Bz) = Y_{i2}(Az)$ for each $z = A, B$, so that Assumption~\ref{asm::nocarryover} holds in addition to Assumption~\ref{asm::noanticipation}.

\item Scenario (c): In addition to the constraints in Scenario (b), we set $Y_{i2}(AA) = Y_{i1}(AA) - Y_{i1}(BA) + Y_{i2}(AB)$ and $Y_{i2}(BA) = Y_{i1}(AA) - Y_{i1}(BA) + Y_{i2}(BB)$ for each unit, so that  Assumption~\ref{asm::invarianteffect} holds in addition to Assumption~\ref{asm::noanticipation}~and~\ref{asm::nocarryover}.
\end{itemize}
We fix the potential outcomes in the simulation and generate the treatment sequences according to the crossover design.
For the design with four treatment sequences, we set $N_{AA}=N_{AB}=N_{BA}=N_{BB}=N/4$; for the design with two treatment sequences, we set $N_{AB}=N_{BA}=N/2$. We set $N=100$.

Our analysis targets five causal effects $\tau_1$, $\tau_2(z)$, and $\tau_2^1(z)$ for $z=A,B$.
We examine all three scenarios under the four-sequence design, and Scenarios (b) and (c) under the two-sequence design, since not all effects are identifiable under Scenario (a). We set $(\beta_{1AA},\beta_{1AB},\beta_{1BA},\beta_{1BB})=(0,0.5,1,0.5)$, $(\beta_{2AA},\beta_{2AB},\beta_{2BA},\beta_{2BB})=(0,0.5,0.5,1)$, and $\rho=0.3$. We compute both the point and the variance estimates using the proposed restricted weighted least squares.

\begin{figure}[!t]
\centering

\begin{minipage}{0.85\textwidth}
  \centering
  \includegraphics[width=\textwidth]{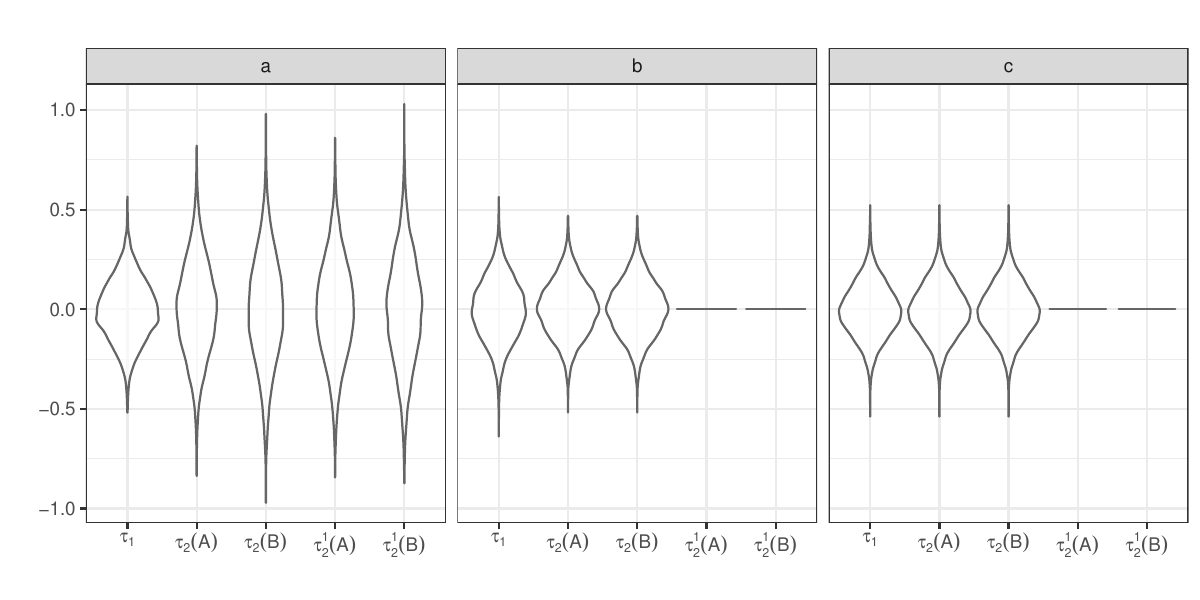}
  
  \textbf{(a)} Two-period crossover design with four treatment sequences with $N=100$. The true values of $\tau_2^1(A)$ and $\tau_2^1(B)$ are restricted to be zero under Scenarios (b) and (c).
\end{minipage}

\vspace{1em}

\begin{minipage}{0.85\textwidth}
  \centering
  \includegraphics[width=\textwidth]{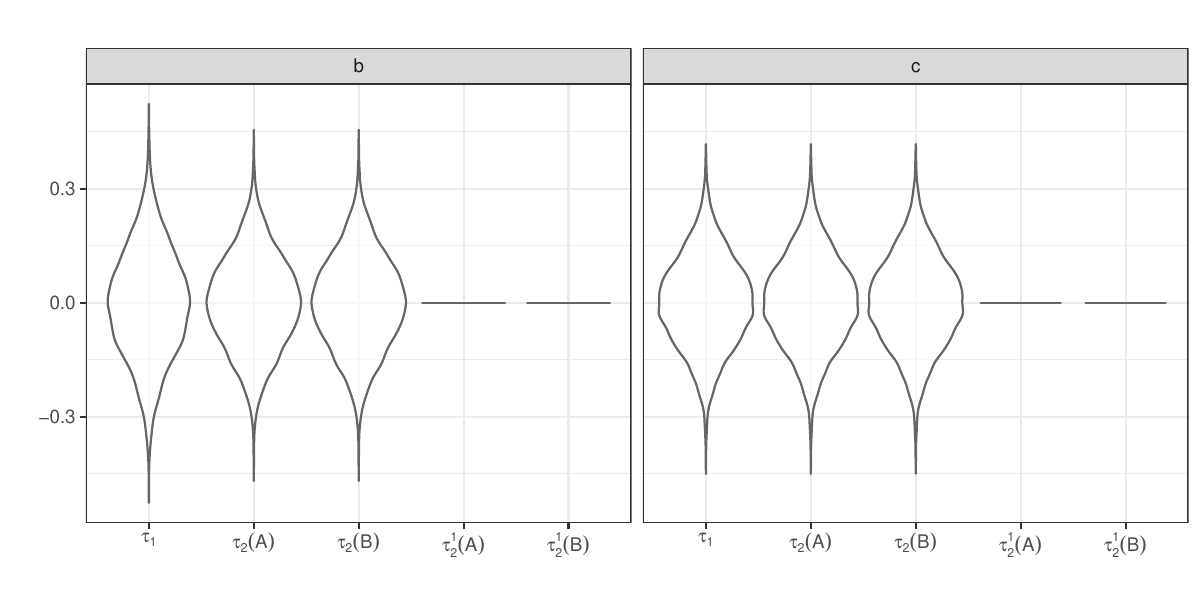}
  
  \textbf{(b)} Two-period crossover design with two treatment sequences.  The true values of $\tau_2^1(A)$ and $\tau_2^1(B)$ are restricted to be zero under both scenarios.
\end{minipage}

\caption{Bias of the regression-based estimator under two crossover designs with $N=100$. Each panel shows violin plots for five causal effects under one scenario.}
\label{fig::sim-combined}
\end{figure}

Figure~\ref{fig::sim-combined}(a) presents the distribution of biases of the estimators across $10,000$ independent complete randomizations under the four-sequence design with $N=100$. The estimates of $\tau_2^1(A)$ and $\tau_2^1(B)$ are exactly zero under Scenarios (b) and (c), reflecting the imposed restrictions. In all cases, the biases center closely around zero, confirming the consistency of our regression-based estimators. Moreover, the variability of the estimators decreases as more restrictions are introduced from Scenarios (a) to (c), which aligns with the intuition that additional assumptions improve precision. Figure~\ref{fig::sim-combined}(b) shows similar patterns for the two-sequence design.

\begin{figure}[!t]
\centering

\begin{minipage}{0.85\textwidth}
  \centering
  \includegraphics[width=\textwidth]{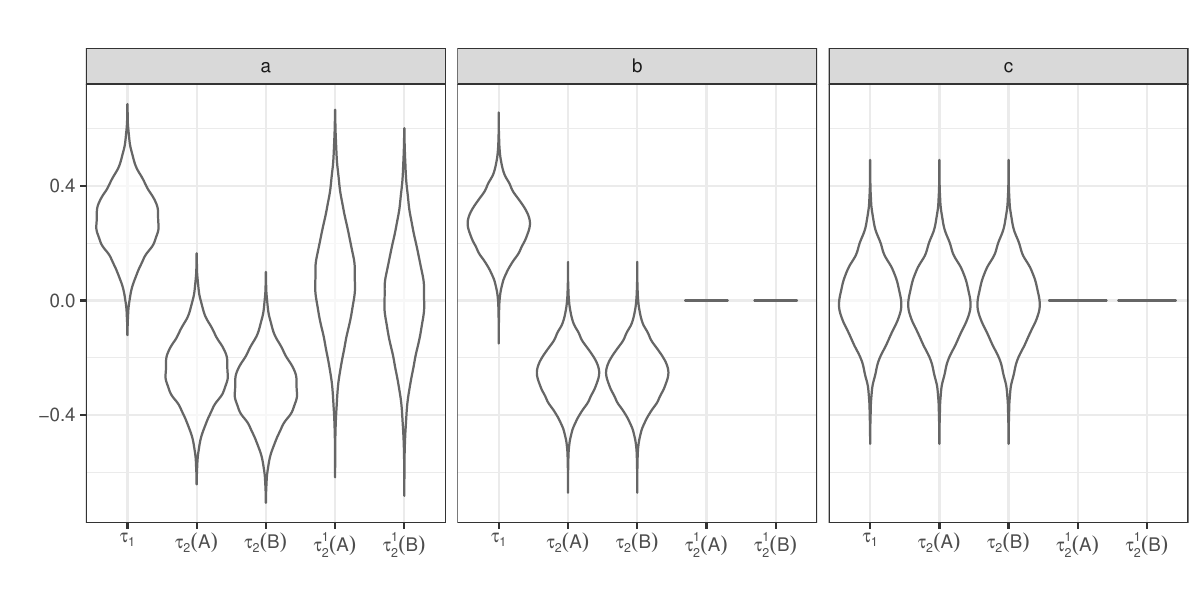}
  
  \textbf{(a)} Two-period crossover design with four treatment sequences with $N=100$. The true values of $\tau_2^1(A)$ and $\tau_2^1(B)$ are restricted to be zero under Scenarios (b) and (c).
\end{minipage}

\vspace{1em}

\begin{minipage}{0.85\textwidth}
  \centering
  \includegraphics[width=\textwidth]{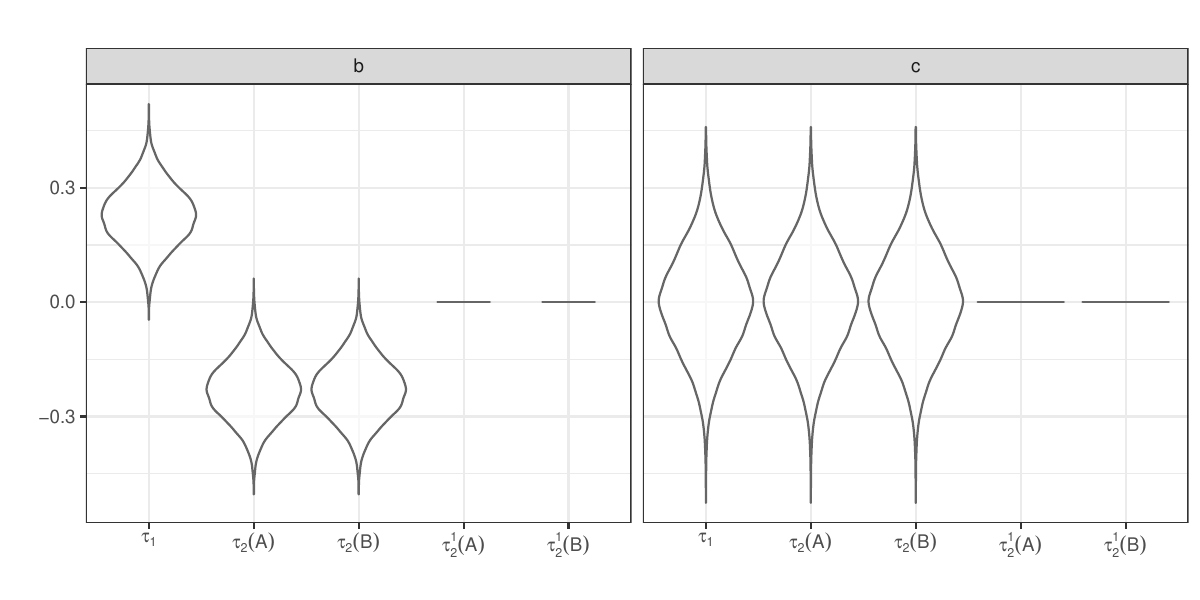}
  
  \textbf{(b)} Two-period crossover design with two treatment sequences.  The true values of $\tau_2^1(A)$ and $\tau_2^1(B)$ are restricted to be zero under both scenarios.
\end{minipage}

\caption{Bias of estimators based on the linear mixed-effects model under two crossover designs with $N=100$. Each panel shows violin plots for five causal effects under one scenario.}
\label{fig::sim-lme}
\end{figure}

We next examine a conventional model-based analysis using a linear mixed-effects model commonly employed in crossover studies \citep{senn2002cross,jones2003design}. We consider the specification
\begin{eqnarray}
\label{eqn::lme}
Y_{it} &=& \pi_t +\beta Z_{it} + \lambda Z_{i,t-1}+ \alpha_i+\epsilon_{it},
\end{eqnarray}
where $Z_{i,0}=0$, $\pi_t$ is the period effect, $\beta$ is the treatment effect, $\lambda$ is the first-order carryover effect, $\alpha_i$ is a random effect for unit $i$, and $\epsilon_{it}$ is the random error. Under Scenario (a), we estimate both $\beta$ and $\lambda$. Under Scenarios (b) and (c), we impose $\lambda=0$ and focus on $\beta$.

Because the model uses a single parameter to represent the treatment effect, we evaluate the estimator of $\beta$ with respect to $\tau_1$, $\tau_2(A)$, and $\tau_2(B)$. Similarly, we evaluate the estimator of $\lambda$ with respect to $\tau_2^1(A)$ and $\tau_2^1(B)$. Figure~\ref{fig::sim-lme} presents the results. The estimators are biased under Scenarios (a) and (b), whereas the biases are centered around zero under Scenario (c). This pattern is consistent with the fact that the linear mixed-effects model implicitly imposes a common treatment effect across periods and treatment histories. 

To assess the variance estimators, we construct 95\% confidence intervals using the estimated variances and compute empirical coverage rates across simulations. Table~\ref{tab::coverage}(I) shows these rates. Since $\tau_2^1(A)$ and $\tau_2^1(B)$ are set to zero in Scenarios (b) and (c), the coverage for these effects is expected to be 1. For the other estimands, the coverage is also close to 1, reflecting the conservative nature of the variance estimators. The degree of conservativeness depends on the amount of heterogeneity in individual causal effects.
We also consider an unweighted restricted least squares estimator, which replaces the estimated weights with the identity matrix. Table~\ref{tab::coverage-constant}(I) reports the corresponding coverage rates, which are similar to those in Table~\ref{tab::coverage}(I).
Finally, Table~\ref{tab::coverage-lme}(I) reports the coverage rates for estimators based on a linear mixed-effects model, which exhibit substantial undercoverage in Scenarios (a) and (b).

\begin{table}[htbp]
\centering
\caption{Coverage rates for 95\% confidence intervals under two data-generating models using the restricted weighted  least squares estimator. 
The true values of $\tau_2^1(A)$ and $\tau_2^1(B)$ are restricted to be zero in Scenarios (b) and (c), so their coverage rates are expected to be 1.}
\label{tab::coverage}

\renewcommand{\arraystretch}{1.1}
\setlength{\tabcolsep}{4pt}

\begin{minipage}{0.48\textwidth}
\centering
\textbf{(I) Data-generating process from~\eqref{eqn::sim-gen1}}\\[0.3em]
\resizebox{\textwidth}{!}{%
\begin{tabular}{lccccc}
\toprule
\textbf{Scenario} & $\tau_1$ & $\tau_2(A)$ & $\tau_2(B)$ & $\tau_2^1(A)$ & $\tau_2^1(B)$ \\
\midrule
\multicolumn{6}{l}{\hspace{1em}\textit{Four-sequence crossover design}} \\
(a) & 0.989 & 0.969 & 0.961 & 0.963 & 0.960 \\
(b) & 0.988 & 0.991 & 0.991 & 1.000 & 1.000 \\
(c) & 0.992 & 0.992 & 0.992 & 1.000 & 1.000 \\
\multicolumn{6}{l}{\hspace{1em}\textit{Two-sequence crossover design}} \\
(b) & 0.993 & 0.997 & 0.997 & 1.000 & 1.000 \\
(c) & 0.997 & 0.997 & 0.997 & 1.000 & 1.000 \\
\bottomrule
\end{tabular}
}
\end{minipage}
\hfill
\begin{minipage}{0.48\textwidth}
\centering
\textbf{(II) Constant treatment effect model}\\[0.3em]
\resizebox{\textwidth}{!}{%
\begin{tabular}{lccccc}
\toprule
\textbf{Scenario} & $\tau_1$ & $\tau_2(A)$ & $\tau_2(B)$ & $\tau_2^1(A)$ & $\tau_2^1(B)$ \\
\midrule
\multicolumn{6}{l}{\hspace{1em}\textit{Four-sequence crossover design}} \\
(a) & 0.935 & 0.934 & 0.934 & 0.934 & 0.935 \\
(b) & 0.928 & 0.924 & 0.924 & 1.000 & 1.000 \\
(c) & 0.922 & 0.922 & 0.922 & 1.000 & 1.000 \\
\multicolumn{6}{l}{\hspace{1em}\textit{Two-sequence crossover design}} \\
(b) & 0.947 & 0.946 & 0.946 & 1.000 & 1.000 \\
(c) & 0.943 & 0.943 & 0.943 & 1.000 & 1.000 \\
\bottomrule
\end{tabular}
}
\end{minipage}

\end{table}

\begin{table}[htbp]
\centering
\caption{Coverage rates for 95\% confidence intervals under two data-generating models using the unweighted restricted least squares estimator. 
The true values of $\tau_2^1(A)$ and $\tau_2^1(B)$ are restricted to be zero in Scenarios (b) and (c), so their coverage rates are expected to be 1.}
\label{tab::coverage-constant}

\renewcommand{\arraystretch}{1.1}
\setlength{\tabcolsep}{4pt}

\begin{minipage}{0.48\textwidth}
\centering
\textbf{(I) Data-generating process from~\eqref{eqn::sim-gen1}}\\[0.3em]
\resizebox{\textwidth}{!}{%
\begin{tabular}{lccccc}
\toprule
\textbf{Scenario} & $\tau_1$ & $\tau_2(A)$ & $\tau_2(B)$ & $\tau_2^1(A)$ & $\tau_2^1(B)$ \\
\midrule
\multicolumn{6}{l}{\hspace{1em}\textit{Four-sequence crossover design}} \\
(a) & 0.992 & 0.973 & 0.967 & 0.966 & 0.965 \\
(b) & 0.994 & 0.996 & 0.996 & 1.000 & 1.000 \\
(c) & 0.995 & 0.995 & 0.995 & 1.000 & 1.000 \\
\multicolumn{6}{l}{\hspace{1em}\textit{Two-sequence crossover design}} \\
(b) & 0.993 & 0.997 & 0.997 & 1.000 & 1.000 \\
(c) & 0.995 & 0.995 & 0.995 & 1.000 & 1.000 \\
\bottomrule
\end{tabular}
}
\end{minipage}
\hfill
\begin{minipage}{0.48\textwidth}
\centering
\textbf{(II) Constant treatment effect model}\\[0.3em]
\resizebox{\textwidth}{!}{%
\begin{tabular}{lccccc}
\toprule
\textbf{Scenario} & $\tau_1$ & $\tau_2(A)$ & $\tau_2(B)$ & $\tau_2^1(A)$ & $\tau_2^1(B)$ \\
\midrule
\multicolumn{6}{l}{\hspace{1em}\textit{Four-sequence crossover design}} \\
(a) & 0.945 & 0.939 & 0.941 & 0.941 & 0.940 \\
(b) & 0.944 & 0.944 & 0.944 & 1.000 & 1.000 \\
(c) & 0.944 & 0.944 & 0.944 & 1.000 & 1.000 \\
\multicolumn{6}{l}{\hspace{1em}\textit{Two-sequence crossover design}} \\
(b) & 0.947 & 0.946 & 0.946 & 1.000 & 1.000 \\
(c) & 0.944 & 0.944 & 0.944 & 1.000 & 1.000 \\
\bottomrule
\end{tabular}
}
\end{minipage}

\end{table}

\begin{table}[htbp]
\centering
\caption{Coverage rates for 95\% confidence intervals under two data-generating models using the linear mixed effects model with $N=100$.  
The true values of $\tau_2^1(A)$ and $\tau_2^1(B)$ are restricted to be zero in Scenarios (b) and (c), so their coverage rates are expected to be 1.}
\label{tab::coverage-lme}

\renewcommand{\arraystretch}{1.1}
\setlength{\tabcolsep}{4pt}

\begin{minipage}{0.48\textwidth}
\centering
\textbf{(I) Data-generating process from~\eqref{eqn::sim-gen1}}\\[0.3em]
\resizebox{\textwidth}{!}{%
\begin{tabular}{lccccc}
\toprule
\textbf{Scenario} & $\tau_1$ & $\tau_2(A)$ & $\tau_2(B)$ & $\tau_2^1(A)$ & $\tau_2^1(B)$ \\
\midrule
\multicolumn{6}{l}{\hspace{1em}\textit{Four-sequence crossover design}} \\
(a) & 0.514 & 0.632 & 0.426 & 0.952 & 0.966 \\
(b) & 0.554 & 0.567 & 0.567 & 1.000 & 1.000 \\
(c) & 0.985 & 0.985 & 0.985 & 1.000 & 1.000 \\
\multicolumn{6}{l}{\hspace{1em}\textit{Two-sequence crossover design}} \\
(b) & 0.613 & 0.599 & 0.599 & 1.000 & 1.000 \\
(c) & 0.988 & 0.988 & 0.988 & 1.000 & 1.000 \\
\bottomrule
\end{tabular}
}
\end{minipage}
\hfill
\begin{minipage}{0.48\textwidth}
\centering
\textbf{(II) Constant treatment effect model}\\[0.3em]
\resizebox{\textwidth}{!}{%
\begin{tabular}{lccccc}
\toprule
\textbf{Scenario} & $\tau_1$ & $\tau_2(A)$ & $\tau_2(B)$ & $\tau_2^1(A)$ & $\tau_2^1(B)$ \\
\midrule
\multicolumn{6}{l}{\hspace{1em}\textit{Four-sequence crossover design}} \\
(a) & 0.950 & 0.950 & 0.950 & 0.938 & 0.938 \\
(b) & 0.947 & 0.947 & 0.947 & 1.000 & 1.000 \\
(c) & 0.944 & 0.944 & 0.944 & 1.000 & 1.000 \\
\multicolumn{6}{l}{\hspace{1em}\textit{Two-sequence crossover design}} \\
(b) & 0.938 & 0.938 & 0.938 & 1.000 & 1.000 \\
(c) & 0.938 & 0.938 & 0.938 & 1.000 & 1.000 \\
\bottomrule
\end{tabular}
}
\end{minipage}

\end{table}

\subsection{Data-generating process with constant individual causal effect} 
We then consider another data-generating mechanism in which individual causal effects are constant across units. Specifically, we independently generate $Y_{i1}(BA)=Y_{i1}(BB)$ and $Y_{i2}(BB)$ from a standard Normal distribution. We then compute the other potential outcomes based on these two potential outcomes by adding fixed treatment effects:
\begin{eqnarray*}
Y_{i1}(AA)&=&Y_{i1}(AB)\ =\ Y_{i1}(BB)+ \tau_1, \\
  Y_{i2}(AA)&=& Y_{i2}(BA)+\tau_2^1(A),\\
 Y_{i2}(AB)& =& Y_{i2}(BB)+\tau_2^1(B), \\ 
 Y_{i2}(BA)&=&Y_{i2}(BB)+\tau_2(B).
\end{eqnarray*}
We set $\tau_1=\tau_2(A)=\tau_2(B)=1$ and $\tau_2^1(A)=\tau_2^1(B)=0$ so that the generated potential outcomes satisfy the conditions under all three scenarios.
Under this setting, the variance estimators are consistent for the true variances.

Table~\ref{tab::coverage}(II) shows the empirical coverage rates of the  restricted weighted least squares estimator   under this second setting. The coverage rates for the four-sequence design are around $0.93$, which are below the nominal 95\% level. This undercoverage is primarily due to finite-sample bias arising from estimating the weights using sample variances within each treatment sequence. In contrast, the coverage rates for the two-sequence design are closer to 95\%, likely because larger group sizes lead to more stable variance estimation. 
In Table~\ref{tab::coverageN500}, we increase the sample size to $N=500$, and the undercoverage issue diminishes.

\begin{table}[!t]
\centering
\caption{Coverage rates for 95\% confidence intervals under two data-generating models using the restricted weighted least squares estimator with $N=500$. 
The true values of $\tau_2^1(A)$ and $\tau_2^1(B)$ are restricted to be zero in Scenarios (b) and (c), so their coverage rates are expected to be 1.}
\label{tab::coverageN500}

\renewcommand{\arraystretch}{1.1}
\setlength{\tabcolsep}{4pt}

\begin{minipage}{0.48\textwidth}
\centering
\textbf{(I) Data-generating process from~\eqref{eqn::sim-gen1}}\\[0.3em]
\resizebox{\textwidth}{!}{%
\begin{tabular}{lccccc}
\toprule
\textbf{Scenario} & $\tau_1$ & $\tau_2(A)$ & $\tau_2(B)$ & $\tau_2^1(A)$ & $\tau_2^1(B)$ \\
\midrule
\multicolumn{6}{l}{\hspace{1em}\textit{Four-sequence crossover design}} \\
(a) & 0.994 & 0.974 & 0.973 & 0.975 & 0.975 \\
(b) & 0.994 & 0.994 & 0.994 & 1.000 & 1.000 \\
(c) & 0.998 & 0.998 & 0.998 & 1.000 & 1.000 \\
\multicolumn{6}{l}{\hspace{1em}\textit{Two-sequence crossover design}} \\
(b) & 0.996 & 0.995 & 0.995 & 1.000 & 1.000 \\
(c) & 0.997 & 0.997 & 0.997 & 1.000 & 1.000 \\
\bottomrule
\end{tabular}
}
\end{minipage}
\hfill
\begin{minipage}{0.48\textwidth}
\centering
\textbf{(II) Constant treatment effect model}\\[0.3em]
\resizebox{\textwidth}{!}{%
\begin{tabular}{lccccc}
\toprule
\textbf{Scenario} & $\tau_1$ & $\tau_2(A)$ & $\tau_2(B)$ & $\tau_2^1(A)$ & $\tau_2^1(B)$ \\
\midrule
\multicolumn{6}{l}{\hspace{1em}\textit{Four-sequence crossover design}} \\
(a) & 0.948 & 0.945 & 0.953 & 0.948 & 0.945 \\
(b) & 0.948 & 0.949 & 0.949 & 1.000 & 1.000 \\
(c) & 0.942 & 0.942 & 0.942 & 1.000 & 1.000 \\
\multicolumn{6}{l}{\hspace{1em}\textit{Two-sequence crossover design}} \\
(b) & 0.951 & 0.951 & 0.951 & 1.000 & 1.000 \\
(c) & 0.946 & 0.946 & 0.946 & 1.000 & 1.000 \\
\bottomrule
\end{tabular}
}
\end{minipage}

\end{table}

 Tables~\ref{tab::coverage-constant}(II)~and~\ref{tab::coverageN500-constant}(II) show the empirical coverage rates for the unweighted restricted least squares estimator under the setting with constant individual treatment effects, with sample size $N=100$ and $N=500$, respectively. The coverage rates are close to 95\%, indicating improved small-sample performance due to the absence of weight estimation.

\begin{table}[!t]
\centering
\caption{Coverage rates for 95\% confidence intervals under two data-generating models using the unweighted restricted least squares estimator with $N=500$. 
The true values of $\tau_2^1(A)$ and $\tau_2^1(B)$ are restricted to be zero in Scenarios (b) and (c), so their coverage rates are expected to be 1.}
\label{tab::coverageN500-constant}

\renewcommand{\arraystretch}{1.1}
\setlength{\tabcolsep}{4pt}

\begin{minipage}{0.48\textwidth}
\centering
\textbf{(I) Data-generating process from~\eqref{eqn::sim-gen1}}\\[0.3em]
\resizebox{\textwidth}{!}{%
\begin{tabular}{lccccc}
\toprule
\textbf{Scenario} & $\tau_1$ & $\tau_2(A)$ & $\tau_2(B)$ & $\tau_2^1(A)$ & $\tau_2^1(B)$ \\
\midrule
\multicolumn{6}{l}{\hspace{1em}\textit{Four-sequence crossover design}} \\
(a) & 0.994 & 0.974 & 0.973 & 0.975 & 0.976 \\
(b) & 0.994 & 0.995 & 0.995 & 1.000 & 1.000 \\
(c) & 0.994 & 0.994 & 0.994 & 1.000 & 1.000 \\
\multicolumn{6}{l}{\hspace{1em}\textit{Two-sequence crossover design}} \\
(b) & 0.996 & 0.995 & 0.995 & 1.000 & 1.000 \\
(c) & 0.994 & 0.994 & 0.994 & 1.000 & 1.000 \\
\bottomrule
\end{tabular}
}
\end{minipage}
\hfill
\begin{minipage}{0.48\textwidth}
\centering
\textbf{(II) Constant treatment effect model}\\[0.3em]
\resizebox{\textwidth}{!}{%
\begin{tabular}{lccccc}
\toprule
\textbf{Scenario} & $\tau_1$ & $\tau_2(A)$ & $\tau_2(B)$ & $\tau_2^1(A)$ & $\tau_2^1(B)$ \\
\midrule
\multicolumn{6}{l}{\hspace{1em}\textit{Four-sequence crossover design}} \\
(a) & 0.950 & 0.946 & 0.954 & 0.948 & 0.946 \\
(b) & 0.950 & 0.952 & 0.952 & 1.000 & 1.000 \\
(c) & 0.944 & 0.944 & 0.944 & 1.000 & 1.000 \\
\multicolumn{6}{l}{\hspace{1em}\textit{Two-sequence crossover design}} \\
(b) & 0.951 & 0.951 & 0.951 & 1.000 & 1.000 \\
(c) & 0.947 & 0.947 & 0.947 & 1.000 & 1.000 \\
\bottomrule
\end{tabular}
}
\end{minipage}

\end{table}

\section{Novel lemmas on restricted weighted least squares}
\label{app::rwls}
To address the vector outcome considered in the main text, we develop the theory for restricted weighted least squares estimation with a vector outcome.
For unit $i$, let $Y_i = (Y_{i1},\ldots,Y_{iT})^\top \in \mathbb{R}^T$ be the outcome vector and $X_i=(X_{i1},\ldots,X_{iT})^\top \in \mathbb{R}^{T\times p}$ the regressor matrix. Consider the following linear model:
\begin{eqnarray*}
Y_i &=& X_i \beta+ \epsilon_i,
\end{eqnarray*}
where $\epsilon_i  \in \mathbb{R}^T$ is the error term with mean $0$ and invertible covariance matrix $\Sigma_i  \in \mathbb{R}^{T\times T}$. 
The parameters satisfy the constraint
 $$C\beta = r,$$ 
 where $C  \in \mathbb{R}^{l\times p}$ has full row rank and $r  \in \mathbb{R}^l$ is a constant vector.
We can stack the outcome and the regressor matrix of the units and write the model in matrix form
\begin{eqnarray}
\label{eqn::model-rwls}
Y &=& X \beta+ \epsilon, \quad \E(\epsilon)\ =\ 0,\quad \cov(\epsilon)\ =\ \Sigma,
\end{eqnarray}
where
\begin{eqnarray*}
Y = \begin{pmatrix}
Y_1\\
Y_2\\
\vdots\\
Y_n
\end{pmatrix},\quad X= \begin{pmatrix}
X_1\\
X_2\\
\vdots\\
X_n
\end{pmatrix},\quad \epsilon = \begin{pmatrix}
\epsilon_1\\
\epsilon_2\\
\vdots\\
\epsilon_n
\end{pmatrix}, \quad
\Sigma = \begin{pmatrix}
\Sigma_1 &  0 & \cdots & 0 \\ 
0 & \Sigma_2 & \cdots  & 0 \\
\vdots & \vdots& \ddots &\vdots\\
0 & \cdots & 0 &   \Sigma_n 
\end{pmatrix}.
\end{eqnarray*}

\subsection{Estimation}
We consider the weighted least squares estimation of~\eqref{eqn::model-rwls} under restriction with an invertible working covariance matrix 
\begin{eqnarray*}
\Omega = \begin{pmatrix}
\Omega_1 &  0 & \cdots & 0 \\ 
0 & \Omega_2 & \cdots  & 0 \\
\vdots & \vdots& \ddots &\vdots\\
0 & \cdots & 0 &   \Omega_n 
\end{pmatrix},
\end{eqnarray*}
which can differ from $\Sigma$.
The restricted weighted least squares estimation solves the following optimization problem:
\begin{eqnarray*}
&&\underset{\beta}{\argmin} (Y-X\beta)^\top \Omega^{-1} (Y-X\beta),\\
&&\text{ subject to} \hspace{0.5cm}C\beta \ =\  r.
\end{eqnarray*}
The Lagrangian for the problem is 
\begin{eqnarray*}
 (Y-X\beta)^\top\Omega^{-1} (Y-X\beta) -2\lambda^\top(C\beta-r).
\end{eqnarray*}
The first order condition is 
\begin{eqnarray*}
2X^\top\Omega^{-1}(Y-X\beta)-2C^\top\lambda &=&0,\\
C\beta &=& r,
\end{eqnarray*}
which can be written as 
\begin{eqnarray}
\label{eqn::firstorder}
\begin{pmatrix}
X^\top \Omega^{-1}X   &C^\top \\
C & 0  
\end{pmatrix} \begin{pmatrix}
\beta \\
\lambda 
\end{pmatrix} &=& \begin{pmatrix}
X^\top \Omega^{-1}Y\\
r
\end{pmatrix}.
\end{eqnarray}
We introduce a linear algebra result to study the uniqueness of the solution in~\eqref{eqn::firstorder}.
\begin{lemma}
\label{lem::matrix}
Assume $\Omega$ is invertible. The matrix 
\begin{eqnarray*}
\begin{pmatrix}
X^\top \Omega^{-1}X   &C^\top \\
C & 0  
\end{pmatrix}
\end{eqnarray*}
is invertible if and only if  $ X^\top X+ C^\top C$ is full  rank.
\end{lemma}
\noindent {\it Proof of Lemma~\ref{lem::matrix}.}  
Because
 \begin{eqnarray*}
X^\top\Omega^{-1} X+ C^\top C &=& (X^\top \Omega^{-1/2} , C^\top) \begin{pmatrix} \Omega^{-1/2}X \\ C\end{pmatrix},
\end{eqnarray*}
the rank of $X^\top\Omega^{-1} X+ C^\top C$  is equal to the column rank of $(X^\top \Omega^{-1/2} , C^\top)^\top$. Furthermore, because $\Omega$ is invertible, $(X^\top \Omega^{-1/2}, C^\top)$ has the same column rank as 
 $(X^\top, C^\top)^\top$. Therefore, we only
need to show that the matrix 
\begin{eqnarray*}
\begin{pmatrix}
X^\top \Omega^{-1}X   &C^\top \\
C & 0  
\end{pmatrix}
\end{eqnarray*}
is invertible if and only if $(X^\top, C^\top)^\top$ is full column rank.

For the ``if'' direction,  let $(\alpha_1^\top,\alpha_2^\top)^\top$ be a vector such that
\begin{eqnarray}
\label{eqn::matrix}
\begin{pmatrix}
X^\top \Omega^{-1}X   &C^\top \\
C & 0  
\end{pmatrix} \begin{pmatrix}
\alpha_1\\
\alpha_2
\end{pmatrix} &=& \begin{pmatrix}
X^\top \Omega^{-1}X \alpha_1+C^\top \alpha_2 \\
C\alpha_1
\end{pmatrix}  \ = \  0.
\end{eqnarray}
We show that $(\alpha_1^\top,\alpha_2^\top)^\top=0$ must hold.
From~\eqref{eqn::matrix}, we have 
\begin{eqnarray*}
 \alpha_1^\top X^\top \Omega^{-1}X \alpha_1+\alpha_1^\top C^\top \alpha_2&=&  \alpha_1^\top X^\top \Omega^{-1}X \alpha_1\ = \ 0,
\end{eqnarray*}
which implies $X \alpha_1=0$. Therefore, we have $(X^\top,C^\top)^\top \alpha_1=0$. Because $(X^\top,C^\top)^\top$ has full column rank, we obtain $\alpha_1=0$.
Plugging $\alpha_1=0$ into~\eqref{eqn::matrix}, we obtain $C^\top \alpha_2=0$. Because $C$ has full row rank, $\alpha_2=0$.

For the ``only if'' direction, let $\beta$ be a vector such that  $(X^\top,C^\top)^\top \beta=0$. We show that $\beta = 0$ must hold. From $(X^\top,C^\top)^\top \beta=0$, we have $X\beta=0$ and $C\beta=0$, which imply that 
\begin{eqnarray*}
\begin{pmatrix}
X^\top \Omega^{-1}X   &C^\top \\
C & 0  
\end{pmatrix}\beta  \ = \  0.
\end{eqnarray*}
Therefore, $\beta=0$ because the coefficient matrix in the above equation is invertible.
\QEDB


\subsection{Statistical properties}
When the matrix in Lemma~\ref{lem::matrix} is invertible, we define 
\begin{eqnarray}
\label{eqn::U}
U &=&\begin{pmatrix}
X^\top \Omega^{-1}X   &C^\top \\
C & 0  
\end{pmatrix}^{-1} \ = \ \begin{pmatrix}
U_{11}   &U_{12} \\
U_{21} & U_{22}
\end{pmatrix}.
\end{eqnarray}
The following lemma extends the restricted least squares results in \cite{greene1991restricted} by allowing for heteroskedasticity.
\begin{lemma}
\label{lem::rwls-unbias}
Suppose that $ X^\top X+ C^\top C$ is of full rank.
\begin{itemize}
\item[(a)] The restricted weighted least squares estimation under restriction with weight matrix $\Omega$  and restriction $C\beta=r$ has a unique solution
\begin{eqnarray*}
\hat \beta_\textup{r}&=& U_{11}X^\top \Omega^{-1}Y+U_{12}r.
\end{eqnarray*}
\item[(b)] Under the model given in~\eqref{eqn::model-rwls}, $\hat \beta_\textup{r}$ is unbiased for $\beta$ with covariance matrix 
\begin{eqnarray*}
\cov(\hat \beta_\textup{r})&=&U_{11}X^\top \Omega^{-1} \Sigma \Omega^{-1}X U_{11}.
\end{eqnarray*}
 \end{itemize}
\end{lemma}
\noindent {\it Proof of Lemma~\ref{lem::rwls-unbias}.} We first prove (a).
From Lemma~\ref{lem::matrix},  the first order condition in~\eqref{eqn::firstorder} has a unique  solution if  $ X^\top X+ C^\top C$ is of full rank.
From~\eqref{eqn::firstorder},
\begin{eqnarray*}
\hat \beta_\textup{r}&=& U_{11}X^\top \Omega^{-1}Y+U_{12}r,
\end{eqnarray*}
where $U_{11}$ and $U_{22}$ are symmetric, and $U_{12}^\top=U_{21}$.

We then prove (b). From 
\begin{eqnarray*}
\begin{pmatrix}
X^\top \Omega^{-1}X   &C^\top \\
C & 0  
\end{pmatrix}  \begin{pmatrix}
U_{11}   &U_{12} \\
U_{21} & U_{22}
\end{pmatrix} &=& \begin{pmatrix}
I_p  &0 \\
0 & I_l
\end{pmatrix},
\end{eqnarray*}
we obtain 
\begin{eqnarray}
\label{eqn::inverseU1}
X^\top \Omega^{-1}XU_{11}+C^\top U_{21} &=& I_p, \quad  CU_{12}\ =\ I_l,\\
\label{eqn::inverseU2} X^\top \Omega^{-1}XU_{12}+C^\top U_{22}&=&0, \quad   CU_{11}\ =\ 0.
\end{eqnarray}
Therefore, under~\eqref{eqn::model-rwls}, we have
\begin{eqnarray*}
C\hat \beta_\textup{r} &=& CU_{11}X^\top \Omega^{-1}Y+CU_{12}r\ = \ r,\\
\E(\hat \beta_\textup{r})&=& U_{11}X^\top \Omega^{-1}X \beta+U_{12}r\ = (I_p-U_{12}C)\beta+U_{12}r\ =\ \beta,\\
\cov(\hat \beta_\textup{r}) &=&U_{11}X^\top \Omega^{-1} \Sigma \Omega^{-1}X U_{11}.
\end{eqnarray*}
\QEDB

Recall in the main paper that we propose to estimate the sampling variance of $\hat \beta_\textup{r}$ under~\eqref{eqn::model-rwls} by
\begin{eqnarray*}
U_{11}X^\top \Omega^{-1} \widehat \Sigma \Omega^{-1}X U_{11},
\end{eqnarray*}
where  $\widehat{\Sigma} = \text{diag}\{\widehat{\Sigma}_i: i=1,\ldots,N\}$ is a diagonal matrix with diagonal entries
 $\widehat{\Sigma}_i = (Y_i -X_i \hat \beta_\textup{r})(Y_i -X_i \hat \beta_\textup{r})^\top$. We refer to this estimator as the EHW variance estimator.

\subsection{Gauss-Markov Theorem}
We present two versions of the Gauss-Markov Theorem for restricted weighted least squares.

\begin{lemma}[Gauss--Markov Theorem]
\label{lem::rwls-blue}
Under~\eqref{eqn::model-rwls}, if $\Omega=\Sigma$ and $ X^\top X+ C^\top C$ is of full rank, then $\hat \beta_\textup{r}$ is the BLUE.  That is, $\cov(\tilde \beta_\textup{r})-\cov(\hat \beta_\textup{r}) $ is positive semi-definite for any estimator $\tilde \beta_\textup{r} = \tilde c + \tilde A_\textup{r} Y$ with $\tilde c \in \mathbb{R}^p$ and $\tilde A_\textup{r} \in \mathbb{R}^{p\times nT}$   such that $\E(\tilde \beta_\textup{r})=\beta$  and $C\tilde \beta_\textup{r} =r$.
\end{lemma}
\noindent {\it Proof of Lemma~\ref{lem::rwls-blue}.} From the proof of Lemma~\ref{lem::rwls-unbias}, we can write $\hat \beta_\textup{r} = \hat c + \hat A_\textup{r} Y$, where $\hat c =U_{12}r$ and $\hat A_\textup{r} = U_{11}X^\top \Omega^{-1}$. We have $\E(\hat \beta_\textup{r})=\beta$ and $C \hat \beta_\textup{r}=r$.
The estimator $\tilde \beta_\textup{r} $ must satisfy 
\begin{eqnarray*}
\E(\tilde \beta_\textup{r} )&=&  \tilde c + \tilde A_\textup{r} X \beta \ = \ \beta
\end{eqnarray*}
for any $\beta$ that satisfies $C \beta = r $. Pick a specific $\beta_0$ that satisfies $C\beta_0=r$.  
For any $\beta'$ that satisfies $C\beta' = 0$, we have both $C\beta_0=r$ and $C(\beta'+\beta_0)=r$. Thus from the property of $\tilde \beta_\textup{r}$, we obtain
\begin{eqnarray*}
\tilde c + \tilde A_\textup{r} X \beta_0 &=&  \beta_0,\\
\tilde c + \tilde A_\textup{r} X(\beta'+ \beta_0) &=&  \beta'+\beta_0.
\end{eqnarray*}
As a result, $  \tilde A_\textup{r} X \beta'=\beta' $ for any $\beta'$ that satisfies $C \beta'=0$, which implies $\tilde A_\textup{r} X- I_p=LC$ for some matrix $L$.
Because
\begin{eqnarray*}
\cov(\tilde \beta_\textup{r})&=&\cov(\tilde \beta_\textup{r}-\hat \beta_\textup{r})+\cov(\hat \beta_\textup{r})+2\cov(\hat \beta_\textup{r},\tilde \beta_\textup{r}-\hat \beta_\textup{r}),
\end{eqnarray*}
it suffices to show $\cov(\hat \beta_\textup{r},\tilde \beta_\textup{r}-\hat \beta_\textup{r})=0$.
Under~\eqref{eqn::model-rwls} with $\Omega=\Sigma$, we have 
\begin{eqnarray*}
\cov(\hat \beta_\textup{r},\tilde \beta_\textup{r}-\hat \beta_\textup{r})&=& \cov(\hat A_\textup{r}Y,\tilde A_\textup{r}Y-\hat A_\textup{r}Y)\\
&=& \hat A_\textup{r} \Omega ( \tilde A_\textup{r}-\hat A_\textup{r})^\top\\
&=& U_{11}X^\top \tilde A_\textup{r} - U_{11}X^\top \Omega^{-1} X U_{11}\\
&=&U_{11} (I_p+C^\top L^\top)-U_{11}(I_p-C^\top U_{21})\\
&=& 0,
\end{eqnarray*}
where the last equality follows from $U_{11}C^\top=(CU_{11})^\top=0$. \QEDB

\begin{lemma}[Pure algebraic form of the Gauss--Markov Theorem]
\label{lem::rwls-blue-algebra}
Suppose that $ X^\top X+ C^\top C$ is of full rank.  Let $A$ be a matrix such that 
$AX\beta = \beta$  for all $\beta$ satisfying $C\beta = 0$. Let $\hat{A} = U_{11}X^\top \Omega^{-1}$, where $U_{11}$ is defined in~\eqref{eqn::U}.
Then, 
\begin{itemize}
\item[(a)]  $\hat{A} X \beta=\beta$ for all $\beta$ satisfying $C\beta = 0$.
\item[(b)]  $A\Omega A^\top - \hat{A}\Omega \hat{A}^\top $ is positive semi-definite.
\end{itemize}
\end{lemma}
Lemma~\ref{lem::rwls-blue-algebra} focuses on the case with restriction $C\beta=0$. This is without loss of generality, as a restriction of the form $C\beta=r$ can be transformed into this case using the same strategy as in the proof of Lemma~\ref{lem::rwls-blue}. 
Lemma~\ref{lem::rwls-blue-algebra}(a) follows directly from the proof of Lemma~\ref{lem::rwls-unbias}, and Lemma~\ref{lem::rwls-blue-algebra}(b) follows directly from the proof of Lemma~\ref{lem::rwls-blue}.
Lemma~\ref{lem::rwls-blue-algebra} is purely algebraic in the sense that it does not rely on the model given in~\eqref{eqn::model-rwls}.

 \subsection{Testing linear restrictions}

We now consider testing the restriction $C\beta = r$. Classical results are available when $X$ has full column rank, so that the regression coefficients are identified without imposing restrictions \citep{chipman1964treatment}. In this setting, one can first fit the unrestricted weighted least squares and then construct Wald-type tests. However, this approach is not applicable when $X$ does not have full column rank, since the unrestricted estimator is not well defined. 
The following lemma provides the basis for the test in the general setting based on the score test.
\begin{lemma}
\label{lem::rwls-testing}
Under the model given in~\eqref{eqn::model-rwls}, the vector $X^\top \Omega^{-1}   (Y-X \hat{\beta}_\textup{r}) $ has mean zero and covariance matrix
\begin{eqnarray*}
C^\top U_{21} X^\top \Omega^{-1} \Sigma  \Omega^{-1} X U_{12}C.
\end{eqnarray*}
Moreover, if a central limit theorem holds for $X^\top \Omega^{-1}   (Y-X \hat{\beta}_\textup{r}) $ and its covariance matrix can be consistently estimated by the EHW-type variance estimator 
\begin{eqnarray*}
C^\top U_{21} X^\top \Omega^{-1} \widehat{\Sigma}  \Omega^{-1} X U_{12}C,
\end{eqnarray*}
 then
\begin{eqnarray*}
(Y-X \hat{\beta}_\textup{r})^\top  \Omega^{-1}  X  (C^\top U_{21} X^\top \Omega^{-1} \widehat{\Sigma}  \Omega^{-1} X U_{12}C)^+  X^\top \Omega^{-1}   (Y-X \hat{\beta}_\textup{r}) 
\end{eqnarray*}
converges in distribution to a chi-squared distribution with degrees of freedom equal to $\text{rank} (X U_{12})$.
\end{lemma}

\noindent {\it Proof of Lemma~\ref{lem::rwls-testing}. }
We have 
\begin{eqnarray*}
X^\top\Omega^{-1} (Y-X \hat{\beta}_\textup{r}) &=&X^\top \Omega^{-1}(Y - XU_{11}X^\top \Omega^{-1}Y-XU_{12}r)\\
&=&X^\top \Omega^{-1} Y - X^\top \Omega^{-1} XU_{11}X^\top \Omega^{-1}Y-X^\top \Omega^{-1} XU_{12}r\\
&=&X^\top \Omega^{-1} Y  -  (I_p - C^\top U_{21})X^\top \Omega^{-1}Y - X^\top \Omega^{-1} XU_{12}r\\
&=&C^\top U_{21} X^\top \Omega^{-1}Y-X^\top \Omega^{-1} XU_{12}r,
\end{eqnarray*}
where the third equality follows from~\eqref{eqn::inverseU1}.

Under the model given in~\eqref{eqn::model-rwls}, we have 
\begin{eqnarray*}
\E\{X^\top\Omega^{-1} (Y-X \hat{\beta}_\textup{r})\}&=& X^\top\Omega^{-1} X\beta -X^\top \Omega^{-1} X \beta \ = \ 0,\\
\cov\{X^\top \Omega^{-1}(Y-X \hat{\beta}_\textup{r})\}&=& \cov\{ C^\top U_{21} X^\top \Omega^{-1}Y\}\\
&=& C^\top U_{21} X^\top \Omega^{-1} \Sigma  \Omega^{-1} X U_{12}C.
\end{eqnarray*}
By the assumed asymptotic normality, we have 
\begin{eqnarray*}
X^\top \Omega^{-1} (Y-X \hat{\beta}_\textup{r})  \stackrel{d}{\rightarrow} N(0,  C^\top U_{21} X^\top \Omega^{-1} \Sigma  \Omega^{-1} X U_{12}C ).
\end{eqnarray*}
It follows from the standard quadratic-form result for asymptotically normal vectors with possibly singular covariance matrices that
\begin{eqnarray*}
(Y-X \hat{\beta}_\textup{r})^\top \Omega^{-1}  X  (C^\top U_{21} X^\top \Omega^{-1} \Sigma  \Omega^{-1} X U_{12}C)^+  X^\top\Omega^{-1}  (Y-X \hat{\beta}_\textup{r}) 
\end{eqnarray*}
converges in distribution  to a chi-squared distribution with degrees of freedom equal to 
\begin{eqnarray*}
\text{rank} (C^\top U_{21} X^\top \Omega^{-1} \Sigma  \Omega^{-1} X U_{12}C ) &= &  \text{rank}(X U_{12}C) \ = \  \text{rank}(X U_{12}),
\end{eqnarray*}
where the first equality follows from the invertibility of $\Omega$ and $\Sigma$, and the second equality follows from the full row rank of $C$.
 Replacing $ X^\top \Omega^{-1} \Sigma  \Omega^{-1} X $ by its consistent estimator $ X^\top \Omega^{-1} \widehat{\Sigma}  \Omega^{-1} X $ yields the stated statistic. \QEDB

We next show that the test based on Lemma~\ref{lem::rwls-testing} is equivalent to the Lagrange multiplier test \citep{silvey1959lagrangian}. The Lagrange multiplier test is based on the estimated multiplier from the first order condition. From~\eqref{eqn::firstorder}, we have
\[
\hat{\lambda}=U_{21}X^\top\Omega^{-1}Y+U_{22}r.
\]
Moreover,
\begin{eqnarray*}
X^\top\Omega^{-1}(Y-X\hat{\beta}_\textup{r})
&=&X^\top\Omega^{-1}Y-X^\top\Omega^{-1}XU_{11}X^\top\Omega^{-1}Y
-X^\top\Omega^{-1}XU_{12}r\\
&=&C^\top U_{21}X^\top\Omega^{-1}Y+C^\top U_{22}r\\
&=&C^\top\hat{\lambda},
\end{eqnarray*}
where the second equality follows from~\eqref{eqn::inverseU1}~and~\eqref{eqn::inverseU2}.
Because  $C^\top$ has full column rank,  the test statistic based on $X^\top\Omega^{-1}(Y-X\hat{\beta}_\textup{r})$ is equivalent to the Lagrange multiplier test based on $\hat{\lambda}$.

\section{Proofs of the results}
\label{app::proof}

\subsection{Reviews of asymptotic tools in the design-based inference}
\label{app::rand-based}
We first review a general result from \cite{li2017general} to obtain the variances of the estimators in the form of~\eqref{eqn::est-form} under the design-based framework.
Consider an experiment with $N$ units and $Q$ treatments, where $N_q$ units receive treatment $q$ with $\sum_{q=1}^QN_q=N$.
For treatment $q$, let $\bY_i(q)$ be the $p$-dimensional potential outcome of unit $i$ under treatment $q$, and $\overline{\bY}(q)=\sum_{i=1}^N \bY_i(q)/N$ be the average potential outcome  vector of the $N$ units. For any $K\geq 1$ and coefficient matrix  $\bW(q) \in \mathbb{R}^{K\times p}$, consider the individual causal effect of the form $\btheta_i(\bW)=\sum_{q=1}^Q \bW(q) \bY_i(q)$ and the population average causal effect of the form
\begin{eqnarray*}
\btheta(\bW)= (\theta_{(1)}(\bW),\ldots,\theta_{(K)}(\bW))^\top =\frac{1}{N}\sum_{i=1}^N \btheta_i(\bW)=\sum_{q=1}^Q \bW(q) \oY(q).
\end{eqnarray*}

 Let $Z_i$ denote the treatment of unit $i$ and $\widehat{\bY}(q) =\sum_{i=1}^N \bY_i\bone(Z_i=q)/N_q$ be the average outcome under treatment $q$. Then an intuitive  estimator for $\btheta(\bW)$ is 
\begin{eqnarray*}
 \widehat{\theta}(\tbW^\textup{wls})=\sum_{q=1}^Q \bW(q) \widehat{\bY}(q).
\end{eqnarray*}
Define  
\begin{eqnarray*}
\bS^2(q) &=&\frac{1}{N-1} \sum_{i=1}^N \{\bY_i(q)-\overline{\bY}(q)\}\{\bY_i(q)-\overline{\bY}(q)\}^\top, \\
  \bS^2(\btheta(\bW)) &=&\frac{1}{N-1} \sum_{i=1}^N \{\btheta_i(\bW)-\btheta(\bW)\}\{\btheta_i(\bW)-\btheta(\bW)\}^\top,
\end{eqnarray*}
as the variances of the potential outcomes and the individual causal effect, respectively.  The following lemma from \cite{li2017general} shows that $ \widehat{\theta}(\tbW^\textup{wls})$ is unbiased for $\btheta(\bW)$ with the variance depending on $\bS^2(q)$ and $\bS^2_{\btheta(\bW)}$.
\begin{lemma}
\label{lem::var}
In a completely randomized experiment with $N$ units and $Q$ treatments, $\E\{ \widehat{\theta}(\tbW^\textup{wls})\}=\btheta(\bW)$ and 
\begin{eqnarray*}
\cov\{ \widehat{\theta}(\tbW^\textup{wls})\}= \sum_{q=1}^Q \frac{1}{N_q} \bW(q) \bS^2(q) \bW(q)^\top -\frac{1}{N}\bS^2(\btheta(\bW)).
\end{eqnarray*}
\end{lemma}

To construct the confidence sets for $\btheta(\bW)$, we need the central limit theorem for $ \widehat{\theta}(\tbW^\textup{wls})$. 
For $q\neq r$, define
\begin{eqnarray*}
\bS^2(q,r) =\frac{1}{N-1} \sum_{i=1}^N \{\bY_i(q)-\overline{\bY}(q)\}\{\bY_i(r)-\overline{\bY}(r)\}^\top
\end{eqnarray*}
as the covariance between potential outcomes. We state Theorem 5 in  \cite{li2017general} in the following lemma.
\begin{lemma}
\label{lem::clt}
In a completely randomized experiment with $N$ units and $Q$ treatments, if for any $1\leq q\neq r\leq Q$, $\bS^2(q)$ and $\bS^2(q,r)$ have limiting values, $N_q/N$ has positive limiting value, and $\max_{1\leq q \leq Q}\max_{1\leq i \leq N}||\bY_i(q)-\obY(q)||^2_2/N \rightarrow 0$, then $N\cov\{ \widehat{\theta}(\tbW^\textup{wls})\}$ has a limiting value, denoted by $\bm{V}$, and 
\begin{eqnarray*}
N\{ \widehat{\theta}(\tbW^\textup{wls})-\btheta(\bW)\} \stackrel{d}{\rightarrow} N(0,\bm{V}).
\end{eqnarray*}
\end{lemma}
The condition $\max_{1\leq i \leq N}||\bY_i(q)-\obY(q)||^2_2/N \rightarrow 0$ holds if the potential outcome $\bY_i(q)$ is bounded.
We also need to estimate the variance $V$ for constructing confidence sets. However, no valid estimator exists for $S^2_{\btheta(\bW)}$ in $\cov\{ \widehat{\theta}(\tbW^\textup{wls})\}$. 
Define
\begin{eqnarray*}
\widehat \bS^2(q) &=&\frac{1}{N_q-1} \sum_{i:Z_i=q} \{\bY_i-\widehat \bY(q)\}\{\bY_i-\widehat\bY(q)\}^\top
\end{eqnarray*}
as the sample variance estimator for the potential outcome, where
\begin{eqnarray*}
\widehat \bY(q)&=&\frac{1}{N_q} \sum_{i:Z_i=q} \bY_i
\end{eqnarray*}
is the sample average outcome under treatment $Z_i=q$.
The following result from \cite{li2017general} shows that we can instead obtain a conservative variance estimator.
\begin{lemma}
\label{lem::ci}
Under the conditions in Lemma~\ref{lem::clt}, $\widehat \bS^2(q) -\bS^2(q) $ converges to zero in probability for each $q$. 
Define the variance estimator 
\begin{eqnarray*}
\widehat V&=& \sum_{q=1}^Q \frac{1}{N_q} \bW(q) \widehat \bS^2(q) \bW(q)^\top. 
\end{eqnarray*}
If the limit of $N \sum_{q=1}^Q N_q^{-1} \bW(q) \bS^2(q) \bW(q)^\top $ is nonsingular, then the probability that $\widehat V$ is nonsingular converges to one.
\end{lemma}
Lemmas~\ref{lem::clt}~and~\ref{lem::ci} imply that the Wald-type confidence region for $\btheta(\bW)$
\begin{eqnarray*}
\{\mu:\{ \widehat \btheta(\bW)-\mu\}^\top\widehat V^{-1}\{ \widehat \btheta(\bW)-\mu\}\leq \chi^2_{K,1-\alpha} \},
\end{eqnarray*}
where  $\chi^2_{K,1-\alpha} $ is the $1-\alpha$ quantile of a chi-squared distribution with degrees of freedom $k$, has asymptotic coverage rate at least as large as $1-\alpha$.


\subsection{Proof of Proposition~\ref{prop::2p4t}}
It is straightforward to prove the result for $\tau_2(z_1)$ and  $\tau_2^1(z_2)$. Thus, 
we only prove that $\widehat \tau_1$ is the BLUE for $\tau_1$.
We will use the results in Section~\ref{app::rand-based} for the proof.
Consider the estimator in~\eqref{eqn::general-est}, where $\mathcal{S}^\textup{obs}=\{AA,AB,BA,BB\}$ in the crossover design with two periods and four sequences.
 We can write the estimator as
  \begin{eqnarray}
  \label{eqn::proof-2p4t}
\bW(AA) \wbY(AA)+\bW(AB) \wbY(AB)+\bW(BA) \wbY(BA)+\bW(BB) \wbY(BB),
\end{eqnarray}
 where 
 \begin{eqnarray*}
\bW(AA)&=&(\tw_{1}(AA),\tw_{2}(AA)), \quad \bW(AB)\ =\ (\tw_{1}(AB),\tw_{2}(AB)),\\
\bW(BA)&=&(\tw_{1}(BA),\tw_{2}(BA)), \quad \bW(BB)\ =\ (\tw_{1}(BB),\tw_{2}(BB)),
\end{eqnarray*}
and $\obY(z_1z_2)=(\obY_1(z_1z_2),\obY_2(z_1z_2))^\top$ for $z_1,z_2=A,B$.

The unbiasedness requires that 
\begin{eqnarray*}
\label{eqn::proof-2p4t-1}
\bW(AA) \obY(AA)+\bW(AB) \obY(AB)+\bW(BA) \obY(BA)+\bW(BB) \obY(BB)&=& \tau_1.
\end{eqnarray*}
for all potential outcomes satisfying $Y_{i1}(zA)=Y_{i1}(zB)$. This implies that the weights have the form
$\bW(AA) =(\tw_1,0)$, $\bW(AB) =(1-\tw_1,0)$, $\bW(BA)=(-\tw_2,0)$ and $\bW(BB)=(-(1-\tw_2),0)$.  

Define
\begin{eqnarray*}
S^2(\tau) &=& \frac{1}{N-1} \sum_{i=1}^N (\tau_i-\tau)^2 
\end{eqnarray*}
as the variance of the individual causal effect $\tau_i$, 
where $\tau$ can be  $\tau_1$, $\tau_2(z_1)$, and $\tau_2^1(z_2)$ and $\tau_i$ is its individual version. 
From  Lemma~\ref{lem::var}, its variance is 
\begin{eqnarray*}
\tw_1^2 \frac{S^2_{1}(AA)}{N_{AA}}+(1-\tw_1)^2 \frac{S^2_{1}(AB)}{N_{AB}}+\tw_2^2 \frac{S^2_{1}(BA)}{N_{BA}}+(1-\tw_2)^2 \frac{S^2_{1}(BB)}{N_{BB}} - \frac{S^2(\tau_1)}{N}.
\end{eqnarray*}
Under Assumption~\ref{asm::noanticipation}, we have $S^2_{1}(AA)=S^2_{1}(AB)=S^2_{1}(A)$ and $S^2_{1}(BA)=S^2_{1}(BB)=S^2_{1}(B)$. Therefore, 
the variance simplifies to
\begin{eqnarray*}
\left\{\frac{\tw_1^2}{N_{AA}}+\frac{(1-\tw_1)^2}{N_{AB}} \right\} S^2_{1}(A)+\left\{\frac{\tw_2^2}{N_{BA}}+\frac{(1-\tw_2)^2}{N_{BB}} \right\} S^2_{1}(B)-\frac{S^2(\tau_1)}{N}.
\end{eqnarray*}
As a result, the optimal choice is $\tw_1=N_{AA}/(N_{AA}+N_{AB})$ and  $\tw_2=N_{BA}/(N_{BA}+N_{BB})$, which leads to the estimator $\widehat \tau_1$ and the corresponding variance
\begin{eqnarray*}
\var(\widehat{\tau}_1 )=\frac{S^2_{1}(A)}{N_{AA}+N_{AB}}+\frac{S^2_{1}(B)}{N_{BA}+N_{BB}} -\frac{S^2(\tau_1) }{N}.
\end{eqnarray*}

To obtain the covariances between the estimators, we can write the vector $(\widehat{\tau}_1,\widehat{\tau}_2(A),\widehat{\tau}_2(B), \widehat{\tau}^1_2(A), \widehat{\tau}^1_2(B) )^\top$ in the form of~\eqref{eqn::proof-2p4t} with
\begin{eqnarray*}
\bW(AA)&=& \begin{pmatrix}
\frac{N_{AA}}{N_{AA}+N_{AB}} & 0 \\
0 & 1\\
0  &  0 \\
0& 1 \\
0  &  0 
\end{pmatrix}, \quad 
\bW(AB)\ =\ \begin{pmatrix}
\frac{N_{AB}}{N_{AA}+N_{AB}}& 0 \\
0 & -1\\
0  &  0 \\
0  &  0 \\
0  & 1
\end{pmatrix}, \\
\bW(BA)&=& \begin{pmatrix}
-\frac{N_{BA}}{N_{BA}+N_{BB}} & 0 \\
0  &  0 \\
0  & 1 \\
0 & -1\\
0  &  0 
\end{pmatrix}, \quad 
\bW(BB)\ =\ \begin{pmatrix}
-\frac{N_{BB}}{N_{BA}+N_{BB}} & 0 \\
0  &  0 \\
0  & -1 \\
0  &  0 \\
0  & -1
\end{pmatrix}
\end{eqnarray*}
 Then, we can apply  Lemma~\ref{lem::var} to obtain the covariance matrix of $(\widehat{\tau}_1,\widehat{\tau}_2(A),\widehat{\tau}_2(B), \widehat{\tau}^1_2(A), \widehat{\tau}^1_2(B))^\top$.
\QEDB

The variances of all estimators depend on  $S^2_{\tau}$, which involves comparisons of potential outcomes that cannot be observed simultaneously. Define 
\begin{eqnarray*}
\widehat{S}^2_{t}(\os) = \frac{1}{N_{\os}-1} \sum_{i=1}^N \bone(\oS_i=\os) \{Y_{it}-  \widehat{Y}_t(\os)\}^2.
\end{eqnarray*}
We can show that $\widehat{S}^2_{t}(\os)$ is unbiased for $S^2_{t}(\os)$. Because $S^2(\tau)\geq 0$, we can replace $S^2_{t}(\os)$ with $\widehat{S}^2_{t}(\os)$ and omit the term $S^2_{\tau}$ to obtain conservative variance estimators. These estimators become unbiased if the corresponding individual causal effects are constant for all units. 
Moreover, because $S^2_{1}(zA)=S^2_{1}(zB)=S^2_{1}(z)$ for $z=A,B$, we can improve the estimation of $S^2_{1}(z)$ by using the following  pooled variance estimator:
\begin{eqnarray*}
\widehat{S}^2_{1}(z) = \frac{1}{N_{zA}+N_{zB}-2}\left[ \sum_{i=1}^N \bone(\oS_i=zA) \{Y_{i1}-  \widehat{Y}_1(zA)\}^2+ \sum_{i=1}^N \bone(\oS_i=zB) \{Y_{i1}-  \widehat{Y}_1(zB)\}^2 \right].
\end{eqnarray*}

\subsection{Proof of Proposition~\ref{prop::2p4t-nocarryover}}
The unbiasedness follows directly from $\E\{\widehat{Y}_t(z_1z_2)\} =\oY_t(z_1,z_2) $ for $z_1,z_2=A,B$.
\QEDB


\subsection{ Proof of Proposition~\ref{prop::2p4t-invariant}}

The unbiasedness requires the weights in~\eqref{eqn::proof-2p4t}
to satisfy 
\begin{eqnarray}
\label{eqn::proof-2p4t-unbiasedness1}
\bW(AA) \obY(AA)+\bW(AB) \obY(AB)+\bW(BA) \obY(BA)+\bW(BB) \obY(BB)&=& \tau
\end{eqnarray}
for all potential outcomes satisfying Assumptions~\ref{asm::nocarryover}~and~\ref{asm::invarianteffect} with $k=1$.
Under Assumption~\ref{asm::noanticipation}, and Assumptions~\ref{asm::nocarryover}, we simplify $ \oY_1(z_1z_2)$ to $\oY_1(z_1)$, and $ \oY_2(z_1z_2)$ to $\oY_2(z_2)$.
Equation~\eqref{eqn::proof-2p4t-unbiasedness1} becomes
\begin{eqnarray}
\label{eqn::proof-2p4t-unbiasedness2}&&(\tw_{1}(AA)+\tw_{1}(AB))\oY_1(A)+(\tw_{1}(BA)+\tw_{1}(BB))\oY_1(B)\\
\nonumber&&+(\tw_{2}(AA)+\tw_{2}(BA))\oY_2(A)+(\tw_{2}(AB)+\tw_{2}(BB))\oY_2(B) =\oY_1(A)-\oY_1(B).
\end{eqnarray}
Under Assumption~\ref{asm::invarianteffect} with $k=1$, we have $\oY_2(A)=\oY_1(A)-\oY_1(B)+\oY_2(B)$. Therefore, Equation~\eqref{eqn::proof-2p4t-unbiasedness2} becomes
\begin{eqnarray}
\label{eqn::proof-2p4t-unbiasedness3}
&&(\tw_{1}(AA)+\tw_{1}(AB)+w_{2}(AA)+\tw_{2}(BA))\oY_1(A)\\
\nonumber &&+ (\tw_{1}(BA)+\tw_{1}(BB)-\tw_{2}(AA)-\tw_{2}(BA))  \oY_1(B)\\
\nonumber &&+( \tw_{2}(AA)+\tw_{2}(BA)+\tw_{2}(AB)+\tw_{2}(BB))\oY_2(B)\ = \ \oY_1(A)-\oY_1(B)
\end{eqnarray}
The unbiasedness requires
Equation~\eqref{eqn::proof-2p4t-unbiasedness3} to hold for any values of $\oY_1(A)$, $\oY_1(B)$, and $\oY_2(B)$, which implies
\begin{eqnarray*}
\tw_{1}(AA)+\tw_{1}(AB)+w_{2}(AA)+\tw_{2}(BA)&=&1,\\
\tw_{1}(BA)+\tw_{1}(BB)-\tw_{2}(AA)-\tw_{2}(BA)&=&-1,\\
\tw_{2}(AA)+\tw_{2}(AB) + \tw_{2}(BA)+ \tw_{2}(BB)&=&0.
\end{eqnarray*}
These three equations are equivalent to those in the proposition.
From  Lemma~\ref{lem::var}, the variance of the estimator  is 
\begin{eqnarray*}
f(\bw)
&=& \sum_{z_1,z_2=A,B}  \left\{ \frac{\tw^2_{1}(z_1z_2)  S^2_{1}(z_1z_2) }{N_{z_1z_2}}+ \frac{2\tw_{1}(z_1z_2)\tw_{2}(z_1z_2)  S_{12}(z_1z_2) }{N_{z_1z_2}}+\frac{\tw^2_{2}(z_1z_2)  S^2_{2}(z_1z_2) }{N_{z_1z_2}}-\frac{S^2_{\tau}}{N}\right\}\\
&=&\sum_{z_1=A,B} \left\{\frac{\tw^2_{1}(z_1A)}{N_{z_1A}}+\frac{\tw^2_{1}(z_1B)}{N_{z_1B}}\right\} S^2_{1}(z_1)+\sum_{z_2=A,B} \left\{\frac{\tw^2_{2}(Az_2)}{N_{Az_2}}+\frac{\tw^2_{2}(Bz_2)}{N_{Bz_2}}\right\} S^2_{2}(z_2)\\
&& +\sum_{z_1,z_2=A,B}\frac{2\tw_{1}(z_1z_2)w_{2}(z_1z_2)  S_{12}(z_1z_2) }{N_{z_1z_2}}-\frac{S^2(\tau)}{N}.
\end{eqnarray*}
This leads to the constrained optimization problem in Proposition~\ref{prop::2p4t-invariant} since $S^2_{\tau}$ does not depend on the weights.
\QEDB

\subsection{Proof of Proposition~\ref{prop::2p2t}}
 Consider the estimator in~\eqref{eqn::general-est}, where $\mathcal{S}^\textup{obs}=\{AB,BA\}$ in the crossover design with two periods and four sequences.
 We can write the estimator as
  \begin{eqnarray}
  \label{eqn::proof-2p2t}
\bW(AB) \wbY(AB)+\bW(BA) \wbY(BA),
\end{eqnarray}
 where 
 \begin{eqnarray*}
 \bW(AB)&=& (\tw_{1}(AA),\tw_{2}(AB)),\quad \bW(BA)\ =\ (\tw_{1}(BA),\tw_{2}(BA)),
\end{eqnarray*}
and $\obY(\os)=(\obY_1(\os),\obY_2(\os))^\top$ for $\os=AB,BA$.

The unbiasedness requires that 
\begin{eqnarray*}
\bW(AB) \obY(AB)+\bW(BA) \obY(BA)&=& \tau_1.
\end{eqnarray*}
for all potential outcomes satisfying $Y_{i1}(zA)=Y_{i1}(zB)$. This implies 
 $\bW(AB) =(1,0)$ and $\bW(BA)=(-1,0)$.  
 
 For $\tau_2(z_1)$, the unbiasedness requires that 
\begin{eqnarray*}
\bW(AB) \obY(AB)+\bW(BA) \obY(BA)&=& \tau_2(z_1).
\end{eqnarray*}
for all potential outcomes satisfying $Y_{i1}(zA)=Y_{i1}(zB)$.  However, there is no choice of weights $\bW(AB)$ and $\bW(BA)$ that satisfies this equality for all possible  potential outcomes. As a result, no unbiased estimator exists for $\tau_2(z_1)$. Similarly, we can show that  no unbiased estimator exists for $\tau_2^1(z_2)$. \QEDB

\subsection{Proof of proposition~\ref{prop::2p2t-nocarryover}}
We only prove that  $\widehat \tau_2$ is the BLUE for $\tau_2=\tau_2(A)=\tau_2(B)$.  Similar to the proof of Proposition~\ref{prop::2p2t}, we write  write the estimator in~\eqref{eqn::general-est} as that in~\eqref{eqn::proof-2p2t}. 
Under Assumption~\ref{asm::nocarryover} with $k=1$, the unbiasedness requires that 
\begin{eqnarray*}
\bW(AB) \obY(AB)+\bW(BA) \obY(BA)&=& \tau_2.
\end{eqnarray*}
for all potential outcomes satisfying $Y_{i1}(zA)=Y_{i1}(zB)$ and $Y_{i2}(Az)=Y_{i2}(Bz) $. 
This implies   $\bW(AB) =(0,-1)$ and $\bW(BA)=(0,1)$.  \QEDB

\subsection{Proof of Proposition~\ref{prop::2p2t-invariant}}
Consider the estimator in~\eqref{eqn::general-est}, where $\mathcal{S}^\textup{obs}=\{AB,BA\}$ in the crossover design with two periods and two sequences.
 We can write the estimator as
  \begin{eqnarray}
  \label{eqn::proof-2p2t}
\bW(AB) \wbY(AB)+\bW(BA) \wbY(BA),
\end{eqnarray}
 where $\bW(AB)=(\tw_{1}(AA),\tw_{2}(AB))$ and $\bW(BA)=(\tw_{1}(BA),\tw_{2}(BA))$.

The unbiasedness requires that 
\begin{eqnarray*}
\label{eqn::proof-2p4t-1}
\bW(AB) \obY(AB)+\bW(BA) \obY(BA)&=& \tau.
\end{eqnarray*}
for all potential outcomes satisfying Assumptions~\ref{asm::nocarryover}~and~\ref{asm::invarianteffect} with $k=1$.
%
%
This leads to the following constraints in the weights:
\begin{eqnarray*}
\tw_{1}(AB)+\tw_{1}(BA)=0,\quad  \tw_{2}(AB)+\tw_{2}(BA)=0, \quad \tw_{1}(AB)+\tw_{2}(BA)=1.
\end{eqnarray*}
The solution to these equations is
$\tw_{1}(AB)=-w_{1}(BA)=p$ and $\tw_{2}(BA)=-\tw_{2}(AB)=1-p$, where $p$ is any constant.
As a result, any unbiased estimator can be written as
\begin{eqnarray*}
\widehat \tau(p) = p \{\wY_1(AB)- \wY_1(BA)\}+ (1-p) \{\wY_2(BA)- \wY_2(AB)\}.
\end{eqnarray*}
From  Lemma~\ref{lem::var}, the variance of the estimator is
\begin{eqnarray*}
\var\left \{\widehat \tau(p)\right\}&=&\frac{1}{N_{AB}}  \left\{p^2 S^2_{1}(AB)+(1-p)^2 S^2_{2}(AB)-2p(1-p)S_{12}(AB) \right\}\\
&&+ \frac{1}{N_{BA}}  \left\{p^2 S^2_{1}(BA)+(1-p)^2 S^2_{2}(BA)-2p(1-p)S_{12}(BA) \right\}-\frac{1}{N}S^2(\btheta(\bW)),
\end{eqnarray*}
where $S^2(\btheta(\bW))= S^2(\tau)$ under Assumptions~\ref{asm::nocarryover}~and~\ref{asm::invarianteffect} with $k=1$.

Therefore, we only need to find $p$ to minimize
\begin{eqnarray*}
f(p)&=&\frac{1}{N_{AB}}  \left\{p^2 S^2_{1}(AB)+(1-p)^2 S^2_{2}(AB)-2p(1-p)S_{12}(AB) \right\}\\
&&+ \frac{1}{N_{BA}}  \left\{p^2 S^2_{1}(BA)+(1-p)^2 S^2_{2}(BA)-2p(1-p)S_{12}(BA) \right\}.
\end{eqnarray*}
We have 
\begin{eqnarray*}
f'(p)&=&\frac{2}{N_{AB}}  \left\{p S^2_{1}(AB)+(p-1) S^2_{2}(AB)+(2p-1)S_{12}(AB) \right\}\\
&&+\frac{2}{N_{BA}}  \left\{p S^2_{1}(BA)+(p-1) S^2_{2}(BA)+(2p-1)S_{12}(BA) \right\}.
\end{eqnarray*}
Solving $f'(p)=0$, we can obtain 
\begin{eqnarray*}
\widehat{p}&=& \frac{  \frac{S^2_{2}(AB)+S_{12}(AB)}{N_{AB}  }+ \frac{S^2_{2}(BA)+S_{12}(BA)}{N_{BA}  } }{  \frac{S^2_{1}(AB)+S^2_{2}(AB)+2S_{12}(AB)}{N_{AB}  }+ \frac{S^2_{1}(BA)+S^2_{2}(BA)+2S_{12}(BA)}{N_{BA}  } }.
\end{eqnarray*}
Because $f'(p)$ is increasing in $p$, $f(p)$ attains its minimum at $\widehat{p}$. Therefore, $\widehat \tau(p)$ has the smallest variance at $p=\widehat{p}$. \QEDB

\begin{remark}
Senn and Richardson \cite{senn1994first} and Hinkelmann
and Kempthorne \cite{hinkelmann2007design} point out that a crossover design can be viewed as a matched-pair design under a certain reinterpretation. Specifically,
 if we treat each unit at the two time periods as  two units, then each treatment sequence corresponds to a matched pair: one unit assigned to treatment $A$ and the other to treatment $B$. The literature on matched-pair design typically targets the average causal effect across all $2N$ units, which, under our notation and Assumptions~\ref{asm::noanticipation}~and~\ref{asm::nocarryover}, is
\begin{eqnarray*}
\frac{1}{2N}\sum_{i=1}^N \left\{Y_{i1}(A)-Y_{i1}(B)+Y_{i2}(A)-Y_{i2}(B) \right\} &=& \frac{\tau_1+\tau_2}{2}.
\end{eqnarray*}
From Proposition~\ref{prop::2p2t-nocarryover}, the unique unbiased estimator for this quantity is 
\begin{eqnarray*}
\frac{\widehat \tau_1+\widehat \tau_2}{2}&=& \frac{\sum_{i=1}\bm{1}(\oS_i=AB) (Y_{i1}-Y_{i2})}{2N_{AB}}+\frac{\sum_{i=1}\bm{1}(\oS_i=BA) (Y_{i2}-Y_{i1})}{2N_{BA}}.
\end{eqnarray*}
When $N_{AB}=N_{BA}$, this estimator coincides with that used in the matched-pair design  \citep{imai2008variance,ding2024first}. 
More generally, the analysis for crossover design with two periods and two sequences under Assumptions~\ref{asm::noanticipation}~and~\ref{asm::nocarryover} is equivalent to that under the matched-pair design conditional on $N_{AB}$ and $N_{BA}$.
\end{remark}

\subsection{Proof of Theorem~\ref{thm::reg-general}}
We first introduce a working linear model of the observed outcome vector $Y_i$ associated with the regression estimation in~\eqref{eqn::reg-matrix}. 
\begin{eqnarray}
\label{eqn::workinglinear-individual}
Y_{i}&=&   \sum_{\os \in \mathcal{S}^\textup{obs}} \bgamma_{\os} \bm{1}(\oS_i=\os)+\epsilon_{i},
\end{eqnarray}
where $\E(\epsilon_{i})=0$ and $\cov(\epsilon_{i})=\bS^2(\os)$ for units with $\oS_i=\os$.
 With slight abuse of notation, we denote $\Omega_i=\cov(\epsilon_{i})$ and write~\eqref{eqn::workinglinear-individual} in a matrix form
\begin{eqnarray*}
Y_i \ = \  X_i \bgamma+\epsilon_i,
\end{eqnarray*}
where $\cov(\epsilon_i)=\Omega_i $ and $X_i$ represents the regressor matrix for unit $i$ in~\eqref{eqn::workinglinear-individual}. Because $X_i=X_{i'}$ and $\Omega_{i}=\Omega_{i'}$ if $\oS_i=\oS_{i'}$ for any $i,i'$, we can simplify the regressor matrix and weight matrix for units with $\oS_i=\os$ as $X_{\os}$ and $\Omega_{\os}$.
Stacking the individual $Y_i$ into an $NT$-length vector $Y$ leads to the following regression model.
\begin{eqnarray}
\label{eqn::workinglinear}
Y \ = \  X \bgamma+\epsilon, \quad \cov(\epsilon)\ = \ \Omega,
\end{eqnarray}
where 
\begin{eqnarray*}
\Omega= \begin{pmatrix}
\Omega_1 &  0 & \cdots & 0 \\ 
0 & \Omega_2 & \cdots  & 0 \\
\vdots & \vdots& \ddots &0\\
0 & \cdots & 0 &   \Omega_N 
\end{pmatrix},\quad
Y = \begin{pmatrix}
Y_1\\
Y_2\\
\vdots\\
Y_N
\end{pmatrix},\quad X= \begin{pmatrix}
X_1\\
X_2\\
\vdots\\
X_N
\end{pmatrix},\quad \epsilon = \begin{pmatrix}
\epsilon_1\\
\epsilon_2\\
\vdots\\
\epsilon_N
\end{pmatrix},
\end{eqnarray*}
and the coefficients satisfy the constraint $C \bgamma =0$. 

It is important to distinguish between the weighted least squares estimation procedure in~\eqref{eqn::reg-matrix} with restriction $C\bgamma=0$ and the working linear model in~\eqref{eqn::workinglinear}. The formulation in~\eqref{eqn::reg-matrix} defines an estimation strategy, without assuming the linear model is correctly specified. In contrast, the working model in~\eqref{eqn::workinglinear} explicitly assumes a linear relationship, along with the corresponding weight matrices and restrictions. Moreover, the estimation procedure in~\eqref{eqn::reg-matrix} operates under the design-based framework where the regressor matrix is treated as random, whereas the working model follows the classical regression setting in which the regressor matrix is fixed.

Recall from~\eqref{eqn::U} that
\begin{eqnarray*}
U &=& \begin{pmatrix}
X^\top \Omega^{-1}X & C^\top\\
C  &  0
\end{pmatrix}^{-1}
 \ = \ \begin{pmatrix}
U_{11}& U_{12}\\
U_{21}  &  U_{22}
\end{pmatrix}.
\end{eqnarray*}
The following lemma establishes key algebraic properties essential for proving Theorem~\ref{thm::reg-general}.
\begin{lemma}
\label{lem::reg-optimal}
The restricted weighted least squares estimator $\widehat \bgamma^\textup{wls}$ can be expressed as the following linear combination of  $ \widehat \bY(\os)$,
\begin{eqnarray*}
\widehat \bgamma^\textup{wls} =\sum_{\os \in \mathcal{S}^\textup{obs}} \tbW^\textup{wls}(\os) \widehat \bY(\os),
\end{eqnarray*}
where  $\tbW^\textup{wls}(\os)= U_{11} N_{\os}  X_{\os}^\top \Omega_{\os}^{-1}$  for all $\os$.  The matrix $\tbW^\textup{wls}(\os)$ satisfies the following properties.
\begin{itemize}
\item[(a)] For any $\bgamma$ such that $C\bgamma = 0$,  we have
\begin{eqnarray*}
\sum_{\os \in \mathcal{S}^\textup{obs}} \tbW^\textup{wls}(\os) \bgamma_{\os} &=& \bgamma. 
\end{eqnarray*}
\item[(b)] Consider another set of constant matrices $\{ W^\prime_{\ast}(\os): \os \in \mathcal{S}^\textup{obs}\}$ that satisfies the same condition in (a), i.e.,
\begin{eqnarray*}
\sum_{\os \in \mathcal{S}^\textup{obs}} \bW^\prime(\os) \bgamma_{\os} &=& \bgamma \ \text{for any} \ \bgamma \ \text{such that } \ C\bgamma = 0. 
\end{eqnarray*}
Then we must have
\begin{eqnarray*}
\label{opt::reg}   \sum_{\os \in \mathcal{S}^\textup{obs}}   \frac{\tbW^\textup{wls}(\os)  \bS^2(\os)\tbW^\textup{wls}(\os)^\top}{N_{\os}}\leq   \sum_{\os \in \mathcal{S}^\textup{obs}}   \frac{ \bW^\prime(\os)  \bS^2(\os) \bW^{\prime}(\os)^\top}{N_{\os}}.
\end{eqnarray*}

\end{itemize}
\end{lemma}
\noindent {\it Proof of Lemma~\ref{lem::reg-optimal}.}
We first show that $\widehat \bgamma^\textup{wls}$ is a linear combination of  the $ \widehat \bY(\os)$'s for $\os \in \mathcal{S}^\textup{obs}$.
From the proof of Lemma~\ref{lem::rwls-unbias}, we have  $\widehat \bgamma^\textup{wls} = U_{11}X^\top \Omega^{-1}Y$.
Because $X_i=X_{\os}$ and $\Omega_i=\Omega_{\os}$  for units with $\oS_i =\os$, we can write
\begin{eqnarray*}
X^\top \Omega^{-1}Y &=& \sum_{i=1}^N  X_i^\top \Omega_i^{-1} Y_i \ = \ \sum_{\os \in \mathcal{S}^\textup{obs}}  N_{\os}  X_{\os}^\top \Omega_{\os}^{-1} \widehat \bY(\os).
\end{eqnarray*}
As a result, we have 
\begin{eqnarray*}
\widehat \bgamma^\textup{wls} &=& U_{11}X^\top \Omega^{-1}Y \ = \   \sum_{\os \in \mathcal{S}^\textup{obs}} U_{11} N_{\os}  X_{\os}^\top \Omega_{\os}^{-1} \widehat \bY(\os)\ =\ \sum_{\os \in \mathcal{S}^\textup{obs}} \tbW^\textup{wls}(\os) \widehat \bY(\os),
\end{eqnarray*}
where  $\tbW^\textup{wls}(\os)= U_{11} N_{\os}  X_{\os}^\top \Omega_{\os}^{-1}$.

We  then prove Lemma~\ref{lem::reg-optimal}(a).   From Lemma~\ref{lem::rwls-blue-algebra}(a), we have 
\begin{eqnarray}
\label{eqn::reg-optimal-a}
U_{11}X^\top \Omega^{-1} X\gamma &=& \gamma 
\end{eqnarray}
for any $\gamma$ such that $C \gamma=0$. Because $X_i=X_{\os}$ and $\Omega_i=\Omega_{\os}$  for units with $\oS_i =\os$, we can simplify~\eqref{eqn::reg-optimal-a} to 
\begin{eqnarray*}
\sum_{\os \in \mathcal{S}^\textup{obs}} \tbW^\textup{wls}(\os) X_{\os}\gamma &=& \gamma. 
\end{eqnarray*}
 Lemma~\ref{lem::reg-optimal}(a) then follows from $X_{\os}\gamma=\gamma_{\os}$.

%

We then prove Lemma~\ref{lem::reg-optimal}(b). Let  $\hat{A} = U_{11}X^\top \Omega^{-1}$. We can write 
\begin{eqnarray*}
\hat{A}  &=& ( U_{11} X_i \Omega_i^{-1}: i=1,\ldots,N )\ = \ (\tbW^\textup{wls}(\oS_i)/N_{\oS_i}: i=1,\ldots,N), 
\end{eqnarray*}
which is the column concatenation of $\tbW^\textup{wls}(\oS_i)/N_{\oS_i}$ across all units.

Define $A =  (\bW^\prime(\oS_i)/N_{\oS_i}: i=1,\ldots,N)$. Then, 
\begin{eqnarray*}
AX\gamma &=& \gamma
\end{eqnarray*}
for any $\gamma$ such that $C\gamma=0$.
From Lemma~\ref{lem::rwls-blue-algebra}(b), we obtain
\begin{eqnarray*}
\hat{A}\Omega \hat{A}^\top  \leq A\Omega A^\top.  
\end{eqnarray*}
By definition, we have 
\begin{eqnarray*}
\hat{A}\Omega \hat{A}^\top &=& \sum_{i=1}^N \frac{\tbW^\textup{wls}(\oS_i) \Omega_i  \tbW^\textup{wls}(\oS_i)^\top  }{N_{\oS_i}^2}
\ =\ \sum_{\os \in \mathcal{S}^\textup{obs}}   \frac{ \tbW^\textup{wls}(\os) \bS^2(\os)\tbW^\textup{wls}(\os)^\top}{N_{\os}},\\
A\Omega A^\top&=& \sum_{i=1}^N \frac{\bW^\prime(\oS_i) \Omega_i  \bW^\prime(\oS_i)^\top  }{N_{\oS_i}^2}
 \ =\ \sum_{\os \in \mathcal{S}^\textup{obs}}   \frac{\bW^\prime(\os)  \bS^2(\os)\bW^{\prime^\top}(\os)^\top}{N_{\os}}.
\end{eqnarray*}
This completes the proof.

    \QEDB

%
\medskip

We then prove Theorem~\ref{thm::reg-general}. Let $\obY=\{\obY(\os): \os \in \mathcal{S}\}$ denote the collection of all average potential outcomes of interest. 
We only need to show that $ \widehat \bgamma^\textup{wls}$ is the BLUE for $\obY$ under the design-based framework. 
From the setup of Constraints 1 to 3, we have $C \obY =0$.
Under complete randomization, we have $\E \{\widehat \bY(\os)\} = \obY(\os)$, which implies that 
\begin{eqnarray*}
\E(\widehat{\bgamma}^\textup{wls})&=&  \sum_{\os \in \mathcal{S}^\textup{obs}} \tbW^\textup{wls}(\os) \obY(\os).
\end{eqnarray*}
From Lemma~\ref{lem::reg-optimal}(a), we obtain that $\E(\widehat{\bgamma}^\textup{wls})= \obY$.
Consider another unbiased estimator $\widehat{\bm{\xi}}=  \sum_{\os \in \mathcal{S}^\textup{obs}} \bW^\prime(\os) \wY(\os)$. The unbiasedness requires $ \bW^\prime(\os)$ to satisfy that
\begin{eqnarray*}
\sum_{\os \in \mathcal{S}^\textup{obs}}  \bW^\prime(\os) \bgamma_{\os} &=& \bgamma \ \text{for any} \ \bgamma \ \text{such that } \ C\bgamma = 0. 
\end{eqnarray*}
From Lemma~\ref{lem::var}, we have 
\begin{eqnarray*}
\cov (\widehat{\bgamma}^\textup{wls}) &=&\sum_{\os \in \mathcal{S}^\textup{obs}}  \frac{\tbW^\textup{wls}(\os) \bS^2(\os)\tbW^\textup{wls}(\os) ^\top}{N_{\os}} -\frac{1}{N}\bS^2(\obY(\cdot)),\\
\cov (\widehat{\bm{\xi}}) &=&\sum_{\os \in \mathcal{S}^\textup{obs}}  \frac{\bW^\prime(\os) \bS^2(\os) \bW^{\prime}(\os)^\top}{N_{\os}} -\frac{1}{N}\bS^2(\obY(\cdot)),
\end{eqnarray*}
where $\obY(\cdot) =\{\obY(\os): \os \in \mathcal{S}\} $  and $\bS^2(\obY(\cdot))$ is the covariance matrix of the individual potential outcomes $\{\bY_i(\os): \os \in \mathcal{S}\}$.
From Lemma~\ref{lem::reg-optimal}(b), we have $\var (\widehat{\bgamma}^\textup{wls}) \leq \var (\widehat{\bm{\xi}}) $. \QEDB

%
%
%
%

\subsection{Proof of Corollary~\ref{thm::inference}}
From Theorem~\ref{thm::reg-general}, $\widehat \btheta(\bW)$ can be written as a linear combination of $\widehat \bY(\os)$. The result then follows from Lemma~\ref{lem::clt}. \QEDB

\subsection{Proof of Theorem~\ref{thm::reg-omegahat}}
From Proposition 3 in \cite{li2017general}, $ \widehat{\bS}^2(\os)- \bS^2(\os)$ converges to zero in probability.
We introduce the following lemma to facilitate the analysis of the inverse of the block matrix
\begin{eqnarray*}
\begin{pmatrix}
X^\top \widehat \Omega^{-1}X & C^\top\\
C  &  0
\end{pmatrix}.
\end{eqnarray*}
Since $X^\top \widehat \Omega^{-1}X $ may not be invertible, the standard block matrix inverse formula using the Schur complement is not directly applicable. We present an alternative matrix inversion formula that does not require the invertibility of the upper-left block.
\begin{lemma}
\label{lem::matrixinverse}
Suppose the following block matrix is invertible
\begin{eqnarray*}
H&=& \begin{pmatrix}
A & B\\
C & 0
\end{pmatrix},
\end{eqnarray*}
where $A$ is a square matrix and  $B$ is full column rank.  We have 
\begin{eqnarray*}
H^{-1} &=&\begin{pmatrix}
S^{-1}  & S^{-1} A^\top B(B^\top B)^{-1} \\
(B^\top B)^{-1} B^\top A & (B^\top B)^{-1}+(B^\top B)^{-1}B^\top AS^{-1} A^\top B (B^\top B)^{-1}
\end{pmatrix} \begin{pmatrix}
A^\top & C^\top\\
B^\top & 0
\end{pmatrix},
\end{eqnarray*}
where $S = A^\top A + C^\top C - A^\top B(B^\top B)^{-1}B^\top A $.
\end{lemma}
{\it Proof of Lemma~\ref{lem::matrixinverse}.}  From $(H^T H)^{-1} H^T H = I$ and the invertibility of $H$, we have
\begin{eqnarray*}
H^{-1} &=& (H^\top H)^{-1} H^\top \ = \ \begin{pmatrix}
A^\top A+C^\top C & A^\top B\\
B^\top A  & B^\top B
\end{pmatrix}^{-1} \begin{pmatrix}
A^\top & C^\top\\
B^\top & 0
\end{pmatrix}.
\end{eqnarray*}
Because $B$ is full column rank, $B^\top B$ is invertible. The formula then follows from the matrix inversion formula via the Schur complement. \QEDB

\begin{lemma}
\label{lem::V-consistency}
Let $\widehat \Omega= \textup{diag}(\widehat{\bS}^2(\os): \os \in \mathcal{S}^\textup{obs})$ denote the consistent estimator for true weight matrix $\Omega$. Recall from~\eqref{eqn::U}~and~\eqref{eqn::U-hat} that
\begin{eqnarray*}
U &=& \begin{pmatrix}
X^\top \Omega^{-1}X & C^\top\\
C  &  0
\end{pmatrix}^{-1} \ = \ \begin{pmatrix}
 U_{11}&U_{12}\\
U_{21}  &   U_{22}
\end{pmatrix}, \\
 \widehat U &=& \begin{pmatrix}
X^\top \widehat \Omega^{-1}X & C^\top\\
C  &  0
\end{pmatrix}^{-1} \ = \ \begin{pmatrix}
\widehat U_{11}& \widehat U_{12}\\
\widehat U_{21}  &  \widehat U_{22}
\end{pmatrix}.
\end{eqnarray*}
If $N_{\os}/N$ has a positive limiting value for each $\os$, then $N \widehat U_{11}= N U_{11} +o_{\mathbb{P}}(1)$.
\end{lemma}
{\it Proof of Lemma~\ref{lem::V-consistency}.}
We have 
\begin{eqnarray}
\label{eqn::V-consistency1}
\frac{1}{N}X^\top \widehat \Omega^{-1}X &=& \frac{1}{N} \sum_{\os \in \mathcal{S}^\textup{obs}} N_{\os} X_{\os}^\top  \widehat \Omega_{\os}  X_{\os}\ =\ \frac{1}{N}X^\top \Omega^{-1}X +o_{\mathbb{P}}(1).
\end{eqnarray}
From Lemma~\ref{lem::matrixinverse}, we have 
\begin{eqnarray*}
\widehat U_{11}&=& \widehat S^{-1} (X^\top \widehat \Omega^{-1}X)^\top +\widehat S^{-1} (X^\top \widehat \Omega^{-1}X)^\top  C^\top(C C^\top)^{-1} C,\\
U_{11}&=&  S^{-1} (X^\top \Omega^{-1}X)^\top + S^{-1} (X^\top \Omega^{-1}X)^\top  C^\top (C C^\top)^{-1} C, 
\end{eqnarray*}
where 
\begin{eqnarray*}
\widehat S &=&(X^\top \widehat \Omega^{-1}X)^\top(X^\top \widehat \Omega^{-1}X) + C^\top C-(X^\top \widehat \Omega^{-1}X)^\top C^\top (C C^\top)^{-1} C (X^\top \widehat \Omega^{-1}X),\\
S &=&(X^\top \Omega^{-1}X)^\top(X^\top \Omega^{-1}X) + C^\top C-(X^\top \Omega^{-1}X)^\top C^\top (C C^\top)^{-1} C (X^\top\Omega^{-1}X).
\end{eqnarray*}
Using~\eqref{eqn::V-consistency1}, we obtain $N \widehat U_{11}= N U_{11} +o_{\mathbb{P}}(1)$. \QEDB

\medskip

We then prove Theorem~\ref{thm::reg-omegahat}.
The consistency of the sample variances follows directly from Lemma~\ref{lem::ci}. 
From the proof of Theorem~\ref{thm::reg-general}(a), we can write
\begin{eqnarray*}
\widehat \bgamma^\textup{wls} &=&\sum_{\os \in \mathcal{S}^\textup{obs}} \tbW^\textup{wls}(\os)  \widehat \bY(\os),\\
\widehat{\bgamma}^\textup{fwls} &=&\sum_{\os \in \mathcal{S}^\textup{obs}} \tbW^\textup{fwls}(\os)  \widehat \bY(\os),
\end{eqnarray*}
where  $\tbW^\textup{wls}(\os)= U_{11} N_{\os}  X_{\os}^\top \Omega_{\os}^{-1}$ and $\tbW^\textup{fwls}(\os) = \widehat U_{11} N_{\os}  X_{\os}^\top \widehat \Omega_{\os}^{-1}$ satisfy
\begin{eqnarray*}
\sum_{\os \in \mathcal{S}^\textup{obs}} \tbW^\textup{wls}(\os) \obY(\os) &=&\sum_{\os \in \mathcal{S}^\textup{obs}} \tbW^\textup{fwls}(\os)  \obY(\os) \ = \  \obY.
\end{eqnarray*}
This yields
\begin{eqnarray}
\nonumber \sqrt{N}(\widehat{\bgamma}^\textup{fwls}-\widehat \bgamma^\textup{wls})&=&\sqrt{N}\left\{\sum_{\os \in \mathcal{S}^\textup{obs}} \tbW^\textup{fwls}(\os) \widehat \bY(\os)-\sum_{\os \in \mathcal{S}^\textup{obs}} \tbW^\textup{wls}(\os)  \widehat \bY(\os)\right\}\\
\nonumber&=&\sqrt{N}\left[\sum_{\os \in \mathcal{S}^\textup{obs}}\tbW^\textup{fwls}(\os)  \{\widehat \bY(\os) - \obY(\os)\} -\sum_{\os \in \mathcal{S}^\textup{obs}}\tbW^\textup{wls}(\os)  \{\widehat \bY(\os) - \obY(\os)\}\right]\\
\label{eqn::omegahat}&=&\sum_{\os \in \mathcal{S}^\textup{obs}} (\tbW^\textup{fwls}(\os) - \tbW^\textup{wls}(\os) ) \sqrt{N} \{\widehat \bY(\os) - \obY(\os)\}.
\end{eqnarray}
 Using Lemma~\ref{lem::V-consistency}, we obtain $\tbW^\textup{fwls}(\os) - \tbW^\textup{wls}(\os) = o_{\mathbb{P}}(1)$.
From Corollary~\ref{thm::inference}, $ \sqrt{N} \{\widehat \bY(\os) - \obY(\os)\}=O_{\mathbb{P}}(1)$. Applying these two results in~\eqref{eqn::omegahat} leads to $\sqrt{N}(\widehat{\bgamma}^\textup{fwls}-\widehat \bgamma^\textup{wls})=o_{\mathbb{P}}(1)$. \QEDB

\subsection{Proof of Theorem~\ref{thm::reg-variance}}
The EHW variance estimator in~\eqref{eqn::fwls-ehw} can be written as 
\begin{eqnarray*}
\widehat V_{\textsc{EHW}}&=& \widehat  U_{11}\left( \sum_{i=1}^N X_i^\top \widehat  \Omega_i^{-1} \widehat \Sigma_i \widehat  \Omega_i^{-1}X_i \right) \widehat  U_{11}. 
\end{eqnarray*}
Because $\widehat \Omega_i =\widehat \Omega_{i'}$  if $\oS_i =\oS_i'$ for any $i,i'$, we can write
\begin{eqnarray*}
\widehat V_{\textsc{EHW}}&=&  \sum_{\os \in \mathcal{S}^\textup{obs}}\left\{ \widehat U_{11}  X_{\os}^\top \widehat \Omega_{\os}^{-1} \left( \sum_{i:\oS_i=\os}  \widehat \Sigma_i\right)   \widehat \Omega_{\os}^{-1} X_{\os} \widehat U_{11}\right\}\\
&=& \sum_{\os \in \mathcal{S}^\textup{obs}}\left\{ \tbW^\textup{fwls}(\os) \left( \frac{1}{N_{\os}^2}\sum_{i:\oS_i=\os}  \widehat \Sigma_i\right)  \tbW^\textup{fwls}(\os)^\top \right\}.
\end{eqnarray*}
For each $\os$, we have 
\begin{eqnarray}
\nonumber && \frac{1}{N_{\os}-1} \sum_{i:\oS_i=\os}  \widehat \Sigma_i - \widehat{\bS}^2(\os) \\
\nonumber &=& \frac{1}{N_{\os}-1} \sum_{i:\oS_i=\os} \left\{ (Y_i -\widehat{\bgamma}_{\os}^\textup{fwls} )(Y_i -\widehat{\bgamma}_{\os}^\textup{fwls} )^\top -  (Y_i -\widehat \bY(\os)) (Y_i -\widehat \bY(\os))^\top  \right\}\\
\nonumber &=& \frac{N_{\os}}{N_{\os}-1}  \{\widehat{\bgamma}_{\os}^\textup{fwls} - \widehat \bY(\os)\}\{\widehat{\bgamma}_{\os}^\textup{fwls} - \widehat \bY(\os)\}^\top\\
\label{eqn::ehw-meatest}&=& o_{\mathbb{P}}(1),
\end{eqnarray}
where the last equality follows from $\widehat{\bgamma}_{\os}^\textup{fwls} - \obY(\os)=o_{\mathbb{P}}(1) $ and $ \widehat \bY(\os)- \obY(\os)=o_{\mathbb{P}}(1)$.
Combining~\eqref{eqn::ehw-meatest} with $ \widehat{\bS}^2(\os) = \bS^2(\os)+o_{\mathbb{P}}(1)$, we obtain that 
\begin{eqnarray*}
\frac{1}{N_{\os}-1} \sum_{i:\oS_i=\os}  \widehat \Sigma_i &=& \bS^2(\os) +o_{\mathbb{P}}(1).
\end{eqnarray*}
From the proof of  Theorem~\ref{thm::reg-omegahat}, we know  $\tbW^\textup{fwls}(\os) - \tbW^\textup{wls}(\os) = o_{\mathbb{P}}(1)$. As a result, we have 
\begin{eqnarray*}
\widehat V_{\textsc{EHW}}&=& \sum_{\os \in \mathcal{S}^\textup{obs}}  \frac{\tbW^\textup{wls}(\os) \bS^2(\os)\tbW^\textup{wls}(\os)^\top}{N_{\os}} +  o_{\mathbb{P}}(N^{-1}).
\end{eqnarray*}
\QEDB

\subsection{Proof of Theorem~\ref{thm::reg-general-constantweight}}
We use $\widehat \bgamma_\ast^\textup{wls2}$ to denote the restricted weighted least squares estimator with weight matrix $\Gamma$.
Similar to the Proof of Theorem~\ref{thm::reg-general}, we can obtain that 
\begin{eqnarray*}
\widehat \bgamma_\ast^\textup{wls2} &=& U_{11}^\textup{wls2}X^\top \Gamma^{-1}Y \ = \   \sum_{\os \in \mathcal{S}^\textup{obs}} U_{11}^\textup{wls2} N_{\os}  X_{\os}^\top \Gamma_{\os}^{-1} \widehat \bY(\os)\ =\ \sum_{\os \in \mathcal{S}^\textup{obs}} \tbW^\textup{wls2}(\os) \widehat \bY(\os),
\end{eqnarray*}
where   $\tbW^\textup{wls2}(\os)= U_{11} N_{\os}  X_{\os}^\top \Gamma_{\os}^{-1}$.
 From Lemma~\ref{lem::rwls-blue-algebra}(a), we have 
\begin{eqnarray*}
U_{11}^\textup{wls2}X^\top \Gamma^{-1} X\gamma &=& \gamma 
\end{eqnarray*}
for any $\gamma$ such that $C \gamma=0$. Because $X_i=X_{\os}$ and $\Gamma_i=\Gamma_{\os}$  for units with $\oS_i =\os$, we have 
\begin{eqnarray*}
\sum_{\os \in \mathcal{S}^\textup{obs}} \tbW^\textup{wls2}(\os) \gamma_{\os} &=& \gamma. 
\end{eqnarray*}
Under complete randomization, we have  $ \E\{\widehat \bY(\os)\} = \obY(\os) $. Therefore,
\begin{eqnarray*}
\E(\widehat{\bgamma}^\textup{wls2})&=&  \sum_{\os \in \mathcal{S}^\textup{obs}} \tbW^\textup{wls2}(\os) \obY(\os)\ = \  \obY.
\end{eqnarray*}
 From Lemma~\ref{lem::ci}, we can obtain the following variance formula
 \begin{eqnarray*}
\cov \{ \widehat{\theta}(\tbW^\textup{wls2})\} &=&\sum_{\os \in \mathcal{S}^\textup{obs}}  \frac{\tbW^\textup{wls2}(\os) \bS^2(\os)\tbW^\textup{wls2}(\os)^\top}{N_{\os}} -\frac{1}{N}S^2(\btheta(\bW)).
\end{eqnarray*}
 Similar to that of Theorem~\ref{thm::reg-variance}, we can obtain the following asymptotically conservative EHW variance estimator:
 \begin{eqnarray*}
\widehat{U}_{11}X^\top \Gamma^{-1}  \widehat{\Sigma} \Gamma^{-1}X \widehat{U}_{11}.
\end{eqnarray*}

\QEDB

\subsection{Proof of Theorem~\ref{thm::reg-general-average}}
Similar to the proof of Theorem~\ref{thm::reg-general}, we can show the unbiasedness of the estimator and write it as 
\begin{eqnarray*}
\widehat{\theta}(\tbW^\textup{wls}) &=& \sum_{\os \in \mathcal{S}^\textup{obs}}\tbW^\textup{wls}(\os) \wY(\os).
\end{eqnarray*}
From Lemma~\ref{lem::var}, we can obtain its variance 
\begin{eqnarray*}
\cov \{ \widehat{\theta}(\tbW^\textup{wls})\} &=&\sum_{\os \in \mathcal{S}^\textup{obs}}  \frac{\tbW^\textup{wls}(\os) \bS^2(\os)\tbW^\textup{wls}(\os)^\top}{N_{\os}} -\frac{1}{N}S^2(\btheta(\tbW^\textup{wls}) ),
\end{eqnarray*}
where
\begin{eqnarray*}
S^2(\btheta(\tbW^\textup{wls}) ) =\frac{1}{N-1} \sum_{i=1}^N \{\btheta_i(\tbW^\textup{wls})-\btheta(\bW)\}\{\btheta_i(\tbW^\textup{wls})-\btheta(\bW)\}^\top
\end{eqnarray*}
is the variance of the individual causal effect $\btheta_i(\tbW^\textup{wls}) =\sum_{\os \in \mathcal{S}^\textup{obs}} \tbW^\textup{wls}(\os)  \bY_i(\os) $. \QEDB

Note that without the individual-level versions of the assumptions, $\theta_i(\tbW^\textup{wls})$ no longer equals $\theta_i(\bW)$. However, because $S^2(\btheta(\tbW^\textup{wls}) )$ is positive semi-definite, we can still use $\widehat V_{\textsc{EHW}}$ as an asymptotically conservative variance estimator.
 
\subsection{Proof of Theorem~\ref{thm::var-compare}}
From the  proof of Theorem~\ref{thm::reg-variance} we have
\begin{eqnarray*}
\widehat V_{\textsc{EHW}}&=& \sum_{\os \in \mathcal{S}^\textup{obs}}  \frac{\tbW^\textup{wls}(\os) \bS^2(\os)\tbW^\textup{wls}(\os)^\top}{N_{\os}} +  o_{\mathbb{P}}(N^{-1}),\\
\widehat V_{\textsc{EHW}}^\textup{wls2}&=& \sum_{\os \in \mathcal{S}^\textup{obs}}  \frac{\tbW^\textup{wls2}(\os) \bS^2(\os)\tbW^\textup{wls2}(\os)^\top}{N_{\os}} +  o_{\mathbb{P}}(N^{-1}).
\end{eqnarray*}
Therefore, $\widehat V_{\textsc{EHW}}^\textup{wls2}-\widehat V_{\textsc{EHW}}=C+o_{\mathbb{P}}(N^{-1})$, where
\begin{eqnarray*}
C&=& \sum_{\os \in \mathcal{S}^\textup{obs}}  \frac{\tbW^\textup{wls2}(\os) \bS^2(\os)\tbW^\textup{wls2}(\os)^\top}{N_{\os}} -\sum_{\os \in \mathcal{S}^\textup{obs}}  \frac{\tbW^\textup{wls}(\os) \bS^2(\os)\tbW^\textup{wls}(\os)^\top}{N_{\os}}. 
\end{eqnarray*}
Lemma~\ref{lem::reg-optimal} implies that $C$ is positive semi-definite. \QEDB

\subsection{Proof of Theorem~\ref{thm::rwls-testing}}
We first consider testing the assumptions based on the restricted weighted least squares estimator with the true weight matrix $S^2(\os)$. In this case, we study the quantity 
 $$X^\top \Omega^{-1} (Y-X \widehat{\bgamma}^\textup{wls}). $$ 
Similar to the proof of Lemma~\ref{lem::rwls-testing}, we have 
\begin{eqnarray}
\nonumber X^\top \Omega^{-1} (Y-X \widehat{\bgamma}^\textup{wls} ) &=&X^\top \Omega^{-1}(Y - XU_{11}X^\top \Omega^{-1}Y)\\
\nonumber&=&X^\top \Omega^{-1} Y - X^\top\Omega^{-1} X U_{11}X^\top \Omega^{-1}Y\\
\nonumber&=&C^\top U_{21} X^\top \Omega^{-1}Y\\
\label{eqn::testingproof1}  &=&C^\top  U_{21} (    N_{\os} \Omega_{\os}^{-1} \wbY(\os): \os  \in \mathcal{S})^\top,
\end{eqnarray}
where the last equality follows from~\eqref{eqn::comp2}, and we set $N_{\os} \Omega_{\os}^{-1} \wbY(\os)=0$ for $\os \notin  \mathcal{S}^\textup{obs}$.
Equation~\eqref{eqn::testingproof1} shows that $X^\top \Omega^{-1} (Y - X \widehat{\bgamma}^\textup{wls})$ is a linear transformation of $(\wbY(\os) : \os \in \mathcal{S}^\textup{obs})^\top$. Therefore, under the conditions in Corollary~\ref{thm::inference}, it satisfies a central limit theorem.

From~\eqref{eqn::testingproof1}, we have 
\begin{eqnarray*}
\E\{X^\top \Omega^{-1} (Y-X \widehat{\bgamma}^\textup{wls} ) \} &=& C^\top  U_{21} (    N_{\os} \Omega_{\os}^{-1} \oY(\os): \os  \in \mathcal{S})^\top\\
&=& C^\top  U_{21} X^\top \Omega^{-1}X \obY\\
&=& -C^\top U_{22}C\obY\\
&=&0,
\end{eqnarray*}
where the second equality follows from~\eqref{eqn::comp1}, the third equality follows from~\eqref{eqn::inverseU2}, and the last equality follows from $C\obY=0$.

Because $X^\top \Omega^{-1} (Y-X \widehat{\bgamma}^\textup{wls} ) $ can be written as a linear transformation of $(\wbY(\os) : \os \in \mathcal{S}^\textup{obs})^\top$, we can construct, similarly to the proof of Theorem~\ref{thm::reg-variance}, a conservative estimator for its covariance matrix:
\begin{eqnarray*}
 C^\top U_{21} X^\top \Omega^{-1} \widehat{\Sigma} \Omega^{-1} X U_{12} C.
\end{eqnarray*}
It then follows that the quadratic form
\begin{eqnarray*}
(Y-X\widehat{\bgamma}^\textup{wls})^\top  \Omega^{-1}  X  (C^\top U_{21} X^\top \Omega^{-1} \widehat{\Sigma}  \Omega^{-1} X U_{12}C)^+  X^\top \Omega^{-1}   (Y-X \widehat{\bgamma}^\textup{wls}) 
\end{eqnarray*}
is asymptotically stochastically dominated by a chi-squared distribution with degrees of freedom
\begin{eqnarray*}
\text{rank} (C^\top U_{21} X^\top \Omega^{-1} \widehat{\Sigma}  \Omega^{-1} X U_{12}C)&=& \text{rank}(XU_{12}).
\end{eqnarray*}

Finally, from Theorem~\ref{thm::reg-variance}, replacing $U_{21}$ and $\Omega$ with their consistent estimators $\widehat{U}_{21}$ and $\widehat{\Omega}$ does not affect the asymptotic distribution. This completes the proof. \QEDB

\end{document}